\documentclass[prl,twocolumn,superscriptaddress,float,aps]{revtex4-2}
\usepackage{graphicx,amsfonts,amssymb,amsmath,hyperref,hypcap,enumerate}
\usepackage{color}
\usepackage{ulem}

\begin{document}
\title{Competing Vortex Topologies in Iron-based Superconductors}
\author{Lun-Hui Hu}
\affiliation{Department of Physics and Astronomy, The University of Tennessee, Knoxville, Tennessee 37996, USA}
\affiliation{Institute for Advanced Materials and Manufacturing, The University of Tennessee, Knoxville, Tennessee 37920, USA}
\affiliation{Department of Physics, The Pennsylvania State University, University Park, Pennsylvania 16802, USA}

\author{Xianxin Wu}
\affiliation{CAS Key Laboratory of Theoretical Physics, Institute of Theoretical Physics, Chinese Academy of Sciences, Beijing 100190, China}
\affiliation{Max-Planck-Institut f\"ur Festk\"orperforschung, Heisenbergstrasse 1, D-70569 Stuttgart, Germany}

\author{Chao-Xing Liu}
\affiliation{Department of Physics, The Pennsylvania State University, University Park, Pennsylvania 16802, USA}

\author{Rui-Xing Zhang}
\email{ruixing@utk.edu}
\affiliation{Department of Physics and Astronomy, The University of Tennessee, Knoxville, Tennessee 37996, USA}
\affiliation{Department of Materials Science and Engineering, The University of Tennessee, Knoxville, Tennessee 37996, USA}
\affiliation{Institute for Advanced Materials and Manufacturing, The University of Tennessee, Knoxville, Tennessee 37920, USA}

\date{\today}

\begin{abstract}
In this work, we establish a new theoretical paradigm for vortex Majorana physics in the recently discovered topological iron-based superconductors (tFeSCs). While tFeSCs are widely accepted as an exemplar of topological insulators (TIs) with intrinsic $s$-wave superconductivity, our theory implies that such common belief could be oversimplified. Our main finding is that the normal-state bulk Dirac nodes, usually ignored in TI-based vortex Majorana theories for tFeSCs, will play a key role of determining the vortex state topology. In particular, the interplay between TI and Dirac nodal bands will lead to multiple competing topological phases for a superconducting vortex line in tFeSCs, including an unprecedented hybrid topological vortex state that carries both Majorana bound states and a gapless dispersion. Remarkably, this exotic hybrid vortex phase generally exists in the vortex phase diagram for our minimal model for tFeSCs and directly relevant to tFeSC candidates such as LiFeAs.When the four-fold rotation symmetry is broken by vortex-line tilting or curving,
the hybrid vortex gets topologically trivialized and becomes Majorana-free, which could
explain the puzzle of ubiquitous trivial vortices observed in LiFeAs. Origin of Majorana signal in other tFeSC candidates such as FeTe$_x$Se$_{1-x}$ and CaKFe$_4$As$_4$ are also interpreted within our theory framework. Our theory sheds new light on theoretically understanding and experimentally engineering Majorana physics in high-temperature iron-based systems.
\end{abstract}

\maketitle

{\it Introduction} - 
The inherent resilience of a topological quantum computer (TQC) makes it one of the most promising paradigms for processing information in a fully quantum mechanical manner~\cite{kitaev_ap_2003,nayak_rmp_2008,sarma_npjqi_2015,elliott_rmp_2015}. 
This inspiring concept of TQC has motivated an intensive experimental search for Majorana zero modes~\cite{mourik_science_2012,rokhinson_np_2012,williams_prl_2012,das_np_2012,deng_nl_2012,finck_prl_2013,lee_nn_2014,nadj_science_2014,xu_prl_2015,albrecht_nature_2016,sun_prl_2016,deng_science_2016,lv_sb_2017,feldman_np_2017,jeon_science_2017,ren_nature_2019,jack_science_2019,fornieri_nature_2019}, the simplest and most feasible building block for TQCs.
While a ``smoking gun" for Majorana modes is still lacking, the recent advent of topological iron-based superconductors~\cite{zhang_science_2018,shi_sb_2017,zhang_np_2019,peng_prb_2019,kong_np_2019,zhu_science_2020,wang_science_2018,machida_nm_2019,liu_prx_2018,chen_cpl_2019,liu_nc_2020,kong_nc_2021} (tFeSCs) sheds light on resolving this long-sought Majorana mystery~\cite{zhang_prl_2019a,zhang_prl_2019b,yan_prl_2018,wang_prl_2018,hu_cp_2019,kawakami_prb_2019,wu_nsr_2021,wu_prx_2020,gray_nl_2019,wu_book_2021,zhang_prl_2021,kheirkhah_prb_2021,qin_arxiv_2021,kheirkhah_arxiv_2021,barik2021signatures}. 
Remarkably, the normal-state band structure of tFeSCs naturally contains two crucial ingredients: (i) a topological insulator (TI) part that provides helical Dirac surface state~\cite{wang_prb_2015,xu_prl_2016,wu_prb_2016}; 
(ii) cylindrical Fermi surfaces that generates intrinsic bulk superconductivity~\cite{kamihara_jacs_2008,fang_prb_2008,hanaguri_science_2010,paglione_np_2010,hirschfeld_rpp_2011,stewart_rmp_2011,wang_science_2011,chubukov_arcmp_2012,chubukov_pt_2015}, which together resemble the well-known Majorana paradigm proposed by Fu and Kane~\cite{fu_prl_2008}. 
Indeed, strong evidence of field-induced vortex Majorana bound states (vMBSs) has been observed in multiple iron-based systems~\cite{wang_science_2018,machida_nm_2019,liu_prx_2018,chen_cpl_2019,liu_nc_2020,kong_nc_2021}.

The tFeSCs, however, are far from being well-understood. For example, unlike other tFeSC candidates, LiFeAs possesses no vMBS signal in any of its free vortices, even though the Fermi level is around the TI gap~\cite{kong_nc_2021}.
This counter-intuitive vortex physics clearly deviates from the naive expectation from the Fu-Kane paradigm, thus calling for a new theoretical interpretation. Meanwhile, most tFeSCs additionally host a pair of massless bulk Dirac nodes in their normal states~\cite{zhang_np_2019,day2021three}, which are energetically above the TI bands with an energy separation dubbed $\delta_{so}$. 
It has been predicted that a Dirac semimetal (DSM), if going superconducting, would feature gapless, vMBS-free magnetic vortices~\cite{konig_prl_2019,qin_prl_2019}, in contrast to the vMBS physics of a superconducting TI~\cite{fu_prl_2008,hosur_prl_2011,chiu_prb_2011,chiu_prl_2012,li_sr_2014,hu_prb_2016,yan_prl_2020,chiu_sa_2020,ghazaryan_prb_2020,li_arxiv_2021}. 
Notably, existing studies on tFeSCs generally adopt the presumption of an infinite $\delta_{so}$ limit, so that only TI or DSM bands are independently studied for simplicity~\cite{qin_sb_2019}. 
However, such presumption remains unjustified for some tFeSC candidates (e.g.~LiFeAs) where $\delta_{so}$ can be as small as 10 meV~\cite{zhang_np_2019,Borisenko2016}.This raises an important open question for the topological nature of vortices in tFeSCs, especially when both TI and DSM bands are highly entangled.

\begin{figure}[t]
	\includegraphics[width=0.48\textwidth]{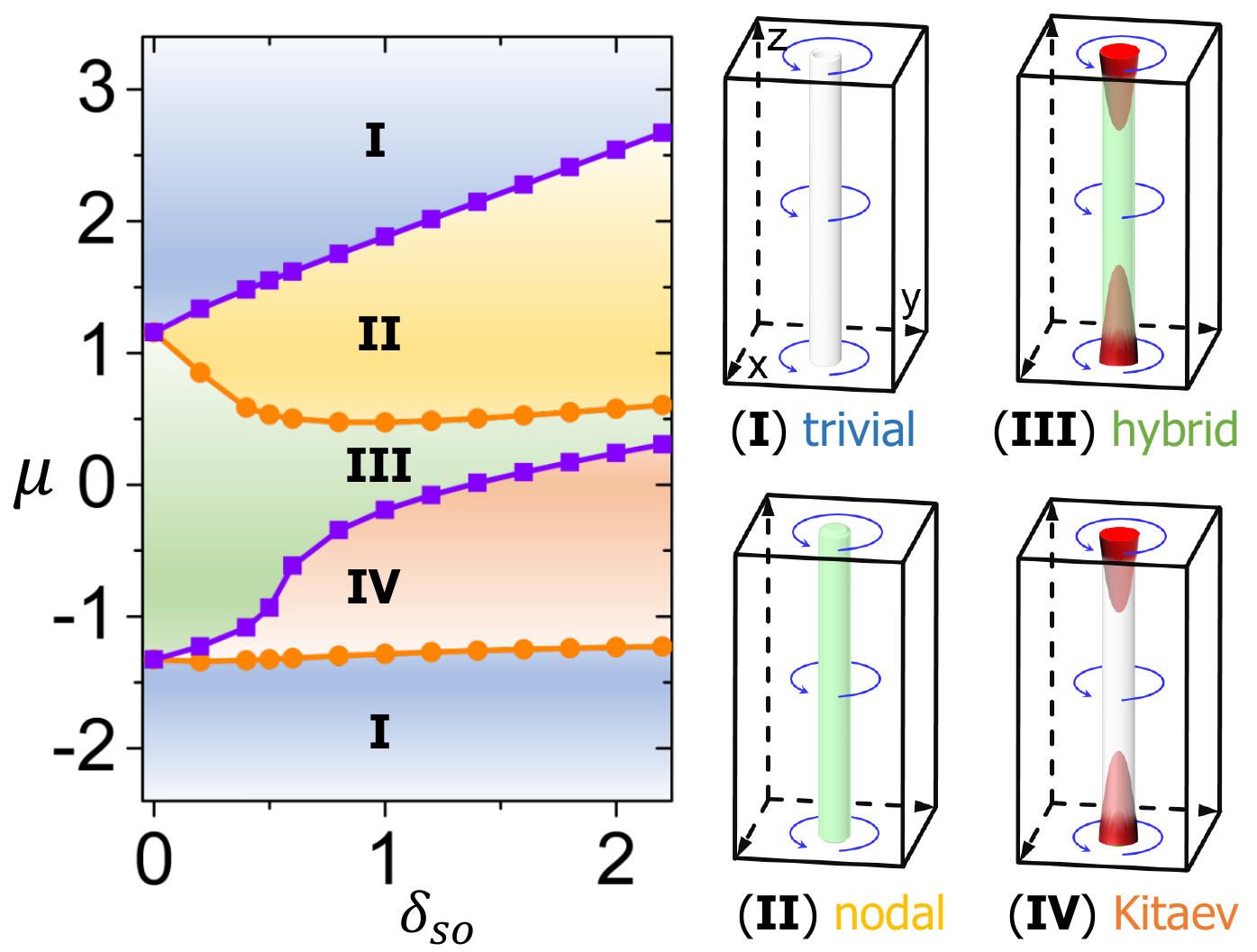}
	\caption{
		The vortex topological phase diagram for a minimal tFeSC model is mapped out as a function of chemical potential $\mu$ and $\delta_{so}$, the separation energy between TI gap and bulk Dirac nodes. It contains four phases: (I) trivial vortex; (II) nodal vortex; (III) hybrid vortex with both 0D end-localized vMBSs (red cone) and 1D gapless channels. (IV)  Kitaev vortex with 0D end-localized vMBSs.
	}
	\label{fig1}
\end{figure}

In this work, we provide a new theoretical paradigm for understanding vortex topological physics in general tFeSCs. To fully incorporate both TI and DSM physics, a minimal 6-band model is constructed to capture the key topological ingredients of general tFeSCs~\cite{xu_prl_2016,zhang_np_2019}. 
For the first time, we have identified the emergence of four competing and topologically distinct vortex states in the vortex phase diagram of tFeSCs, as shown in Fig.~\ref{fig1}. Remarkably, a new exotic ``hybrid topological" vortex phase manifests as the most probable vortex state for small $\delta_{so}$ systems, which features both well-defined vMBS and a one-dimensional (1D) nodal band structure along its $k_z$ dispersion. The stability of the hybrid vortex phase relies on the protection of four-fold rotation symmetry $C_4$, and upon $C_4$ breaking, the hybrid vortex can be easily trivialized to become vMBS-free. This offers a natural explanation for the observed missing-Majorana puzzle in LiFeAs~\cite{kong_nc_2021}. Applications of our theory to other tFeSCs and new experimental signatures are also discussed.

{\it Model Hamiltonian} - 
We start by defining a minimal ${\bf k \cdot p}$ Hamiltonian that captures the key low-energy topological physics of general tFeSCs~\cite{xu_prl_2016,zhang_np_2019}. We first consider a normal-state basis $\Psi_{\mathbf{k}} =  (\vert p_z,\uparrow\rangle, \vert p_z,\downarrow\rangle, \vert d_{xz+iyz},\downarrow\rangle, \vert d_{xz-iyz},\uparrow\rangle, \vert d_{xz+iyz},\uparrow\rangle, \vert d_{xz-iyz},\downarrow\rangle)^T$, which, in the angular momentum basis, becomes $\Psi_{\mathbf{k}} = (\vert p_-, \tfrac{1}{2}\rangle, \vert p_-, -\tfrac{1}{2}\rangle,  \vert d_+, \tfrac{1}{2}\rangle, \vert d_+, -\tfrac{1}{2}\rangle, \vert d_+, \tfrac{3}{2}\rangle, \vert d_+, -\tfrac{3}{2}\rangle)^T$. Here, $|\alpha_{\eta}, J_z\rangle$ denote a basis originating from $\alpha=p,d$ atomic orbitals that carries both a $z$-component angular momentum $J_z$ and an even/odd parity $\eta=\pm$. Then the normal-state Hamiltonian is
\begin{eqnarray}
	{\cal H}_0({\bf k}) = \begin{pmatrix}
	h^{(\frac{1}{2})}({\bf k}) & T({\bf k}) \\
	T^\dagger({\bf k}) & h^{(\frac{3}{2})} ({\bf k}) \\
	\end{pmatrix}
\end{eqnarray}
which consists of $h^{\frac{3}{2}}({\bf k}) = (M_2({\bf k}) + \delta_{so})\sigma_0$ and
\begin{eqnarray}
	h^{(\frac{1}{2})} ({\bf k}) &=& \begin{pmatrix}
	M_1(\mathbf{k}) & 0 & A_2k_z & -A_1k_- \\
	0 & M_1(\mathbf{k}) & A_1k_+ & A_2k_z \\
	A_2k_z & A_1k_- & M_2(\mathbf{k}) & 0  \\
	-A_1k_+ & A_2k_z & 0 & M_2(\mathbf{k}) \\
	\end{pmatrix} \nonumber \\
	T^\dagger ({\bf k}) &=& \begin{pmatrix}
	A_1k_- & 0 & 0 & D^\ast(\mathbf{k}) \\
	0 & -A_1k_+ & D(\mathbf{k}) & 0 \\
	\end{pmatrix}
\end{eqnarray}
Here $\sigma$ denotes the Pauli matrix for spin degree of freedom and we have defined $k_\pm = k_x \pm ik_y$, $M_i(\mathbf{k})=M_0^{(i)} + M_1^{(i)}(k_x^2+k_y^2)+M_2^{(i)} k_z^2$ and $D(\mathbf{k})=D_1(k_x^2-k_y^2)-iD_2k_xk_y$. $H_0({\bf k})$ features four-fold rotation symmetry $C_{4} = e^{i\tfrac{\pi}{2}J_z}$, inversion symmetry ${\cal P} = \text{diag}[-1,-1,1,1,1,1]$, and time-reversal symmetry $\Theta = \text{diag}[-i,i,-i]\otimes \sigma_y {\cal K}$, with $J_z=\text{diag}[\frac{1}{2},-\frac{1}{2},\frac{1}{2},-\frac{1}{2},\frac{3}{2},-\frac{3}{2}]$ and ${\cal K}$ denoting the complex conjugation operation. For our purpose, we choose $M_0^{(1)} = M_1^{(1)} =-2 M_2^{(1)}=-1, A_1=0.5, A_2=0.1,D_{1,2}=0$ and $M_{0,1,2}^{(2)}=-M_{0,1,2}^{(1)}$ throughout this work. This ensures an inverted band structure among the $p_z$ and $d_{xz, yz}$ bands.

\begin{figure}[t]
	\includegraphics[width=0.48\textwidth]{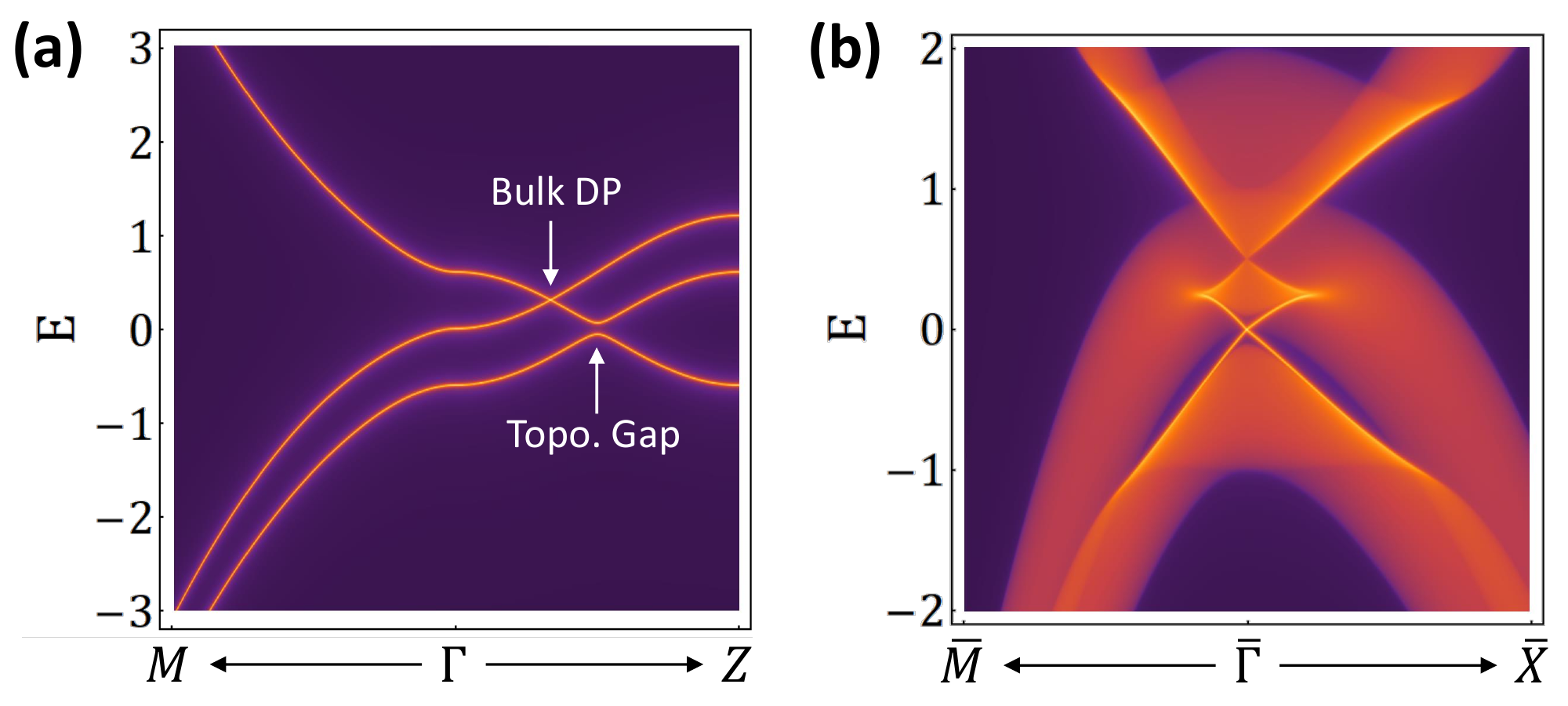}
	\caption{Bulk and (001) surface energy spectra are shown in (a) and (b), respectively. Both TI and DSM physics coexist and are energetically separated by an energy scale of $\delta_{so}$.
	}
	\label{fig2}
\end{figure}

Notably, $h^{(\frac{1}{2})}({\bf k})$ by itself manifests as a standard Hamiltonian for a 3D time-reversal invariant TI~\cite{zhang_np_2009,liu_prb_2010}. 
Besides, as shown in Fig.~\ref{fig2} (a), a second band inversion between $|p_-,\pm\frac{1}{2}\rangle$ and $|d_+,\pm\frac{3}{2}\rangle$ generates an additional 3D Dirac semimetal phase with a pair of four-fold-degenerate bulk Dirac nodes~\cite{armitage_rmp_2018,lv_rmp_2021}. The energy separation between the TI and DSM bands is controlled by $\delta_{so}$, the spin-orbit splitting among the $d$-orbital bands. Remarkably, the robustness of the bulk Dirac points are guaranteed by the combination of $C_4$, ${\cal P}$, and $\Theta$. In Fig.~\ref{fig2} (b), we exploit iterative Green function method to map out the energy spectrum for $(001)$ surface in a semi-infinite geometry. This clearly reveals the coexistence of a 2D Dirac surface state and the 3D bulk Dirac nodes, a common topological feature shared by most tFeSCs.

{\it Kitaev, Nodal, \& Hybrid Topological Vortices} - 
When the system becomes superconducting, the corresponding Bogoliubov-de Gennes (BdG) Hamiltonian is
\begin{align}\label{eq:BdG Ham}
\mathcal{H}_{BdG} = \begin{pmatrix}
\mathcal{H}_0(\mathbf{r})  & \mathcal{H}_\Delta \\
\mathcal{H}_\Delta^\dagger & -\mathcal{H}_0^\ast(\mathbf{r})
\end{pmatrix},
\end{align}
under the Nambu basis $\{ \Psi_{\mathbf{r}}^\dagger, \Psi_{\mathbf{r}}^T  \}$. We consider an isotropic $s$-wave pairing Hamiltonian ${\cal H}_{\Delta} = i\Delta({\bf r}) \text{diag}[1,-1,1]\otimes \sigma_y$~\cite{footnote}. A single magnetic vortex line can be described by $\Delta({\bf r}) = \Delta_0\tanh(r/\xi_0) e^{i\theta}$, where $\Delta_0=0.2$, $(r, \theta)$ are the in-plane polar coordinates with respect to the vortex core and $\xi_0$ denotes the superconducting coherence length. Physically, a vortex line is a 1D class D system~\cite{chiu_rmp_2016}, which, if being gapped, admits a $Z_2$ band topology that determines the presence of vMBS.

To understand possible vortex topologies in tFeSCs, it is suggestive to start with a $\delta_{so}\rightarrow \infty$ limit, where the normal-state TI phase and DSM phase can be viewed as two independent systems~\cite{qin_sb_2019}. When the Fermi level lies around the TI gap, the system enters the ``TI limit" and its vortex physics is well captured by the Fu-Kane model~\cite{fu_prl_2008,hosur_prl_2011}. Specifically, the lowest-energy $k_z$-dispersing vortex line modes carries an angular momentum of $l_z=0$ and there exist two critical chemical potentials $\mu=\mu_{c0,\pm}$ where the vortex modes close their energy gap at $k_z=0$ or $\pi$. Such gap closing signals a change of the 1D vortex topology and thus serves as the phase boundaries for two topologically distinct phases: (i) a Majorana-free trivial phase and (ii) a gapped topological phase with end-localized vMBS (dubbed the ``Kitaev vortex"), which lives within $\mu\in (\mu_{c0,-}, \mu_{c0,+})$.

Meanwhile, a ``DSM limit" with $\delta_{so}\rightarrow \infty$ is reached when the Fermi level is near the bulk Dirac nodes. Similarly, there exist two critical chemical potentials $\mu_{c1,\pm}$ where the vortex gap vanishes at $k_z=0$ or $\pi$.
For $\mu\in (\mu_{c1,-}, \mu_{c1,+})$, however, the lowest-energy vortex modes necessarily carry $l_z=\pm1$ and further form a pair of $C_4$-protected band crossings at zero energy along $k_z$. While such a nodal vortex is NOT vMBS-carrying, it can be turned into a gapped Kitaev vortex with vMBS by simply spoiling the protecting $C_4$ symmetry, as we will show later.

In realistic FeSC systems, $\delta_{so}$ can be small enough such that neither the TI limit nor the DSM limit applies. In this case, the aforementioned phase diagrams for both TI and DSM limits will now mix and interact with each other.
Nonetheless, the notions of $\mu_{c0,\pm}$ and $\mu_{c1,\pm}$ remain well-defined and thus still decide the vortex topological phase boundaries even for a small $\delta_{so}$ system. Remarkably, as we show analytically in the SM~\cite{supplementary}, the energy range for each vortex phase, i.e., $\Delta\mu_{c0/c1}=\mu_{c0/c1,+}-\mu_{c0/c1,-}$, will get significantly enhanced by reducing the value of $\delta_{so}$. This fact is crucial for small-$\delta_{so}$ systems, where the Kitaev vortex phase and nodal vortex phase tend to have a finite overlap around $\mu=0$ in the phase diagram, leading to a new {\it hybrid topological vortex phase}. This hybrid vortex state inherits two key topological features from its parent vortex states: (i) it features a $C_4$-protected nodal dispersion along $k_z$; (ii) it hosts vMBS with a finite localization length. Notably, the 1D nodal bands and the vMBSs are living in different $C_4$ symmetry sectors and thus will not hybridize with each other. Similar Majorana-carrying gapless topological phase has also been reported in certain 1D Luttinger-liquid systems~\cite{cheng2011Majorana,zhang2018crystalline}.

\begin{figure}[t]
	\includegraphics[width=0.45\textwidth]{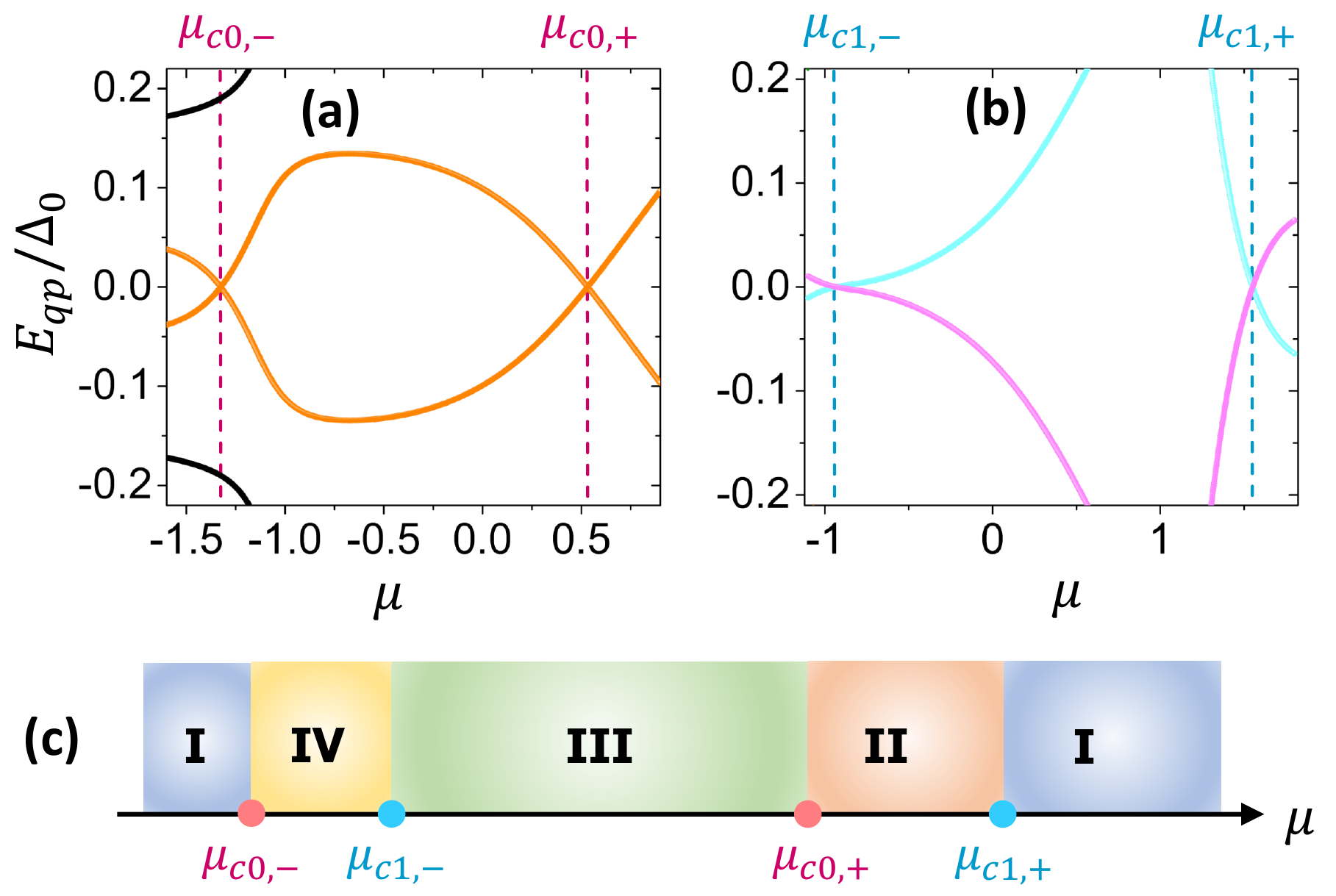}
	\caption{The 1D vortex spectrum at $k_z=\pi$ is shown in (a) for $l_z=0$ and (b) for $\vert l_z\vert=1$. All the other subspaces with $\vert l_z\vert >1$ are fully gapped. The critical chemical potentials, with which the energy gap closes, determine the phase boundaries for the vortex phase diagram schematically shown in (c). 
	}
	\label{fig3}
\end{figure}

We now proceed to numerically map out the vortex topological phase diagram (VTPD) for our six-band tFeSC model in Eq.~\eqref{eq:BdG Ham} as a function of both $\mu$ and $\delta_{so}$.
The topological phase boundaries in Fig.~\ref{fig1} are determined by $\mu_{c0,\pm}$ (orange line) and $\mu_{c1,\pm}$ (purple line) for a fixed $\delta_{so}$, which can be extracted by calculating the vortex mode spectrum. We therefore consider a cylindrical geometry~\cite{gygi_prb_1991} for our tFeSC model, with $k_z$ being a good quantum number and open boundary conditions imposed for in-plane directions. Details of numerical calculations are discussed in the SM~\cite{supplementary}. As shown in Fig.~\ref{fig3} (a) and (b), we plot the vortex mode spectrum as a function of $\mu$ with fixed $\delta_{so}=0.5$ and $k_z=\pi$ for both $l_z=0$ and $|l_z|=1$ sectors, respectively, where the gap closing points indicate the topological phase boundaries as defined in the insets. The vortex spectra with $\vert l_z\vert\ge2$ are fully gapped.
Four topologically inequivalent regions are found to show up, as shown in Fig.~\ref{fig3} (c): (I) trivial vortex for either $\mu>\mu_{c1,+}$ or $\mu<\mu_{c0,-}$; (II) nodal vortex for $\mu_{c0,+}<\mu<\mu_{c1,+}$; (III) hybrid vortex for $\mu_{c1,-}<\mu<\mu_{c0,+}$; (IV) Kitaev vortex for $\mu_{c0,-}<\mu<\mu_{c1,-}$. Further varying the value of $\delta_{so}$, we eventually obtain the complete 
$\mu$-$\delta_{so}$
VTPD in Fig.~\ref{fig1}. Just as we expect, the hybrid vortex is indeed the dominating phase for clean tFeSCs with small $\mu$ and $\delta_{so}$.

The hybrid vortex physics is captured by a minimal four-band effective Hamiltonian,
\begin{align}
	h_{\text{hybrid}}(k_z) = h_{\text{Kitaev}}(k_z) \oplus h_{\text{nodal}}(k_z).
	\label{eq-hybrid-ham}
\end{align}
The decoupled $2\times2$ blocks $ h_{\text{Kitaev}}=(m_0 + m_1 k_z^2) \tau_z + m_2 k_z \tau_x$ and $h_{\text{nodal}}= (m_0'+m_1'k_z^2) \tau_z$ correspond to the gapped Kitaev vortex part with $l_z=0$ and the gapless nodal vortex part with $l_z=\pm 1$, respectively. While it can be easily constructed with the symmetry principle, we also provide an analytical derivation of $h_\text{hybrid}$ in the SM~\cite{supplementary}. The band parameters for $h_\text{hybrid}$ can be extracted numerically. For example,  with $\mu=0,\delta_{so}=0.5$, we find $m_0=0.018, m_1=-0.0022, m_2=0.032$ and $m_0'=0.025, m_1'=-0.011$ for the above model. The sign reversal between $m_0$ ($m_0'$) and $m_1$ ($m_1'$) signatures the topological band inversion for $h_\text{Kitaev}$ ($h_\text{nodal}$).

\begin{figure}[t]
	\includegraphics[width=0.48\textwidth]{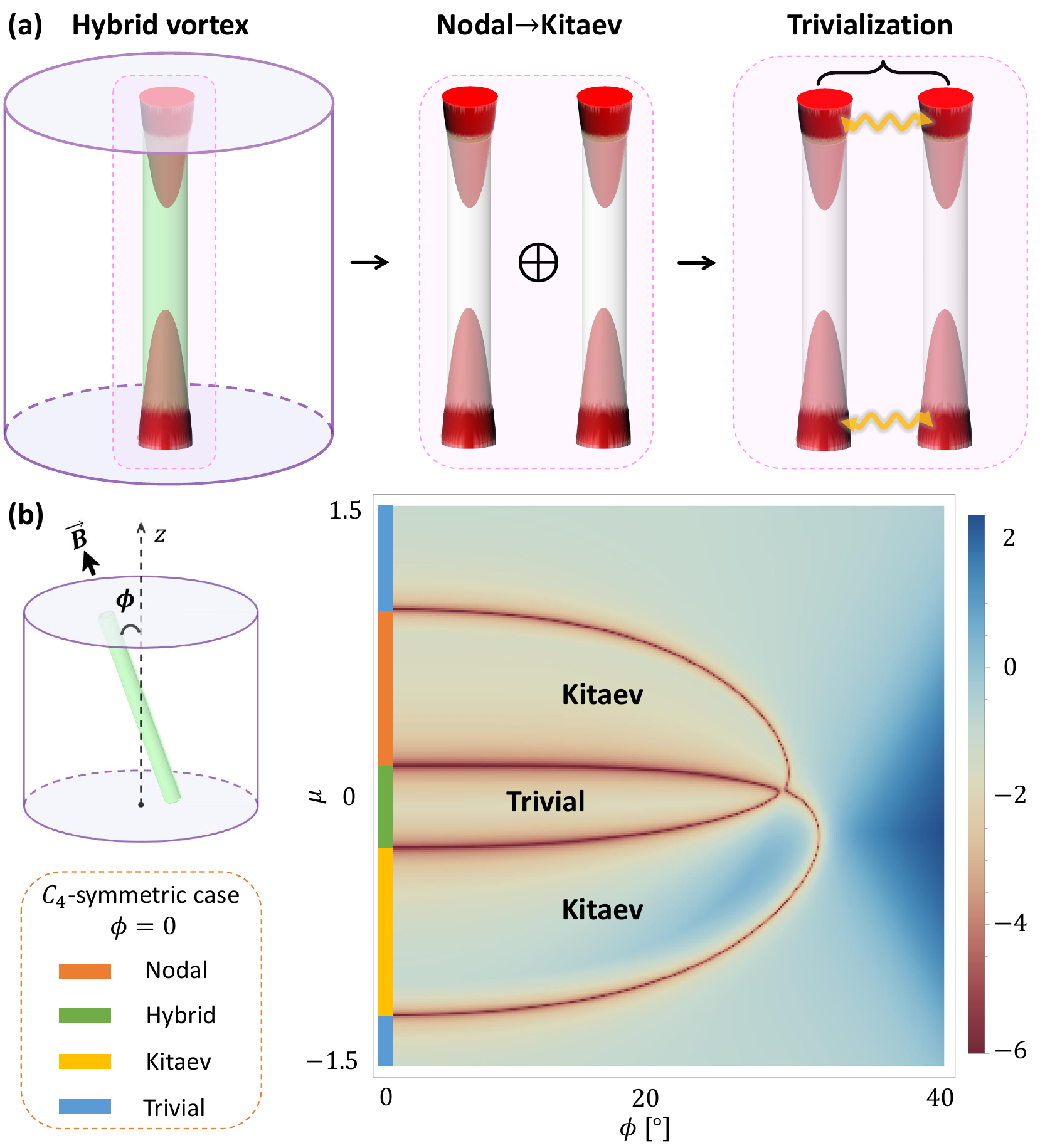}
	\caption{
	(a) Vortex line tilting gaps out the nodal componenet of a hybrid vortex, leading to two copies of Kitaev vortex states that can be together trivialized topologically.  
	(b) The vortex topological phase diagram as a function of $\mu$ and $\phi$ for $\delta_{so}=0.5$. The logarithmic value of the vortex energy gap at $k_z'=\pi$ is shown by the colors. The red lines denote vortex topological phase transitions.
	}
	\label{fig4}
\end{figure}

The topological nature of the hybrid vortex state delicately relies on the around-axis $C_4$ symmetry. As schematically shown in Fig.~\ref{fig4}~(a), spoiling $C_4$ drives the nodal component of the hybrid vortex into additional Kitaev vortex degrees of freedom, which further interacts with the original Kitaev vortex component and gets trivialized as a whole. Therefore, a $C_4$-broken ``hybrid vortex" is essentially a trivial vortex state with no Majorana physics. In practice, local $C_4$ breaking at nanometer scale appears generally unavoidable and can arise from a plethora of mechanisms in tFeSCs. These scenarios include (i) the applied magnetic field tilts away from $\hat{z}$ axis~\cite{zhangST_prb_2019,giwa_prl_2021,giwa_arxiv_2022}; (ii) bulk impurities bends the vortex line. In the following, we will focus on the effect of vortex line tilting in tFeSCs and a detailed discussion of the vortex line bending can be found in the SM~\cite{supplementary}.

{\it Vortex Line Tilting} - For a small tilting angle $\phi \ll \frac{\pi}{2}$ [as defined in Fig.~\ref{fig4}~(b)], we can adopt the second-order perturbation theory to analytically rederive the hybrid vortex Hamiltonian $h_{\phi} (k_z')$, where $k_z'$ is aligned with the vortex-line orientation. Details of the perturbation theory can be found in the SM~\cite{supplementary}. We find that formally,
\begin{align} \label{eq-ham-vortex-kz-phi}
	h_{\phi}(k_z') =  h_\text{hybrid}(k_z',\phi) + h_\text{SB}(k_z',\phi) + {\cal O}(\phi^3).
\end{align}
Here $h_\text{hybrid}(k_z',\phi)$ resembles the original $C_4$-preserving hybrid vortex Hamiltonian in Eq.~\ref{eq-hybrid-ham}, but with a set of renormalized parameters $m_{0,1}\to m_{0,1}+m_{5,6}\phi^2$ and $m_{0,1}'\to m_{0,1}'+m_{5,6}'\phi^2$. We thus expect turning on $\phi$ to quantitatively change our VTPD (i.e., $\mu_{c0,\pm}$ and $\mu_{c1,\pm}$).

Meanwhile, $h_\text{SB}(k_z', \phi)$ describes the geometry-induced $C_4$ breaking terms and its effects on the vortex-state topology is two-fold. First, it generates a topological gap for the nodal vortex bands, with $h_\text{nodal}\rightarrow h_{\text{nodal}}+m_2'\phi^2k_z' \tau_x$. The linear-$k_z$ dependence here is required by the particle-hole symmetry $\Xi = \tau_x \mathcal{K}$, with $\mathcal{K}$ the complex conjugation. Second, the Kitaev and nodal vortex degrees of freedom get hybridized via a coupling matrix that is linearly proportional to $\phi$ (see the SM~\cite{supplementary} for details). Therefore, it is exactly the above two contributions of $h_\text{SB}(k_z', \phi)$ that lead to  the ``hybrid $\rightarrow$ trivial" scenario described in Fig.~\ref{fig4}~(a).

Fig.~\ref{fig4}~(b) is a numerical map of VTPD as a function of $\mu$ and $\phi$ based on a lattice-regularized tight-binding model of Eq.~\ref{eq:BdG Ham}, by calculating the logarithmic value of the vortex band gap at $k_z'=\pi$. In the $\phi=0$ limit, the $\mu$-$\phi$ VTPD reproduces the $C_4$-symmetric phase diagram in Fig.~\ref{fig3} (c), up to some quantitative differences from the lattice regularization procedure. The topological vortex phase boundaries (red lines) manifest a $\phi$-dependence due to the vortex band renormalization, agreeing with our perturbation theory. Interestingly, we also find that the Kitaev vortex phase will terminate at $\phi\sim \pi/6$, which can be feasily checked in experiments by mapping out the local density of states near the vortex core.

{\it Discussions on tFeSCs Candidates} - We first note that FeTe$_{x}$Se$_{1-x}$, a paradigmatic tFeSC candidate, features a strong spin-orbital coupling effect with $\delta_{so}\sim35$ meV. We believe that such $\delta_{so}$ is large enough for FeTe$_x$Se$_{1-x}$ to approach the ``TI limit", as justified by earlier first-principles-based calculations~\cite{xu_prl_2016}. Thus, FeTe$_{x}$Se$_{1-x}$ should manifest as a standard Fu-Kane superconductor with vMBS signals, agreeing with the experimental observations~\cite{wang_science_2018,zhang_science_2018}.

LiFeAs, however, features a small $\delta_{so}\sim$10.7 meV~\cite{zhang_np_2019}, three times smaller than that of FeTe$_x$Se$_{1-x}$. Based on the $\mu$-$\delta_{so}$ VTPD in Fig.~\ref{fig1}, we expect LiFeAs to carry the hybrid vortex topology for hosting both small $\delta_{so}$ and $\mu$. 
As we discussed earlier, a $C_4$-breaking perturbation such as the ${\bf B}$ field tilting in Fig.~\ref{fig4} can break the hybrid vortex down to a trivial one with no MZM signal, which is likely the reason behind the disappearance of vortex Majorana signals in LiFeAs~\cite{kong_nc_2021}. In particular, even when the ${\bf B}$ field is carefully aligned with the crystalline rotation axis, the tilting angle $\phi$ of the vortex line can still be greatly enhanced, when a near-surface impurity locally distorts the vortex-line geometry. Notably, these atomic impurities could be completely invisible to surface-sensitive probes, such as scanning tunneling microscopy (STM). In the SM~\cite{supplementary}, we numerically simulate such impurity-induced vortex-line bending effect and have confirmed its crucial role of trivializing vortex topology.

Ref.~\cite{kong_nc_2021} also reports the appearance of vMBS signal in LiFeAs due to surface-impurity-induced electron doping. The reported levitation of Fermi level is around $5$ meV for the so-called strong impurities. Given that $\delta_{so}\sim 10$ meV for LiFeAs, this effect would be capable of driving a transition from a trivialized hybrid vortex to a Kitaev vortex, following our $\mu$-$\phi$ VTPD in Fig.~\ref{fig4} (b). We further predict that if we continuously lower the Fermi level via hole doping, {\it the vMBS signal will reemerge at a critical $\mu_{c1,-}$} [i.e., regime IV in Fig. \ref{fig3} (c)] {\it and eventually disappear at a negatively large $\mu_{c0,-}$}. Such an exotic ``reentrant Majorana signal" serves as an experimental ``smoking gun" for our theory. 
We also predict a similar but more complex reentrant vortex Majorana phenomenon in CaKFe$_4$As$_4$~\cite{liu_nc_2020}, where a detailed discussion can be found in the SM~\cite{supplementary}.

{\it Conclusion} - 
To summarize, the entanglement between TI and DSM physics has a decisive impact on the topological nature of vortex lines in tFeSCs. A direct outcome of the entangled bulk topological bands is the competition among multiple topologically distinct vortex states in the VTPD, including trivial, Kitaev, nodal and hybrid vortex phases. Notably, the unprecendented hybrid vortex topology naturally explains the puzzling absence of vMBS signal in LiFeAs. Our theory can also be feasibly tested in both LiFeAs and CaKFe$_4$As$_4$ with the state-of-the-art Fermi level engineering and scanning tunneling microscopy. Besides, by replacing the Te/Se/As atoms in tFeSCs with other atoms with different spin-orbital coupling, the value of $\delta_{so}$ can be continuously tuned to manipulate the vortex topology. 
An interesting future direction is to explore other symmetry breaking effects and their impact on the vortex topology in tFeSCs. For example, breaking inversion symmetry by strain can split the bulk Dirac nodes into pairs of Weyl nodes, which is expected to further complicate the VTPD~\cite{yan_prl_2020,giwa_prl_2021,giwa_arxiv_2022}. We leave a detailed study on these possibilities of engineering vortex topological physics to future works.

{\it Acknowledgement} - 
We acknowledge S. Das Sarma, C.-K. Chiu, and especially L. Kong for helpful discussions. C.X.L. acknowledges the support of the Office of Naval Research (Grant No. N00014-18-1-2793) and Kaufman New Initiative research (Grant No. KA2018-98553) of the Pittsburgh Foundation.
L.H.H. and R.X.Z. acknowledge the start-up fund at the University of Tennessee.

{\it Note Added} - 
After the submission of our manuscript, we became aware of two experimental papers that reported strain-induced hole-doping effect achieved in LiFeAs~\cite{li_nature_2022,liu_arxiv_2021}, where reappearance of vortex-bound zero-bias peak signals is also found. These observations agree with our ``reentrant Majorana signal'' prediction.

\bibliographystyle{apsrev4-2}
\bibliography{vortex}

\begin{thebibliography}{99}%
\makeatletter
\providecommand \@ifxundefined [1]{%
 \@ifx{#1\undefined}
}%
\providecommand \@ifnum [1]{%
 \ifnum #1\expandafter \@firstoftwo
 \else \expandafter \@secondoftwo
 \fi
}%
\providecommand \@ifx [1]{%
 \ifx #1\expandafter \@firstoftwo
 \else \expandafter \@secondoftwo
 \fi
}%
\providecommand \natexlab [1]{#1}%
\providecommand \enquote  [1]{``#1''}%
\providecommand \bibnamefont  [1]{#1}%
\providecommand \bibfnamefont [1]{#1}%
\providecommand \citenamefont [1]{#1}%
\providecommand \href@noop [0]{\@secondoftwo}%
\providecommand \href [0]{\begingroup \@sanitize@url \@href}%
\providecommand \@href[1]{\@@startlink{#1}\@@href}%
\providecommand \@@href[1]{\endgroup#1\@@endlink}%
\providecommand \@sanitize@url [0]{\catcode `\\12\catcode `\$12\catcode
  `\&12\catcode `\#12\catcode `\^12\catcode `\_12\catcode `\%12\relax}%
\providecommand \@@startlink[1]{}%
\providecommand \@@endlink[0]{}%
\providecommand \url  [0]{\begingroup\@sanitize@url \@url }%
\providecommand \@url [1]{\endgroup\@href {#1}{\urlprefix }}%
\providecommand \urlprefix  [0]{URL }%
\providecommand \Eprint [0]{\href }%
\providecommand \doibase [0]{https://doi.org/}%
\providecommand \selectlanguage [0]{\@gobble}%
\providecommand \bibinfo  [0]{\@secondoftwo}%
\providecommand \bibfield  [0]{\@secondoftwo}%
\providecommand \translation [1]{[#1]}%
\providecommand \BibitemOpen [0]{}%
\providecommand \bibitemStop [0]{}%
\providecommand \bibitemNoStop [0]{.\EOS\space}%
\providecommand \EOS [0]{\spacefactor3000\relax}%
\providecommand \BibitemShut  [1]{\csname bibitem#1\endcsname}%
\let\auto@bib@innerbib\@empty
\bibitem [{\citenamefont {Kitaev}(2003)}]{kitaev_ap_2003}%
  \BibitemOpen
  \bibfield  {author} {\bibinfo {author} {\bibfnamefont {A.~Y.}\ \bibnamefont
  {Kitaev}},\ }\href {https://doi.org/10.1016/S0003-4916(02)00018-0} {\bibfield
   {journal} {\bibinfo  {journal} {Annals of Physics}\ }\textbf {\bibinfo
  {volume} {303}},\ \bibinfo {pages} {2} (\bibinfo {year} {2003})}\BibitemShut
  {NoStop}%
\bibitem [{\citenamefont {Nayak}\ \emph {et~al.}(2008)\citenamefont {Nayak},
  \citenamefont {Simon}, \citenamefont {Stern}, \citenamefont {Freedman},\ and\
  \citenamefont {Das~Sarma}}]{nayak_rmp_2008}%
  \BibitemOpen
  \bibfield  {author} {\bibinfo {author} {\bibfnamefont {C.}~\bibnamefont
  {Nayak}}, \bibinfo {author} {\bibfnamefont {S.~H.}\ \bibnamefont {Simon}},
  \bibinfo {author} {\bibfnamefont {A.}~\bibnamefont {Stern}}, \bibinfo
  {author} {\bibfnamefont {M.}~\bibnamefont {Freedman}},\ and\ \bibinfo
  {author} {\bibfnamefont {S.}~\bibnamefont {Das~Sarma}},\ }\href
  {https://doi.org/10.1103/RevModPhys.80.1083} {\bibfield  {journal} {\bibinfo
  {journal} {Rev. Mod. Phys.}\ }\textbf {\bibinfo {volume} {80}},\ \bibinfo
  {pages} {1083} (\bibinfo {year} {2008})}\BibitemShut {NoStop}%
\bibitem [{\citenamefont {Sarma}\ \emph {et~al.}(2015)\citenamefont {Sarma},
  \citenamefont {Freedman},\ and\ \citenamefont {Nayak}}]{sarma_npjqi_2015}%
  \BibitemOpen
  \bibfield  {author} {\bibinfo {author} {\bibfnamefont {S.~D.}\ \bibnamefont
  {Sarma}}, \bibinfo {author} {\bibfnamefont {M.}~\bibnamefont {Freedman}},\
  and\ \bibinfo {author} {\bibfnamefont {C.}~\bibnamefont {Nayak}},\ }\href
  {https://doi.org/10.1038/npjqi.2015.1} {\bibfield  {journal} {\bibinfo
  {journal} {npj Quantum Information}\ }\textbf {\bibinfo {volume} {1}},\
  \bibinfo {pages} {1} (\bibinfo {year} {2015})}\BibitemShut {NoStop}%
\bibitem [{\citenamefont {Elliott}\ and\ \citenamefont
  {Franz}(2015)}]{elliott_rmp_2015}%
  \BibitemOpen
  \bibfield  {author} {\bibinfo {author} {\bibfnamefont {S.~R.}\ \bibnamefont
  {Elliott}}\ and\ \bibinfo {author} {\bibfnamefont {M.}~\bibnamefont
  {Franz}},\ }\href {https://doi.org/10.1103/RevModPhys.87.137} {\bibfield
  {journal} {\bibinfo  {journal} {Rev. Mod. Phys.}\ }\textbf {\bibinfo {volume}
  {87}},\ \bibinfo {pages} {137} (\bibinfo {year} {2015})}\BibitemShut
  {NoStop}%
\bibitem [{\citenamefont {Mourik}\ \emph {et~al.}(2012)\citenamefont {Mourik},
  \citenamefont {Zuo}, \citenamefont {Frolov}, \citenamefont {Plissard},
  \citenamefont {Bakkers},\ and\ \citenamefont
  {Kouwenhoven}}]{mourik_science_2012}%
  \BibitemOpen
  \bibfield  {author} {\bibinfo {author} {\bibfnamefont {V.}~\bibnamefont
  {Mourik}}, \bibinfo {author} {\bibfnamefont {K.}~\bibnamefont {Zuo}},
  \bibinfo {author} {\bibfnamefont {S.~M.}\ \bibnamefont {Frolov}}, \bibinfo
  {author} {\bibfnamefont {S.}~\bibnamefont {Plissard}}, \bibinfo {author}
  {\bibfnamefont {E.~P.}\ \bibnamefont {Bakkers}},\ and\ \bibinfo {author}
  {\bibfnamefont {L.~P.}\ \bibnamefont {Kouwenhoven}},\ }\href
  {https://doi.org/10.1126/science.1222360} {\bibfield  {journal} {\bibinfo
  {journal} {Science}\ }\textbf {\bibinfo {volume} {336}},\ \bibinfo {pages}
  {1003} (\bibinfo {year} {2012})}\BibitemShut {NoStop}%
\bibitem [{\citenamefont {Rokhinson}\ \emph {et~al.}(2012)\citenamefont
  {Rokhinson}, \citenamefont {Liu},\ and\ \citenamefont
  {Furdyna}}]{rokhinson_np_2012}%
  \BibitemOpen
  \bibfield  {author} {\bibinfo {author} {\bibfnamefont {L.~P.}\ \bibnamefont
  {Rokhinson}}, \bibinfo {author} {\bibfnamefont {X.}~\bibnamefont {Liu}},\
  and\ \bibinfo {author} {\bibfnamefont {J.~K.}\ \bibnamefont {Furdyna}},\
  }\href {https://doi.org/10.1038/nphys2429} {\bibfield  {journal} {\bibinfo
  {journal} {Nature Physics}\ }\textbf {\bibinfo {volume} {8}},\ \bibinfo
  {pages} {795} (\bibinfo {year} {2012})}\BibitemShut {NoStop}%
\bibitem [{\citenamefont {Williams}\ \emph {et~al.}(2012)\citenamefont
  {Williams}, \citenamefont {Bestwick}, \citenamefont {Gallagher},
  \citenamefont {Hong}, \citenamefont {Cui}, \citenamefont {Bleich},
  \citenamefont {Analytis}, \citenamefont {Fisher},\ and\ \citenamefont
  {Goldhaber-Gordon}}]{williams_prl_2012}%
  \BibitemOpen
  \bibfield  {author} {\bibinfo {author} {\bibfnamefont {J.~R.}\ \bibnamefont
  {Williams}}, \bibinfo {author} {\bibfnamefont {A.~J.}\ \bibnamefont
  {Bestwick}}, \bibinfo {author} {\bibfnamefont {P.}~\bibnamefont {Gallagher}},
  \bibinfo {author} {\bibfnamefont {S.~S.}\ \bibnamefont {Hong}}, \bibinfo
  {author} {\bibfnamefont {Y.}~\bibnamefont {Cui}}, \bibinfo {author}
  {\bibfnamefont {A.~S.}\ \bibnamefont {Bleich}}, \bibinfo {author}
  {\bibfnamefont {J.~G.}\ \bibnamefont {Analytis}}, \bibinfo {author}
  {\bibfnamefont {I.~R.}\ \bibnamefont {Fisher}},\ and\ \bibinfo {author}
  {\bibfnamefont {D.}~\bibnamefont {Goldhaber-Gordon}},\ }\href
  {https://doi.org/10.1103/PhysRevLett.109.056803} {\bibfield  {journal}
  {\bibinfo  {journal} {Phys. Rev. Lett.}\ }\textbf {\bibinfo {volume} {109}},\
  \bibinfo {pages} {056803} (\bibinfo {year} {2012})}\BibitemShut {NoStop}%
\bibitem [{\citenamefont {Das}\ \emph {et~al.}(2012)\citenamefont {Das},
  \citenamefont {Ronen}, \citenamefont {Most}, \citenamefont {Oreg},
  \citenamefont {Heiblum},\ and\ \citenamefont {Shtrikman}}]{das_np_2012}%
  \BibitemOpen
  \bibfield  {author} {\bibinfo {author} {\bibfnamefont {A.}~\bibnamefont
  {Das}}, \bibinfo {author} {\bibfnamefont {Y.}~\bibnamefont {Ronen}}, \bibinfo
  {author} {\bibfnamefont {Y.}~\bibnamefont {Most}}, \bibinfo {author}
  {\bibfnamefont {Y.}~\bibnamefont {Oreg}}, \bibinfo {author} {\bibfnamefont
  {M.}~\bibnamefont {Heiblum}},\ and\ \bibinfo {author} {\bibfnamefont
  {H.}~\bibnamefont {Shtrikman}},\ }\href {https://doi.org/10.1038/nphys2479}
  {\bibfield  {journal} {\bibinfo  {journal} {Nature Physics}\ }\textbf
  {\bibinfo {volume} {8}},\ \bibinfo {pages} {887} (\bibinfo {year}
  {2012})}\BibitemShut {NoStop}%
\bibitem [{\citenamefont {Deng}\ \emph {et~al.}(2012)\citenamefont {Deng},
  \citenamefont {Yu}, \citenamefont {Huang}, \citenamefont {Larsson},
  \citenamefont {Caroff},\ and\ \citenamefont {Xu}}]{deng_nl_2012}%
  \BibitemOpen
  \bibfield  {author} {\bibinfo {author} {\bibfnamefont {M.}~\bibnamefont
  {Deng}}, \bibinfo {author} {\bibfnamefont {C.}~\bibnamefont {Yu}}, \bibinfo
  {author} {\bibfnamefont {G.}~\bibnamefont {Huang}}, \bibinfo {author}
  {\bibfnamefont {M.}~\bibnamefont {Larsson}}, \bibinfo {author} {\bibfnamefont
  {P.}~\bibnamefont {Caroff}},\ and\ \bibinfo {author} {\bibfnamefont
  {H.}~\bibnamefont {Xu}},\ }\href {https://doi.org/10.1021/nl303758w}
  {\bibfield  {journal} {\bibinfo  {journal} {Nano letters}\ }\textbf {\bibinfo
  {volume} {12}},\ \bibinfo {pages} {6414} (\bibinfo {year}
  {2012})}\BibitemShut {NoStop}%
\bibitem [{\citenamefont {Finck}\ \emph {et~al.}(2013)\citenamefont {Finck},
  \citenamefont {Van~Harlingen}, \citenamefont {Mohseni}, \citenamefont
  {Jung},\ and\ \citenamefont {Li}}]{finck_prl_2013}%
  \BibitemOpen
  \bibfield  {author} {\bibinfo {author} {\bibfnamefont {A.~D.~K.}\
  \bibnamefont {Finck}}, \bibinfo {author} {\bibfnamefont {D.~J.}\ \bibnamefont
  {Van~Harlingen}}, \bibinfo {author} {\bibfnamefont {P.~K.}\ \bibnamefont
  {Mohseni}}, \bibinfo {author} {\bibfnamefont {K.}~\bibnamefont {Jung}},\ and\
  \bibinfo {author} {\bibfnamefont {X.}~\bibnamefont {Li}},\ }\href
  {https://doi.org/10.1103/PhysRevLett.110.126406} {\bibfield  {journal}
  {\bibinfo  {journal} {Phys. Rev. Lett.}\ }\textbf {\bibinfo {volume} {110}},\
  \bibinfo {pages} {126406} (\bibinfo {year} {2013})}\BibitemShut {NoStop}%
\bibitem [{\citenamefont {Lee}\ \emph {et~al.}(2014)\citenamefont {Lee},
  \citenamefont {Jiang}, \citenamefont {Houzet}, \citenamefont {Aguado},
  \citenamefont {Lieber},\ and\ \citenamefont {De~Franceschi}}]{lee_nn_2014}%
  \BibitemOpen
  \bibfield  {author} {\bibinfo {author} {\bibfnamefont {E.~J.}\ \bibnamefont
  {Lee}}, \bibinfo {author} {\bibfnamefont {X.}~\bibnamefont {Jiang}}, \bibinfo
  {author} {\bibfnamefont {M.}~\bibnamefont {Houzet}}, \bibinfo {author}
  {\bibfnamefont {R.}~\bibnamefont {Aguado}}, \bibinfo {author} {\bibfnamefont
  {C.~M.}\ \bibnamefont {Lieber}},\ and\ \bibinfo {author} {\bibfnamefont
  {S.}~\bibnamefont {De~Franceschi}},\ }\href
  {https://doi.org/10.1038/nnano.2013.267} {\bibfield  {journal} {\bibinfo
  {journal} {Nature nanotechnology}\ }\textbf {\bibinfo {volume} {9}},\
  \bibinfo {pages} {79} (\bibinfo {year} {2014})}\BibitemShut {NoStop}%
\bibitem [{\citenamefont {Nadj-Perge}\ \emph {et~al.}(2014)\citenamefont
  {Nadj-Perge}, \citenamefont {Drozdov}, \citenamefont {Li}, \citenamefont
  {Chen}, \citenamefont {Jeon}, \citenamefont {Seo}, \citenamefont {MacDonald},
  \citenamefont {Bernevig},\ and\ \citenamefont {Yazdani}}]{nadj_science_2014}%
  \BibitemOpen
  \bibfield  {author} {\bibinfo {author} {\bibfnamefont {S.}~\bibnamefont
  {Nadj-Perge}}, \bibinfo {author} {\bibfnamefont {I.~K.}\ \bibnamefont
  {Drozdov}}, \bibinfo {author} {\bibfnamefont {J.}~\bibnamefont {Li}},
  \bibinfo {author} {\bibfnamefont {H.}~\bibnamefont {Chen}}, \bibinfo {author}
  {\bibfnamefont {S.}~\bibnamefont {Jeon}}, \bibinfo {author} {\bibfnamefont
  {J.}~\bibnamefont {Seo}}, \bibinfo {author} {\bibfnamefont {A.~H.}\
  \bibnamefont {MacDonald}}, \bibinfo {author} {\bibfnamefont {B.~A.}\
  \bibnamefont {Bernevig}},\ and\ \bibinfo {author} {\bibfnamefont
  {A.}~\bibnamefont {Yazdani}},\ }\href
  {https://doi.org/10.1126/science.1259327} {\bibfield  {journal} {\bibinfo
  {journal} {Science}\ }\textbf {\bibinfo {volume} {346}},\ \bibinfo {pages}
  {602} (\bibinfo {year} {2014})}\BibitemShut {NoStop}%
\bibitem [{\citenamefont {Xu}\ \emph {et~al.}(2015)\citenamefont {Xu},
  \citenamefont {Wang}, \citenamefont {Liu}, \citenamefont {Ge}, \citenamefont
  {Yang}, \citenamefont {Liu}, \citenamefont {Xu}, \citenamefont {Guan},
  \citenamefont {Gao}, \citenamefont {Qian}, \citenamefont {Liu}, \citenamefont
  {Wang}, \citenamefont {Zhang}, \citenamefont {Xue},\ and\ \citenamefont
  {Jia}}]{xu_prl_2015}%
  \BibitemOpen
  \bibfield  {author} {\bibinfo {author} {\bibfnamefont {J.-P.}\ \bibnamefont
  {Xu}}, \bibinfo {author} {\bibfnamefont {M.-X.}\ \bibnamefont {Wang}},
  \bibinfo {author} {\bibfnamefont {Z.~L.}\ \bibnamefont {Liu}}, \bibinfo
  {author} {\bibfnamefont {J.-F.}\ \bibnamefont {Ge}}, \bibinfo {author}
  {\bibfnamefont {X.}~\bibnamefont {Yang}}, \bibinfo {author} {\bibfnamefont
  {C.}~\bibnamefont {Liu}}, \bibinfo {author} {\bibfnamefont {Z.~A.}\
  \bibnamefont {Xu}}, \bibinfo {author} {\bibfnamefont {D.}~\bibnamefont
  {Guan}}, \bibinfo {author} {\bibfnamefont {C.~L.}\ \bibnamefont {Gao}},
  \bibinfo {author} {\bibfnamefont {D.}~\bibnamefont {Qian}}, \bibinfo {author}
  {\bibfnamefont {Y.}~\bibnamefont {Liu}}, \bibinfo {author} {\bibfnamefont
  {Q.-H.}\ \bibnamefont {Wang}}, \bibinfo {author} {\bibfnamefont {F.-C.}\
  \bibnamefont {Zhang}}, \bibinfo {author} {\bibfnamefont {Q.-K.}\ \bibnamefont
  {Xue}},\ and\ \bibinfo {author} {\bibfnamefont {J.-F.}\ \bibnamefont {Jia}},\
  }\href {https://doi.org/10.1103/PhysRevLett.114.017001} {\bibfield  {journal}
  {\bibinfo  {journal} {Phys. Rev. Lett.}\ }\textbf {\bibinfo {volume} {114}},\
  \bibinfo {pages} {017001} (\bibinfo {year} {2015})}\BibitemShut {NoStop}%
\bibitem [{\citenamefont {Albrecht}\ \emph {et~al.}(2016)\citenamefont
  {Albrecht}, \citenamefont {Higginbotham}, \citenamefont {Madsen},
  \citenamefont {Kuemmeth}, \citenamefont {Jespersen}, \citenamefont
  {Nyg{\aa}rd}, \citenamefont {Krogstrup},\ and\ \citenamefont
  {Marcus}}]{albrecht_nature_2016}%
  \BibitemOpen
  \bibfield  {author} {\bibinfo {author} {\bibfnamefont {S.~M.}\ \bibnamefont
  {Albrecht}}, \bibinfo {author} {\bibfnamefont {A.~P.}\ \bibnamefont
  {Higginbotham}}, \bibinfo {author} {\bibfnamefont {M.}~\bibnamefont
  {Madsen}}, \bibinfo {author} {\bibfnamefont {F.}~\bibnamefont {Kuemmeth}},
  \bibinfo {author} {\bibfnamefont {T.~S.}\ \bibnamefont {Jespersen}}, \bibinfo
  {author} {\bibfnamefont {J.}~\bibnamefont {Nyg{\aa}rd}}, \bibinfo {author}
  {\bibfnamefont {P.}~\bibnamefont {Krogstrup}},\ and\ \bibinfo {author}
  {\bibfnamefont {C.}~\bibnamefont {Marcus}},\ }\href
  {https://doi.org/10.1038/nature17162} {\bibfield  {journal} {\bibinfo
  {journal} {Nature}\ }\textbf {\bibinfo {volume} {531}},\ \bibinfo {pages}
  {206} (\bibinfo {year} {2016})}\BibitemShut {NoStop}%
\bibitem [{\citenamefont {Sun}\ \emph {et~al.}(2016)\citenamefont {Sun},
  \citenamefont {Zhang}, \citenamefont {Hu}, \citenamefont {Li}, \citenamefont
  {Wang}, \citenamefont {Ma}, \citenamefont {Xu}, \citenamefont {Gao},
  \citenamefont {Guan}, \citenamefont {Li}, \citenamefont {Liu}, \citenamefont
  {Qian}, \citenamefont {Zhou}, \citenamefont {Fu}, \citenamefont {Li},
  \citenamefont {Zhang},\ and\ \citenamefont {Jia}}]{sun_prl_2016}%
  \BibitemOpen
  \bibfield  {author} {\bibinfo {author} {\bibfnamefont {H.-H.}\ \bibnamefont
  {Sun}}, \bibinfo {author} {\bibfnamefont {K.-W.}\ \bibnamefont {Zhang}},
  \bibinfo {author} {\bibfnamefont {L.-H.}\ \bibnamefont {Hu}}, \bibinfo
  {author} {\bibfnamefont {C.}~\bibnamefont {Li}}, \bibinfo {author}
  {\bibfnamefont {G.-Y.}\ \bibnamefont {Wang}}, \bibinfo {author}
  {\bibfnamefont {H.-Y.}\ \bibnamefont {Ma}}, \bibinfo {author} {\bibfnamefont
  {Z.-A.}\ \bibnamefont {Xu}}, \bibinfo {author} {\bibfnamefont {C.-L.}\
  \bibnamefont {Gao}}, \bibinfo {author} {\bibfnamefont {D.-D.}\ \bibnamefont
  {Guan}}, \bibinfo {author} {\bibfnamefont {Y.-Y.}\ \bibnamefont {Li}},
  \bibinfo {author} {\bibfnamefont {C.}~\bibnamefont {Liu}}, \bibinfo {author}
  {\bibfnamefont {D.}~\bibnamefont {Qian}}, \bibinfo {author} {\bibfnamefont
  {Y.}~\bibnamefont {Zhou}}, \bibinfo {author} {\bibfnamefont {L.}~\bibnamefont
  {Fu}}, \bibinfo {author} {\bibfnamefont {S.-C.}\ \bibnamefont {Li}}, \bibinfo
  {author} {\bibfnamefont {F.-C.}\ \bibnamefont {Zhang}},\ and\ \bibinfo
  {author} {\bibfnamefont {J.-F.}\ \bibnamefont {Jia}},\ }\href
  {https://doi.org/10.1103/PhysRevLett.116.257003} {\bibfield  {journal}
  {\bibinfo  {journal} {Phys. Rev. Lett.}\ }\textbf {\bibinfo {volume} {116}},\
  \bibinfo {pages} {257003} (\bibinfo {year} {2016})}\BibitemShut {NoStop}%
\bibitem [{\citenamefont {Deng}\ \emph {et~al.}(2016)\citenamefont {Deng},
  \citenamefont {Vaitiek{\.e}nas}, \citenamefont {Hansen}, \citenamefont
  {Danon}, \citenamefont {Leijnse}, \citenamefont {Flensberg}, \citenamefont
  {Nyg{\aa}rd}, \citenamefont {Krogstrup},\ and\ \citenamefont
  {Marcus}}]{deng_science_2016}%
  \BibitemOpen
  \bibfield  {author} {\bibinfo {author} {\bibfnamefont {M.}~\bibnamefont
  {Deng}}, \bibinfo {author} {\bibfnamefont {S.}~\bibnamefont
  {Vaitiek{\.e}nas}}, \bibinfo {author} {\bibfnamefont {E.~B.}\ \bibnamefont
  {Hansen}}, \bibinfo {author} {\bibfnamefont {J.}~\bibnamefont {Danon}},
  \bibinfo {author} {\bibfnamefont {M.}~\bibnamefont {Leijnse}}, \bibinfo
  {author} {\bibfnamefont {K.}~\bibnamefont {Flensberg}}, \bibinfo {author}
  {\bibfnamefont {J.}~\bibnamefont {Nyg{\aa}rd}}, \bibinfo {author}
  {\bibfnamefont {P.}~\bibnamefont {Krogstrup}},\ and\ \bibinfo {author}
  {\bibfnamefont {C.~M.}\ \bibnamefont {Marcus}},\ }\href
  {https://doi.org/10.1126/science.aaf3961} {\bibfield  {journal} {\bibinfo
  {journal} {Science}\ }\textbf {\bibinfo {volume} {354}},\ \bibinfo {pages}
  {1557} (\bibinfo {year} {2016})}\BibitemShut {NoStop}%
\bibitem [{\citenamefont {Lv}\ \emph {et~al.}(2017)\citenamefont {Lv},
  \citenamefont {Wang}, \citenamefont {Zhang}, \citenamefont {Ding},
  \citenamefont {Li}, \citenamefont {Wang}, \citenamefont {He}, \citenamefont
  {Song}, \citenamefont {Ma},\ and\ \citenamefont {Xue}}]{lv_sb_2017}%
  \BibitemOpen
  \bibfield  {author} {\bibinfo {author} {\bibfnamefont {Y.-F.}\ \bibnamefont
  {Lv}}, \bibinfo {author} {\bibfnamefont {W.-L.}\ \bibnamefont {Wang}},
  \bibinfo {author} {\bibfnamefont {Y.-M.}\ \bibnamefont {Zhang}}, \bibinfo
  {author} {\bibfnamefont {H.}~\bibnamefont {Ding}}, \bibinfo {author}
  {\bibfnamefont {W.}~\bibnamefont {Li}}, \bibinfo {author} {\bibfnamefont
  {L.}~\bibnamefont {Wang}}, \bibinfo {author} {\bibfnamefont {K.}~\bibnamefont
  {He}}, \bibinfo {author} {\bibfnamefont {C.-L.}\ \bibnamefont {Song}},
  \bibinfo {author} {\bibfnamefont {X.-C.}\ \bibnamefont {Ma}},\ and\ \bibinfo
  {author} {\bibfnamefont {Q.-K.}\ \bibnamefont {Xue}},\ }\href
  {https://doi.org/10.1016/j.scib.2017.05.008} {\bibfield  {journal} {\bibinfo
  {journal} {Science bulletin}\ }\textbf {\bibinfo {volume} {62}},\ \bibinfo
  {pages} {852} (\bibinfo {year} {2017})}\BibitemShut {NoStop}%
\bibitem [{\citenamefont {Feldman}\ \emph {et~al.}(2017)\citenamefont
  {Feldman}, \citenamefont {Randeria}, \citenamefont {Li}, \citenamefont
  {Jeon}, \citenamefont {Xie}, \citenamefont {Wang}, \citenamefont {Drozdov},
  \citenamefont {Bernevig},\ and\ \citenamefont {Yazdani}}]{feldman_np_2017}%
  \BibitemOpen
  \bibfield  {author} {\bibinfo {author} {\bibfnamefont {B.~E.}\ \bibnamefont
  {Feldman}}, \bibinfo {author} {\bibfnamefont {M.~T.}\ \bibnamefont
  {Randeria}}, \bibinfo {author} {\bibfnamefont {J.}~\bibnamefont {Li}},
  \bibinfo {author} {\bibfnamefont {S.}~\bibnamefont {Jeon}}, \bibinfo {author}
  {\bibfnamefont {Y.}~\bibnamefont {Xie}}, \bibinfo {author} {\bibfnamefont
  {Z.}~\bibnamefont {Wang}}, \bibinfo {author} {\bibfnamefont {I.~K.}\
  \bibnamefont {Drozdov}}, \bibinfo {author} {\bibfnamefont {B.~A.}\
  \bibnamefont {Bernevig}},\ and\ \bibinfo {author} {\bibfnamefont
  {A.}~\bibnamefont {Yazdani}},\ }\href {https://doi.org/10.1038/nphys3947}
  {\bibfield  {journal} {\bibinfo  {journal} {Nature Physics}\ }\textbf
  {\bibinfo {volume} {13}},\ \bibinfo {pages} {286} (\bibinfo {year}
  {2017})}\BibitemShut {NoStop}%
\bibitem [{\citenamefont {Jeon}\ \emph {et~al.}(2017)\citenamefont {Jeon},
  \citenamefont {Xie}, \citenamefont {Li}, \citenamefont {Wang}, \citenamefont
  {Bernevig},\ and\ \citenamefont {Yazdani}}]{jeon_science_2017}%
  \BibitemOpen
  \bibfield  {author} {\bibinfo {author} {\bibfnamefont {S.}~\bibnamefont
  {Jeon}}, \bibinfo {author} {\bibfnamefont {Y.}~\bibnamefont {Xie}}, \bibinfo
  {author} {\bibfnamefont {J.}~\bibnamefont {Li}}, \bibinfo {author}
  {\bibfnamefont {Z.}~\bibnamefont {Wang}}, \bibinfo {author} {\bibfnamefont
  {B.~A.}\ \bibnamefont {Bernevig}},\ and\ \bibinfo {author} {\bibfnamefont
  {A.}~\bibnamefont {Yazdani}},\ }\href
  {https://doi.org/10.1126/science.aan3670} {\bibfield  {journal} {\bibinfo
  {journal} {Science}\ }\textbf {\bibinfo {volume} {358}},\ \bibinfo {pages}
  {772} (\bibinfo {year} {2017})}\BibitemShut {NoStop}%
\bibitem [{\citenamefont {Ren}\ \emph {et~al.}(2019)\citenamefont {Ren},
  \citenamefont {Pientka}, \citenamefont {Hart}, \citenamefont {Pierce},
  \citenamefont {Kosowsky}, \citenamefont {Lunczer}, \citenamefont {Schlereth},
  \citenamefont {Scharf}, \citenamefont {Hankiewicz}, \citenamefont {Molenkamp}
  \emph {et~al.}}]{ren_nature_2019}%
  \BibitemOpen
  \bibfield  {author} {\bibinfo {author} {\bibfnamefont {H.}~\bibnamefont
  {Ren}}, \bibinfo {author} {\bibfnamefont {F.}~\bibnamefont {Pientka}},
  \bibinfo {author} {\bibfnamefont {S.}~\bibnamefont {Hart}}, \bibinfo {author}
  {\bibfnamefont {A.~T.}\ \bibnamefont {Pierce}}, \bibinfo {author}
  {\bibfnamefont {M.}~\bibnamefont {Kosowsky}}, \bibinfo {author}
  {\bibfnamefont {L.}~\bibnamefont {Lunczer}}, \bibinfo {author} {\bibfnamefont
  {R.}~\bibnamefont {Schlereth}}, \bibinfo {author} {\bibfnamefont
  {B.}~\bibnamefont {Scharf}}, \bibinfo {author} {\bibfnamefont {E.~M.}\
  \bibnamefont {Hankiewicz}}, \bibinfo {author} {\bibfnamefont {L.~W.}\
  \bibnamefont {Molenkamp}}, \emph {et~al.},\ }\href
  {https://doi.org/10.1038/s41586-019-1148-9} {\bibfield  {journal} {\bibinfo
  {journal} {Nature}\ }\textbf {\bibinfo {volume} {569}},\ \bibinfo {pages}
  {93} (\bibinfo {year} {2019})}\BibitemShut {NoStop}%
\bibitem [{\citenamefont {J{\"a}ck}\ \emph {et~al.}(2019)\citenamefont
  {J{\"a}ck}, \citenamefont {Xie}, \citenamefont {Li}, \citenamefont {Jeon},
  \citenamefont {Bernevig},\ and\ \citenamefont {Yazdani}}]{jack_science_2019}%
  \BibitemOpen
  \bibfield  {author} {\bibinfo {author} {\bibfnamefont {B.}~\bibnamefont
  {J{\"a}ck}}, \bibinfo {author} {\bibfnamefont {Y.}~\bibnamefont {Xie}},
  \bibinfo {author} {\bibfnamefont {J.}~\bibnamefont {Li}}, \bibinfo {author}
  {\bibfnamefont {S.}~\bibnamefont {Jeon}}, \bibinfo {author} {\bibfnamefont
  {B.~A.}\ \bibnamefont {Bernevig}},\ and\ \bibinfo {author} {\bibfnamefont
  {A.}~\bibnamefont {Yazdani}},\ }\href
  {https://doi.org/10.1126/science.aax1444} {\bibfield  {journal} {\bibinfo
  {journal} {Science}\ }\textbf {\bibinfo {volume} {364}},\ \bibinfo {pages}
  {1255} (\bibinfo {year} {2019})}\BibitemShut {NoStop}%
\bibitem [{\citenamefont {Fornieri}\ \emph {et~al.}(2019)\citenamefont
  {Fornieri}, \citenamefont {Whiticar}, \citenamefont {Setiawan}, \citenamefont
  {Portol{\'e}s}, \citenamefont {Drachmann}, \citenamefont {Keselman},
  \citenamefont {Gronin}, \citenamefont {Thomas}, \citenamefont {Wang},
  \citenamefont {Kallaher} \emph {et~al.}}]{fornieri_nature_2019}%
  \BibitemOpen
  \bibfield  {author} {\bibinfo {author} {\bibfnamefont {A.}~\bibnamefont
  {Fornieri}}, \bibinfo {author} {\bibfnamefont {A.~M.}\ \bibnamefont
  {Whiticar}}, \bibinfo {author} {\bibfnamefont {F.}~\bibnamefont {Setiawan}},
  \bibinfo {author} {\bibfnamefont {E.}~\bibnamefont {Portol{\'e}s}}, \bibinfo
  {author} {\bibfnamefont {A.~C.}\ \bibnamefont {Drachmann}}, \bibinfo {author}
  {\bibfnamefont {A.}~\bibnamefont {Keselman}}, \bibinfo {author}
  {\bibfnamefont {S.}~\bibnamefont {Gronin}}, \bibinfo {author} {\bibfnamefont
  {C.}~\bibnamefont {Thomas}}, \bibinfo {author} {\bibfnamefont
  {T.}~\bibnamefont {Wang}}, \bibinfo {author} {\bibfnamefont {R.}~\bibnamefont
  {Kallaher}}, \emph {et~al.},\ }\href
  {https://doi.org/10.1038/s41586-019-1068-8} {\bibfield  {journal} {\bibinfo
  {journal} {Nature}\ }\textbf {\bibinfo {volume} {569}},\ \bibinfo {pages}
  {89} (\bibinfo {year} {2019})}\BibitemShut {NoStop}%
\bibitem [{\citenamefont {Zhang}\ \emph {et~al.}(2018)\citenamefont {Zhang},
  \citenamefont {Yaji}, \citenamefont {Hashimoto}, \citenamefont {Ota},
  \citenamefont {Kondo}, \citenamefont {Okazaki}, \citenamefont {Wang},
  \citenamefont {Wen}, \citenamefont {Gu}, \citenamefont {Ding} \emph
  {et~al.}}]{zhang_science_2018}%
  \BibitemOpen
  \bibfield  {author} {\bibinfo {author} {\bibfnamefont {P.}~\bibnamefont
  {Zhang}}, \bibinfo {author} {\bibfnamefont {K.}~\bibnamefont {Yaji}},
  \bibinfo {author} {\bibfnamefont {T.}~\bibnamefont {Hashimoto}}, \bibinfo
  {author} {\bibfnamefont {Y.}~\bibnamefont {Ota}}, \bibinfo {author}
  {\bibfnamefont {T.}~\bibnamefont {Kondo}}, \bibinfo {author} {\bibfnamefont
  {K.}~\bibnamefont {Okazaki}}, \bibinfo {author} {\bibfnamefont
  {Z.}~\bibnamefont {Wang}}, \bibinfo {author} {\bibfnamefont {J.}~\bibnamefont
  {Wen}}, \bibinfo {author} {\bibfnamefont {G.}~\bibnamefont {Gu}}, \bibinfo
  {author} {\bibfnamefont {H.}~\bibnamefont {Ding}}, \emph {et~al.},\ }\href
  {https://doi.org/10.1126/science.aan4596} {\bibfield  {journal} {\bibinfo
  {journal} {Science}\ }\textbf {\bibinfo {volume} {360}},\ \bibinfo {pages}
  {182} (\bibinfo {year} {2018})}\BibitemShut {NoStop}%
\bibitem [{\citenamefont {Shi}\ \emph {et~al.}(2017)\citenamefont {Shi},
  \citenamefont {Han}, \citenamefont {Richard}, \citenamefont {Wu},
  \citenamefont {Peng}, \citenamefont {Qian}, \citenamefont {Wang},
  \citenamefont {Hu}, \citenamefont {Sun},\ and\ \citenamefont
  {Ding}}]{shi_sb_2017}%
  \BibitemOpen
  \bibfield  {author} {\bibinfo {author} {\bibfnamefont {X.}~\bibnamefont
  {Shi}}, \bibinfo {author} {\bibfnamefont {Z.-Q.}\ \bibnamefont {Han}},
  \bibinfo {author} {\bibfnamefont {P.}~\bibnamefont {Richard}}, \bibinfo
  {author} {\bibfnamefont {X.-X.}\ \bibnamefont {Wu}}, \bibinfo {author}
  {\bibfnamefont {X.-L.}\ \bibnamefont {Peng}}, \bibinfo {author}
  {\bibfnamefont {T.}~\bibnamefont {Qian}}, \bibinfo {author} {\bibfnamefont
  {S.-C.}\ \bibnamefont {Wang}}, \bibinfo {author} {\bibfnamefont {J.-P.}\
  \bibnamefont {Hu}}, \bibinfo {author} {\bibfnamefont {Y.-J.}\ \bibnamefont
  {Sun}},\ and\ \bibinfo {author} {\bibfnamefont {H.}~\bibnamefont {Ding}},\
  }\href {https://doi.org/10.1016/j.scib.2017.03.010} {\bibfield  {journal}
  {\bibinfo  {journal} {Science bulletin}\ }\textbf {\bibinfo {volume} {62}},\
  \bibinfo {pages} {503} (\bibinfo {year} {2017})}\BibitemShut {NoStop}%
\bibitem [{\citenamefont {Zhang}\ \emph
  {et~al.}(2019{\natexlab{a}})\citenamefont {Zhang}, \citenamefont {Wang},
  \citenamefont {Wu}, \citenamefont {Yaji}, \citenamefont {Ishida},
  \citenamefont {Kohama}, \citenamefont {Dai}, \citenamefont {Sun},
  \citenamefont {Bareille}, \citenamefont {Kuroda} \emph
  {et~al.}}]{zhang_np_2019}%
  \BibitemOpen
  \bibfield  {author} {\bibinfo {author} {\bibfnamefont {P.}~\bibnamefont
  {Zhang}}, \bibinfo {author} {\bibfnamefont {Z.}~\bibnamefont {Wang}},
  \bibinfo {author} {\bibfnamefont {X.}~\bibnamefont {Wu}}, \bibinfo {author}
  {\bibfnamefont {K.}~\bibnamefont {Yaji}}, \bibinfo {author} {\bibfnamefont
  {Y.}~\bibnamefont {Ishida}}, \bibinfo {author} {\bibfnamefont
  {Y.}~\bibnamefont {Kohama}}, \bibinfo {author} {\bibfnamefont
  {G.}~\bibnamefont {Dai}}, \bibinfo {author} {\bibfnamefont {Y.}~\bibnamefont
  {Sun}}, \bibinfo {author} {\bibfnamefont {C.}~\bibnamefont {Bareille}},
  \bibinfo {author} {\bibfnamefont {K.}~\bibnamefont {Kuroda}}, \emph
  {et~al.},\ }\href {https://doi.org/10.1038/s41567-018-0280-z} {\bibfield
  {journal} {\bibinfo  {journal} {Nature Physics}\ }\textbf {\bibinfo {volume}
  {15}},\ \bibinfo {pages} {41} (\bibinfo {year}
  {2019}{\natexlab{a}})}\BibitemShut {NoStop}%
\bibitem [{\citenamefont {Peng}\ \emph {et~al.}(2019)\citenamefont {Peng},
  \citenamefont {Li}, \citenamefont {Wu}, \citenamefont {Deng}, \citenamefont
  {Shi}, \citenamefont {Fan}, \citenamefont {Li}, \citenamefont {Huang},
  \citenamefont {Qian}, \citenamefont {Richard}, \citenamefont {Hu},
  \citenamefont {Pan}, \citenamefont {Mao}, \citenamefont {Sun},\ and\
  \citenamefont {Ding}}]{peng_prb_2019}%
  \BibitemOpen
  \bibfield  {author} {\bibinfo {author} {\bibfnamefont {X.-L.}\ \bibnamefont
  {Peng}}, \bibinfo {author} {\bibfnamefont {Y.}~\bibnamefont {Li}}, \bibinfo
  {author} {\bibfnamefont {X.-X.}\ \bibnamefont {Wu}}, \bibinfo {author}
  {\bibfnamefont {H.-B.}\ \bibnamefont {Deng}}, \bibinfo {author}
  {\bibfnamefont {X.}~\bibnamefont {Shi}}, \bibinfo {author} {\bibfnamefont
  {W.-H.}\ \bibnamefont {Fan}}, \bibinfo {author} {\bibfnamefont
  {M.}~\bibnamefont {Li}}, \bibinfo {author} {\bibfnamefont {Y.-B.}\
  \bibnamefont {Huang}}, \bibinfo {author} {\bibfnamefont {T.}~\bibnamefont
  {Qian}}, \bibinfo {author} {\bibfnamefont {P.}~\bibnamefont {Richard}},
  \bibinfo {author} {\bibfnamefont {J.-P.}\ \bibnamefont {Hu}}, \bibinfo
  {author} {\bibfnamefont {S.-H.}\ \bibnamefont {Pan}}, \bibinfo {author}
  {\bibfnamefont {H.-Q.}\ \bibnamefont {Mao}}, \bibinfo {author} {\bibfnamefont
  {Y.-J.}\ \bibnamefont {Sun}},\ and\ \bibinfo {author} {\bibfnamefont
  {H.}~\bibnamefont {Ding}},\ }\href
  {https://doi.org/10.1103/PhysRevB.100.155134} {\bibfield  {journal} {\bibinfo
   {journal} {Phys. Rev. B}\ }\textbf {\bibinfo {volume} {100}},\ \bibinfo
  {pages} {155134} (\bibinfo {year} {2019})}\BibitemShut {NoStop}%
\bibitem [{\citenamefont {Kong}\ \emph {et~al.}(2019)\citenamefont {Kong},
  \citenamefont {Zhu}, \citenamefont {Papaj}, \citenamefont {Chen},
  \citenamefont {Cao}, \citenamefont {Isobe}, \citenamefont {Xing},
  \citenamefont {Liu}, \citenamefont {Wang}, \citenamefont {Fan} \emph
  {et~al.}}]{kong_np_2019}%
  \BibitemOpen
  \bibfield  {author} {\bibinfo {author} {\bibfnamefont {L.}~\bibnamefont
  {Kong}}, \bibinfo {author} {\bibfnamefont {S.}~\bibnamefont {Zhu}}, \bibinfo
  {author} {\bibfnamefont {M.}~\bibnamefont {Papaj}}, \bibinfo {author}
  {\bibfnamefont {H.}~\bibnamefont {Chen}}, \bibinfo {author} {\bibfnamefont
  {L.}~\bibnamefont {Cao}}, \bibinfo {author} {\bibfnamefont {H.}~\bibnamefont
  {Isobe}}, \bibinfo {author} {\bibfnamefont {Y.}~\bibnamefont {Xing}},
  \bibinfo {author} {\bibfnamefont {W.}~\bibnamefont {Liu}}, \bibinfo {author}
  {\bibfnamefont {D.}~\bibnamefont {Wang}}, \bibinfo {author} {\bibfnamefont
  {P.}~\bibnamefont {Fan}}, \emph {et~al.},\ }\href
  {https://doi.org/10.1038/s41567-019-0630-5} {\bibfield  {journal} {\bibinfo
  {journal} {Nature Physics}\ }\textbf {\bibinfo {volume} {15}},\ \bibinfo
  {pages} {1181} (\bibinfo {year} {2019})}\BibitemShut {NoStop}%
\bibitem [{\citenamefont {Zhu}\ \emph {et~al.}(2020)\citenamefont {Zhu},
  \citenamefont {Kong}, \citenamefont {Cao}, \citenamefont {Chen},
  \citenamefont {Papaj}, \citenamefont {Du}, \citenamefont {Xing},
  \citenamefont {Liu}, \citenamefont {Wang}, \citenamefont {Shen} \emph
  {et~al.}}]{zhu_science_2020}%
  \BibitemOpen
  \bibfield  {author} {\bibinfo {author} {\bibfnamefont {S.}~\bibnamefont
  {Zhu}}, \bibinfo {author} {\bibfnamefont {L.}~\bibnamefont {Kong}}, \bibinfo
  {author} {\bibfnamefont {L.}~\bibnamefont {Cao}}, \bibinfo {author}
  {\bibfnamefont {H.}~\bibnamefont {Chen}}, \bibinfo {author} {\bibfnamefont
  {M.}~\bibnamefont {Papaj}}, \bibinfo {author} {\bibfnamefont
  {S.}~\bibnamefont {Du}}, \bibinfo {author} {\bibfnamefont {Y.}~\bibnamefont
  {Xing}}, \bibinfo {author} {\bibfnamefont {W.}~\bibnamefont {Liu}}, \bibinfo
  {author} {\bibfnamefont {D.}~\bibnamefont {Wang}}, \bibinfo {author}
  {\bibfnamefont {C.}~\bibnamefont {Shen}}, \emph {et~al.},\ }\href
  {https://doi.org/10.1126/science.aax0274} {\bibfield  {journal} {\bibinfo
  {journal} {Science}\ }\textbf {\bibinfo {volume} {367}},\ \bibinfo {pages}
  {189} (\bibinfo {year} {2020})}\BibitemShut {NoStop}%
\bibitem [{\citenamefont {Wang}\ \emph
  {et~al.}(2018{\natexlab{a}})\citenamefont {Wang}, \citenamefont {Kong},
  \citenamefont {Fan}, \citenamefont {Chen}, \citenamefont {Zhu}, \citenamefont
  {Liu}, \citenamefont {Cao}, \citenamefont {Sun}, \citenamefont {Du},
  \citenamefont {Schneeloch} \emph {et~al.}}]{wang_science_2018}%
  \BibitemOpen
  \bibfield  {author} {\bibinfo {author} {\bibfnamefont {D.}~\bibnamefont
  {Wang}}, \bibinfo {author} {\bibfnamefont {L.}~\bibnamefont {Kong}}, \bibinfo
  {author} {\bibfnamefont {P.}~\bibnamefont {Fan}}, \bibinfo {author}
  {\bibfnamefont {H.}~\bibnamefont {Chen}}, \bibinfo {author} {\bibfnamefont
  {S.}~\bibnamefont {Zhu}}, \bibinfo {author} {\bibfnamefont {W.}~\bibnamefont
  {Liu}}, \bibinfo {author} {\bibfnamefont {L.}~\bibnamefont {Cao}}, \bibinfo
  {author} {\bibfnamefont {Y.}~\bibnamefont {Sun}}, \bibinfo {author}
  {\bibfnamefont {S.}~\bibnamefont {Du}}, \bibinfo {author} {\bibfnamefont
  {J.}~\bibnamefont {Schneeloch}}, \emph {et~al.},\ }\href@noop {} {\bibfield
  {journal} {\bibinfo  {journal} {Science}\ }\textbf {\bibinfo {volume}
  {362}},\ \bibinfo {pages} {333} (\bibinfo {year}
  {2018}{\natexlab{a}})}\BibitemShut {NoStop}%
\bibitem [{\citenamefont {Machida}\ \emph {et~al.}(2019)\citenamefont
  {Machida}, \citenamefont {Sun}, \citenamefont {Pyon}, \citenamefont {Takeda},
  \citenamefont {Kohsaka}, \citenamefont {Hanaguri}, \citenamefont {Sasagawa},\
  and\ \citenamefont {Tamegai}}]{machida_nm_2019}%
  \BibitemOpen
  \bibfield  {author} {\bibinfo {author} {\bibfnamefont {T.}~\bibnamefont
  {Machida}}, \bibinfo {author} {\bibfnamefont {Y.}~\bibnamefont {Sun}},
  \bibinfo {author} {\bibfnamefont {S.}~\bibnamefont {Pyon}}, \bibinfo {author}
  {\bibfnamefont {S.}~\bibnamefont {Takeda}}, \bibinfo {author} {\bibfnamefont
  {Y.}~\bibnamefont {Kohsaka}}, \bibinfo {author} {\bibfnamefont
  {T.}~\bibnamefont {Hanaguri}}, \bibinfo {author} {\bibfnamefont
  {T.}~\bibnamefont {Sasagawa}},\ and\ \bibinfo {author} {\bibfnamefont
  {T.}~\bibnamefont {Tamegai}},\ }\href
  {https://doi.org/10.1038/s41563-019-0397-1} {\bibfield  {journal} {\bibinfo
  {journal} {Nature materials}\ }\textbf {\bibinfo {volume} {18}},\ \bibinfo
  {pages} {811} (\bibinfo {year} {2019})}\BibitemShut {NoStop}%
\bibitem [{\citenamefont {Liu}\ \emph {et~al.}(2018)\citenamefont {Liu},
  \citenamefont {Chen}, \citenamefont {Zhang}, \citenamefont {Peng},
  \citenamefont {Yan}, \citenamefont {Wen}, \citenamefont {Lou}, \citenamefont
  {Huang}, \citenamefont {Tian}, \citenamefont {Dong}, \citenamefont {Wang},
  \citenamefont {Bao}, \citenamefont {Wang}, \citenamefont {Yin}, \citenamefont
  {Zhao},\ and\ \citenamefont {Feng}}]{liu_prx_2018}%
  \BibitemOpen
  \bibfield  {author} {\bibinfo {author} {\bibfnamefont {Q.}~\bibnamefont
  {Liu}}, \bibinfo {author} {\bibfnamefont {C.}~\bibnamefont {Chen}}, \bibinfo
  {author} {\bibfnamefont {T.}~\bibnamefont {Zhang}}, \bibinfo {author}
  {\bibfnamefont {R.}~\bibnamefont {Peng}}, \bibinfo {author} {\bibfnamefont
  {Y.-J.}\ \bibnamefont {Yan}}, \bibinfo {author} {\bibfnamefont {C.-H.-P.}\
  \bibnamefont {Wen}}, \bibinfo {author} {\bibfnamefont {X.}~\bibnamefont
  {Lou}}, \bibinfo {author} {\bibfnamefont {Y.-L.}\ \bibnamefont {Huang}},
  \bibinfo {author} {\bibfnamefont {J.-P.}\ \bibnamefont {Tian}}, \bibinfo
  {author} {\bibfnamefont {X.-L.}\ \bibnamefont {Dong}}, \bibinfo {author}
  {\bibfnamefont {G.-W.}\ \bibnamefont {Wang}}, \bibinfo {author}
  {\bibfnamefont {W.-C.}\ \bibnamefont {Bao}}, \bibinfo {author} {\bibfnamefont
  {Q.-H.}\ \bibnamefont {Wang}}, \bibinfo {author} {\bibfnamefont {Z.-P.}\
  \bibnamefont {Yin}}, \bibinfo {author} {\bibfnamefont {Z.-X.}\ \bibnamefont
  {Zhao}},\ and\ \bibinfo {author} {\bibfnamefont {D.-L.}\ \bibnamefont
  {Feng}},\ }\href {https://doi.org/10.1103/PhysRevX.8.041056} {\bibfield
  {journal} {\bibinfo  {journal} {Phys. Rev. X}\ }\textbf {\bibinfo {volume}
  {8}},\ \bibinfo {pages} {041056} (\bibinfo {year} {2018})}\BibitemShut
  {NoStop}%
\bibitem [{\citenamefont {Chen}\ \emph {et~al.}(2019)\citenamefont {Chen},
  \citenamefont {Liu}, \citenamefont {Zhang}, \citenamefont {Li}, \citenamefont
  {Shen}, \citenamefont {Dong}, \citenamefont {Zhao}, \citenamefont {Zhang},\
  and\ \citenamefont {Feng}}]{chen_cpl_2019}%
  \BibitemOpen
  \bibfield  {author} {\bibinfo {author} {\bibfnamefont {C.}~\bibnamefont
  {Chen}}, \bibinfo {author} {\bibfnamefont {Q.}~\bibnamefont {Liu}}, \bibinfo
  {author} {\bibfnamefont {T.}~\bibnamefont {Zhang}}, \bibinfo {author}
  {\bibfnamefont {D.}~\bibnamefont {Li}}, \bibinfo {author} {\bibfnamefont
  {P.}~\bibnamefont {Shen}}, \bibinfo {author} {\bibfnamefont {X.}~\bibnamefont
  {Dong}}, \bibinfo {author} {\bibfnamefont {Z.-X.}\ \bibnamefont {Zhao}},
  \bibinfo {author} {\bibfnamefont {T.}~\bibnamefont {Zhang}},\ and\ \bibinfo
  {author} {\bibfnamefont {D.}~\bibnamefont {Feng}},\ }\href
  {https://doi.org/10.1088/0256-307X/36/5/057403} {\bibfield  {journal}
  {\bibinfo  {journal} {Chinese Physics Letters}\ }\textbf {\bibinfo {volume}
  {36}},\ \bibinfo {pages} {057403} (\bibinfo {year} {2019})}\BibitemShut
  {NoStop}%
\bibitem [{\citenamefont {Liu}\ \emph {et~al.}(2020)\citenamefont {Liu},
  \citenamefont {Cao}, \citenamefont {Zhu}, \citenamefont {Kong}, \citenamefont
  {Wang}, \citenamefont {Papaj}, \citenamefont {Zhang}, \citenamefont {Liu},
  \citenamefont {Chen}, \citenamefont {Li} \emph {et~al.}}]{liu_nc_2020}%
  \BibitemOpen
  \bibfield  {author} {\bibinfo {author} {\bibfnamefont {W.}~\bibnamefont
  {Liu}}, \bibinfo {author} {\bibfnamefont {L.}~\bibnamefont {Cao}}, \bibinfo
  {author} {\bibfnamefont {S.}~\bibnamefont {Zhu}}, \bibinfo {author}
  {\bibfnamefont {L.}~\bibnamefont {Kong}}, \bibinfo {author} {\bibfnamefont
  {G.}~\bibnamefont {Wang}}, \bibinfo {author} {\bibfnamefont {M.}~\bibnamefont
  {Papaj}}, \bibinfo {author} {\bibfnamefont {P.}~\bibnamefont {Zhang}},
  \bibinfo {author} {\bibfnamefont {Y.-B.}\ \bibnamefont {Liu}}, \bibinfo
  {author} {\bibfnamefont {H.}~\bibnamefont {Chen}}, \bibinfo {author}
  {\bibfnamefont {G.}~\bibnamefont {Li}}, \emph {et~al.},\ }\href
  {https://doi.org/10.1038/s41467-020-19487-1} {\bibfield  {journal} {\bibinfo
  {journal} {Nature communications}\ }\textbf {\bibinfo {volume} {11}},\
  \bibinfo {pages} {1} (\bibinfo {year} {2020})}\BibitemShut {NoStop}%
\bibitem [{\citenamefont {Kong}\ \emph {et~al.}(2021)\citenamefont {Kong},
  \citenamefont {Cao}, \citenamefont {Zhu}, \citenamefont {Papaj},
  \citenamefont {Dai}, \citenamefont {Li}, \citenamefont {Fan}, \citenamefont
  {Liu}, \citenamefont {Yang}, \citenamefont {Wang} \emph
  {et~al.}}]{kong_nc_2021}%
  \BibitemOpen
  \bibfield  {author} {\bibinfo {author} {\bibfnamefont {L.}~\bibnamefont
  {Kong}}, \bibinfo {author} {\bibfnamefont {L.}~\bibnamefont {Cao}}, \bibinfo
  {author} {\bibfnamefont {S.}~\bibnamefont {Zhu}}, \bibinfo {author}
  {\bibfnamefont {M.}~\bibnamefont {Papaj}}, \bibinfo {author} {\bibfnamefont
  {G.}~\bibnamefont {Dai}}, \bibinfo {author} {\bibfnamefont {G.}~\bibnamefont
  {Li}}, \bibinfo {author} {\bibfnamefont {P.}~\bibnamefont {Fan}}, \bibinfo
  {author} {\bibfnamefont {W.}~\bibnamefont {Liu}}, \bibinfo {author}
  {\bibfnamefont {F.}~\bibnamefont {Yang}}, \bibinfo {author} {\bibfnamefont
  {X.}~\bibnamefont {Wang}}, \emph {et~al.},\ }\href
  {https://doi.org/10.1038/s41467-021-24372-6} {\bibfield  {journal} {\bibinfo
  {journal} {Nature Communications}\ }\textbf {\bibinfo {volume} {12}},\
  \bibinfo {pages} {1} (\bibinfo {year} {2021})}\BibitemShut {NoStop}%
\bibitem [{\citenamefont {Zhang}\ \emph
  {et~al.}(2019{\natexlab{b}})\citenamefont {Zhang}, \citenamefont {Cole},\
  and\ \citenamefont {Das~Sarma}}]{zhang_prl_2019a}%
  \BibitemOpen
  \bibfield  {author} {\bibinfo {author} {\bibfnamefont {R.-X.}\ \bibnamefont
  {Zhang}}, \bibinfo {author} {\bibfnamefont {W.~S.}\ \bibnamefont {Cole}},\
  and\ \bibinfo {author} {\bibfnamefont {S.}~\bibnamefont {Das~Sarma}},\ }\href
  {https://doi.org/10.1103/PhysRevLett.122.187001} {\bibfield  {journal}
  {\bibinfo  {journal} {Phys. Rev. Lett.}\ }\textbf {\bibinfo {volume} {122}},\
  \bibinfo {pages} {187001} (\bibinfo {year} {2019}{\natexlab{b}})}\BibitemShut
  {NoStop}%
\bibitem [{\citenamefont {Zhang}\ \emph
  {et~al.}(2019{\natexlab{c}})\citenamefont {Zhang}, \citenamefont {Cole},
  \citenamefont {Wu},\ and\ \citenamefont {Das~Sarma}}]{zhang_prl_2019b}%
  \BibitemOpen
  \bibfield  {author} {\bibinfo {author} {\bibfnamefont {R.-X.}\ \bibnamefont
  {Zhang}}, \bibinfo {author} {\bibfnamefont {W.~S.}\ \bibnamefont {Cole}},
  \bibinfo {author} {\bibfnamefont {X.}~\bibnamefont {Wu}},\ and\ \bibinfo
  {author} {\bibfnamefont {S.}~\bibnamefont {Das~Sarma}},\ }\href
  {https://doi.org/10.1103/PhysRevLett.123.167001} {\bibfield  {journal}
  {\bibinfo  {journal} {Phys. Rev. Lett.}\ }\textbf {\bibinfo {volume} {123}},\
  \bibinfo {pages} {167001} (\bibinfo {year} {2019}{\natexlab{c}})}\BibitemShut
  {NoStop}%
\bibitem [{\citenamefont {Yan}\ \emph {et~al.}(2018)\citenamefont {Yan},
  \citenamefont {Song},\ and\ \citenamefont {Wang}}]{yan_prl_2018}%
  \BibitemOpen
  \bibfield  {author} {\bibinfo {author} {\bibfnamefont {Z.}~\bibnamefont
  {Yan}}, \bibinfo {author} {\bibfnamefont {F.}~\bibnamefont {Song}},\ and\
  \bibinfo {author} {\bibfnamefont {Z.}~\bibnamefont {Wang}},\ }\href
  {https://doi.org/10.1103/PhysRevLett.121.096803} {\bibfield  {journal}
  {\bibinfo  {journal} {Phys. Rev. Lett.}\ }\textbf {\bibinfo {volume} {121}},\
  \bibinfo {pages} {096803} (\bibinfo {year} {2018})}\BibitemShut {NoStop}%
\bibitem [{\citenamefont {Wang}\ \emph
  {et~al.}(2018{\natexlab{b}})\citenamefont {Wang}, \citenamefont {Liu},
  \citenamefont {Lu},\ and\ \citenamefont {Zhang}}]{wang_prl_2018}%
  \BibitemOpen
  \bibfield  {author} {\bibinfo {author} {\bibfnamefont {Q.}~\bibnamefont
  {Wang}}, \bibinfo {author} {\bibfnamefont {C.-C.}\ \bibnamefont {Liu}},
  \bibinfo {author} {\bibfnamefont {Y.-M.}\ \bibnamefont {Lu}},\ and\ \bibinfo
  {author} {\bibfnamefont {F.}~\bibnamefont {Zhang}},\ }\href
  {https://doi.org/10.1103/PhysRevLett.121.186801} {\bibfield  {journal}
  {\bibinfo  {journal} {Phys. Rev. Lett.}\ }\textbf {\bibinfo {volume} {121}},\
  \bibinfo {pages} {186801} (\bibinfo {year} {2018}{\natexlab{b}})}\BibitemShut
  {NoStop}%
\bibitem [{\citenamefont {Hu}\ \emph {et~al.}(2019)\citenamefont {Hu},
  \citenamefont {Liu},\ and\ \citenamefont {Zhang}}]{hu_cp_2019}%
  \BibitemOpen
  \bibfield  {author} {\bibinfo {author} {\bibfnamefont {L.-H.}\ \bibnamefont
  {Hu}}, \bibinfo {author} {\bibfnamefont {C.-X.}\ \bibnamefont {Liu}},\ and\
  \bibinfo {author} {\bibfnamefont {F.-C.}\ \bibnamefont {Zhang}},\ }\href
  {https://doi.org/10.1038/s42005-019-0126-8} {\bibfield  {journal} {\bibinfo
  {journal} {Communications Physics}\ }\textbf {\bibinfo {volume} {2}},\
  \bibinfo {pages} {1} (\bibinfo {year} {2019})}\BibitemShut {NoStop}%
\bibitem [{\citenamefont {Kawakami}\ and\ \citenamefont
  {Sato}(2019)}]{kawakami_prb_2019}%
  \BibitemOpen
  \bibfield  {author} {\bibinfo {author} {\bibfnamefont {T.}~\bibnamefont
  {Kawakami}}\ and\ \bibinfo {author} {\bibfnamefont {M.}~\bibnamefont
  {Sato}},\ }\href {https://doi.org/10.1103/PhysRevB.100.094520} {\bibfield
  {journal} {\bibinfo  {journal} {Phys. Rev. B}\ }\textbf {\bibinfo {volume}
  {100}},\ \bibinfo {pages} {094520} (\bibinfo {year} {2019})}\BibitemShut
  {NoStop}%
\bibitem [{\citenamefont {Wu}\ \emph {et~al.}(2021)\citenamefont {Wu},
  \citenamefont {Liu}, \citenamefont {Thomale},\ and\ \citenamefont
  {Liu}}]{wu_nsr_2021}%
  \BibitemOpen
  \bibfield  {author} {\bibinfo {author} {\bibfnamefont {X.}~\bibnamefont
  {Wu}}, \bibinfo {author} {\bibfnamefont {X.}~\bibnamefont {Liu}}, \bibinfo
  {author} {\bibfnamefont {R.}~\bibnamefont {Thomale}},\ and\ \bibinfo {author}
  {\bibfnamefont {C.-X.}\ \bibnamefont {Liu}},\ }\bibfield  {journal} {\bibinfo
   {journal} {National Science Review}\ }\href
  {https://doi.org/10.1093/nsr/nwab087} {10.1093/nsr/nwab087} (\bibinfo {year}
  {2021}),\ \bibinfo {note} {nwab087}\BibitemShut {NoStop}%
\bibitem [{\citenamefont {Wu}\ \emph {et~al.}(2020)\citenamefont {Wu},
  \citenamefont {Benalcazar}, \citenamefont {Li}, \citenamefont {Thomale},
  \citenamefont {Liu},\ and\ \citenamefont {Hu}}]{wu_prx_2020}%
  \BibitemOpen
  \bibfield  {author} {\bibinfo {author} {\bibfnamefont {X.}~\bibnamefont
  {Wu}}, \bibinfo {author} {\bibfnamefont {W.~A.}\ \bibnamefont {Benalcazar}},
  \bibinfo {author} {\bibfnamefont {Y.}~\bibnamefont {Li}}, \bibinfo {author}
  {\bibfnamefont {R.}~\bibnamefont {Thomale}}, \bibinfo {author} {\bibfnamefont
  {C.-X.}\ \bibnamefont {Liu}},\ and\ \bibinfo {author} {\bibfnamefont
  {J.}~\bibnamefont {Hu}},\ }\href {https://doi.org/10.1103/PhysRevX.10.041014}
  {\bibfield  {journal} {\bibinfo  {journal} {Phys. Rev. X}\ }\textbf {\bibinfo
  {volume} {10}},\ \bibinfo {pages} {041014} (\bibinfo {year}
  {2020})}\BibitemShut {NoStop}%
\bibitem [{\citenamefont {Gray}\ \emph {et~al.}(2019)\citenamefont {Gray},
  \citenamefont {Freudenstein}, \citenamefont {Zhao}, \citenamefont
  {O'~Connor}, \citenamefont {Jenkins}, \citenamefont {Kumar}, \citenamefont
  {Hoek}, \citenamefont {Kopec}, \citenamefont {Huh}, \citenamefont {Taniguchi}
  \emph {et~al.}}]{gray_nl_2019}%
  \BibitemOpen
  \bibfield  {author} {\bibinfo {author} {\bibfnamefont {M.~J.}\ \bibnamefont
  {Gray}}, \bibinfo {author} {\bibfnamefont {J.}~\bibnamefont {Freudenstein}},
  \bibinfo {author} {\bibfnamefont {S.~Y.~F.}\ \bibnamefont {Zhao}}, \bibinfo
  {author} {\bibfnamefont {R.}~\bibnamefont {O'~Connor}}, \bibinfo {author}
  {\bibfnamefont {S.}~\bibnamefont {Jenkins}}, \bibinfo {author} {\bibfnamefont
  {N.}~\bibnamefont {Kumar}}, \bibinfo {author} {\bibfnamefont
  {M.}~\bibnamefont {Hoek}}, \bibinfo {author} {\bibfnamefont {A.}~\bibnamefont
  {Kopec}}, \bibinfo {author} {\bibfnamefont {S.}~\bibnamefont {Huh}}, \bibinfo
  {author} {\bibfnamefont {T.}~\bibnamefont {Taniguchi}}, \emph {et~al.},\
  }\href {https://doi.org/10.1021/acs.nanolett.9b00844} {\bibfield  {journal}
  {\bibinfo  {journal} {Nano letters}\ }\textbf {\bibinfo {volume} {19}},\
  \bibinfo {pages} {4890} (\bibinfo {year} {2019})}\BibitemShut {NoStop}%
\bibitem [{\citenamefont {Wu}\ \emph {et~al.}()\citenamefont {Wu},
  \citenamefont {Zhang}, \citenamefont {Xu}, \citenamefont {Hu},\ and\
  \citenamefont {Liu}}]{wu_book_2021}%
  \BibitemOpen
  \bibfield  {author} {\bibinfo {author} {\bibfnamefont {X.}~\bibnamefont
  {Wu}}, \bibinfo {author} {\bibfnamefont {R.-X.}\ \bibnamefont {Zhang}},
  \bibinfo {author} {\bibfnamefont {G.}~\bibnamefont {Xu}}, \bibinfo {author}
  {\bibfnamefont {J.}~\bibnamefont {Hu}},\ and\ \bibinfo {author}
  {\bibfnamefont {C.-X.}\ \bibnamefont {Liu}},\ }\bibinfo {title} {In the
  pursuit of majorana modes in iron-based high-$t_c$ superconductors},\ in\
  \href {https://doi.org/10.1142/9789811231711_0005} {\emph {\bibinfo
  {booktitle} {Memorial Volume for Shoucheng Zhang}}},\ Chap.\ \bibinfo
  {chapter} {Chapter 3}, pp.\ \bibinfo {pages} {35--60}\BibitemShut {NoStop}%
\bibitem [{\citenamefont {Zhang}\ and\ \citenamefont
  {Das~Sarma}(2021)}]{zhang_prl_2021}%
  \BibitemOpen
  \bibfield  {author} {\bibinfo {author} {\bibfnamefont {R.-X.}\ \bibnamefont
  {Zhang}}\ and\ \bibinfo {author} {\bibfnamefont {S.}~\bibnamefont
  {Das~Sarma}},\ }\href {https://doi.org/10.1103/PhysRevLett.126.137001}
  {\bibfield  {journal} {\bibinfo  {journal} {Phys. Rev. Lett.}\ }\textbf
  {\bibinfo {volume} {126}},\ \bibinfo {pages} {137001} (\bibinfo {year}
  {2021})}\BibitemShut {NoStop}%
\bibitem [{\citenamefont {Kheirkhah}\ \emph
  {et~al.}(2021{\natexlab{a}})\citenamefont {Kheirkhah}, \citenamefont {Yan},\
  and\ \citenamefont {Marsiglio}}]{kheirkhah_prb_2021}%
  \BibitemOpen
  \bibfield  {author} {\bibinfo {author} {\bibfnamefont {M.}~\bibnamefont
  {Kheirkhah}}, \bibinfo {author} {\bibfnamefont {Z.}~\bibnamefont {Yan}},\
  and\ \bibinfo {author} {\bibfnamefont {F.}~\bibnamefont {Marsiglio}},\ }\href
  {https://doi.org/10.1103/PhysRevB.103.L140502} {\bibfield  {journal}
  {\bibinfo  {journal} {Phys. Rev. B}\ }\textbf {\bibinfo {volume} {103}},\
  \bibinfo {pages} {L140502} (\bibinfo {year}
  {2021}{\natexlab{a}})}\BibitemShut {NoStop}%
\bibitem [{\citenamefont {Qin}\ \emph {et~al.}(2021)\citenamefont {Qin},
  \citenamefont {Fang}, \citenamefont {Zhang},\ and\ \citenamefont
  {Hu}}]{qin_arxiv_2021}%
  \BibitemOpen
  \bibfield  {author} {\bibinfo {author} {\bibfnamefont {S.}~\bibnamefont
  {Qin}}, \bibinfo {author} {\bibfnamefont {C.}~\bibnamefont {Fang}}, \bibinfo
  {author} {\bibfnamefont {F.-C.}\ \bibnamefont {Zhang}},\ and\ \bibinfo
  {author} {\bibfnamefont {J.}~\bibnamefont {Hu}},\ }\href
  {https://ui.adsabs.harvard.edu/abs/2021arXiv210604200Q} {\bibfield  {journal}
  {\bibinfo  {journal} {arXiv preprint arXiv:2106.04200}\ } (\bibinfo {year}
  {2021})}\BibitemShut {NoStop}%
\bibitem [{\citenamefont {Kheirkhah}\ \emph
  {et~al.}(2021{\natexlab{b}})\citenamefont {Kheirkhah}, \citenamefont
  {Zhuang}, \citenamefont {Maciejko},\ and\ \citenamefont
  {Yan}}]{kheirkhah_arxiv_2021}%
  \BibitemOpen
  \bibfield  {author} {\bibinfo {author} {\bibfnamefont {M.}~\bibnamefont
  {Kheirkhah}}, \bibinfo {author} {\bibfnamefont {Z.-Y.}\ \bibnamefont
  {Zhuang}}, \bibinfo {author} {\bibfnamefont {J.}~\bibnamefont {Maciejko}},\
  and\ \bibinfo {author} {\bibfnamefont {Z.}~\bibnamefont {Yan}},\ }\href
  {https://ui.adsabs.harvard.edu/abs/2021arXiv210702811K} {\bibfield  {journal}
  {\bibinfo  {journal} {arXiv preprint arXiv:2107.02811}\ } (\bibinfo {year}
  {2021}{\natexlab{b}})}\BibitemShut {NoStop}%
\bibitem [{\citenamefont {Barik}\ and\ \citenamefont
  {Sau}(2021)}]{barik2021signatures}%
  \BibitemOpen
  \bibfield  {author} {\bibinfo {author} {\bibfnamefont {T.}~\bibnamefont
  {Barik}}\ and\ \bibinfo {author} {\bibfnamefont {J.~D.}\ \bibnamefont
  {Sau}},\ }\href {https://arxiv.org/abs/2108.13385} {\bibfield  {journal}
  {\bibinfo  {journal} {arXiv preprint arXiv:2108.13385}\ } (\bibinfo {year}
  {2021})}\BibitemShut {NoStop}%
\bibitem [{\citenamefont {Wang}\ \emph {et~al.}(2015)\citenamefont {Wang},
  \citenamefont {Zhang}, \citenamefont {Xu}, \citenamefont {Zeng},
  \citenamefont {Miao}, \citenamefont {Xu}, \citenamefont {Qian}, \citenamefont
  {Weng}, \citenamefont {Richard}, \citenamefont {Fedorov}, \citenamefont
  {Ding}, \citenamefont {Dai},\ and\ \citenamefont {Fang}}]{wang_prb_2015}%
  \BibitemOpen
  \bibfield  {author} {\bibinfo {author} {\bibfnamefont {Z.}~\bibnamefont
  {Wang}}, \bibinfo {author} {\bibfnamefont {P.}~\bibnamefont {Zhang}},
  \bibinfo {author} {\bibfnamefont {G.}~\bibnamefont {Xu}}, \bibinfo {author}
  {\bibfnamefont {L.~K.}\ \bibnamefont {Zeng}}, \bibinfo {author}
  {\bibfnamefont {H.}~\bibnamefont {Miao}}, \bibinfo {author} {\bibfnamefont
  {X.}~\bibnamefont {Xu}}, \bibinfo {author} {\bibfnamefont {T.}~\bibnamefont
  {Qian}}, \bibinfo {author} {\bibfnamefont {H.}~\bibnamefont {Weng}}, \bibinfo
  {author} {\bibfnamefont {P.}~\bibnamefont {Richard}}, \bibinfo {author}
  {\bibfnamefont {A.~V.}\ \bibnamefont {Fedorov}}, \bibinfo {author}
  {\bibfnamefont {H.}~\bibnamefont {Ding}}, \bibinfo {author} {\bibfnamefont
  {X.}~\bibnamefont {Dai}},\ and\ \bibinfo {author} {\bibfnamefont
  {Z.}~\bibnamefont {Fang}},\ }\href
  {https://doi.org/10.1103/PhysRevB.92.115119} {\bibfield  {journal} {\bibinfo
  {journal} {Phys. Rev. B}\ }\textbf {\bibinfo {volume} {92}},\ \bibinfo
  {pages} {115119} (\bibinfo {year} {2015})}\BibitemShut {NoStop}%
\bibitem [{\citenamefont {Xu}\ \emph {et~al.}(2016)\citenamefont {Xu},
  \citenamefont {Lian}, \citenamefont {Tang}, \citenamefont {Qi},\ and\
  \citenamefont {Zhang}}]{xu_prl_2016}%
  \BibitemOpen
  \bibfield  {author} {\bibinfo {author} {\bibfnamefont {G.}~\bibnamefont
  {Xu}}, \bibinfo {author} {\bibfnamefont {B.}~\bibnamefont {Lian}}, \bibinfo
  {author} {\bibfnamefont {P.}~\bibnamefont {Tang}}, \bibinfo {author}
  {\bibfnamefont {X.-L.}\ \bibnamefont {Qi}},\ and\ \bibinfo {author}
  {\bibfnamefont {S.-C.}\ \bibnamefont {Zhang}},\ }\href
  {https://doi.org/10.1103/PhysRevLett.117.047001} {\bibfield  {journal}
  {\bibinfo  {journal} {Phys. Rev. Lett.}\ }\textbf {\bibinfo {volume} {117}},\
  \bibinfo {pages} {047001} (\bibinfo {year} {2016})}\BibitemShut {NoStop}%
\bibitem [{\citenamefont {Wu}\ \emph {et~al.}(2016)\citenamefont {Wu},
  \citenamefont {Qin}, \citenamefont {Liang}, \citenamefont {Fan},\ and\
  \citenamefont {Hu}}]{wu_prb_2016}%
  \BibitemOpen
  \bibfield  {author} {\bibinfo {author} {\bibfnamefont {X.}~\bibnamefont
  {Wu}}, \bibinfo {author} {\bibfnamefont {S.}~\bibnamefont {Qin}}, \bibinfo
  {author} {\bibfnamefont {Y.}~\bibnamefont {Liang}}, \bibinfo {author}
  {\bibfnamefont {H.}~\bibnamefont {Fan}},\ and\ \bibinfo {author}
  {\bibfnamefont {J.}~\bibnamefont {Hu}},\ }\href
  {https://doi.org/10.1103/PhysRevB.93.115129} {\bibfield  {journal} {\bibinfo
  {journal} {Phys. Rev. B}\ }\textbf {\bibinfo {volume} {93}},\ \bibinfo
  {pages} {115129} (\bibinfo {year} {2016})}\BibitemShut {NoStop}%
\bibitem [{\citenamefont {Kamihara}\ \emph {et~al.}(2008)\citenamefont
  {Kamihara}, \citenamefont {Watanabe}, \citenamefont {Hirano},\ and\
  \citenamefont {Hosono}}]{kamihara_jacs_2008}%
  \BibitemOpen
  \bibfield  {author} {\bibinfo {author} {\bibfnamefont {Y.}~\bibnamefont
  {Kamihara}}, \bibinfo {author} {\bibfnamefont {T.}~\bibnamefont {Watanabe}},
  \bibinfo {author} {\bibfnamefont {M.}~\bibnamefont {Hirano}},\ and\ \bibinfo
  {author} {\bibfnamefont {H.}~\bibnamefont {Hosono}},\ }\href
  {https://doi.org/10.1021/ja800073m} {\bibfield  {journal} {\bibinfo
  {journal} {Journal of the American Chemical Society}\ }\textbf {\bibinfo
  {volume} {130}},\ \bibinfo {pages} {3296} (\bibinfo {year}
  {2008})}\BibitemShut {NoStop}%
\bibitem [{\citenamefont {Fang}\ \emph {et~al.}(2008)\citenamefont {Fang},
  \citenamefont {Pham}, \citenamefont {Qian}, \citenamefont {Liu},
  \citenamefont {Vehstedt}, \citenamefont {Liu}, \citenamefont {Spinu},\ and\
  \citenamefont {Mao}}]{fang_prb_2008}%
  \BibitemOpen
  \bibfield  {author} {\bibinfo {author} {\bibfnamefont {M.~H.}\ \bibnamefont
  {Fang}}, \bibinfo {author} {\bibfnamefont {H.~M.}\ \bibnamefont {Pham}},
  \bibinfo {author} {\bibfnamefont {B.}~\bibnamefont {Qian}}, \bibinfo {author}
  {\bibfnamefont {T.~J.}\ \bibnamefont {Liu}}, \bibinfo {author} {\bibfnamefont
  {E.~K.}\ \bibnamefont {Vehstedt}}, \bibinfo {author} {\bibfnamefont
  {Y.}~\bibnamefont {Liu}}, \bibinfo {author} {\bibfnamefont {L.}~\bibnamefont
  {Spinu}},\ and\ \bibinfo {author} {\bibfnamefont {Z.~Q.}\ \bibnamefont
  {Mao}},\ }\href {https://doi.org/10.1103/PhysRevB.78.224503} {\bibfield
  {journal} {\bibinfo  {journal} {Phys. Rev. B}\ }\textbf {\bibinfo {volume}
  {78}},\ \bibinfo {pages} {224503} (\bibinfo {year} {2008})}\BibitemShut
  {NoStop}%
\bibitem [{\citenamefont {Hanaguri}\ \emph {et~al.}(2010)\citenamefont
  {Hanaguri}, \citenamefont {Niitaka}, \citenamefont {Kuroki},\ and\
  \citenamefont {Takagi}}]{hanaguri_science_2010}%
  \BibitemOpen
  \bibfield  {author} {\bibinfo {author} {\bibfnamefont {T.}~\bibnamefont
  {Hanaguri}}, \bibinfo {author} {\bibfnamefont {S.}~\bibnamefont {Niitaka}},
  \bibinfo {author} {\bibfnamefont {K.}~\bibnamefont {Kuroki}},\ and\ \bibinfo
  {author} {\bibfnamefont {H.}~\bibnamefont {Takagi}},\ }\href
  {https://doi.org/10.1126/science.1187399} {\bibfield  {journal} {\bibinfo
  {journal} {Science}\ }\textbf {\bibinfo {volume} {328}},\ \bibinfo {pages}
  {474} (\bibinfo {year} {2010})}\BibitemShut {NoStop}%
\bibitem [{\citenamefont {Paglione}\ and\ \citenamefont
  {Greene}(2010)}]{paglione_np_2010}%
  \BibitemOpen
  \bibfield  {author} {\bibinfo {author} {\bibfnamefont {J.}~\bibnamefont
  {Paglione}}\ and\ \bibinfo {author} {\bibfnamefont {R.~L.}\ \bibnamefont
  {Greene}},\ }\href {https://doi.org/10.1038/nphys1759} {\bibfield  {journal}
  {\bibinfo  {journal} {Nature physics}\ }\textbf {\bibinfo {volume} {6}},\
  \bibinfo {pages} {645} (\bibinfo {year} {2010})}\BibitemShut {NoStop}%
\bibitem [{\citenamefont {Hirschfeld}\ \emph
  {et~al.}(2011{\natexlab{a}})\citenamefont {Hirschfeld}, \citenamefont
  {Korshunov},\ and\ \citenamefont {Mazin}}]{hirschfeld_rpp_2011}%
  \BibitemOpen
  \bibfield  {author} {\bibinfo {author} {\bibfnamefont {P.}~\bibnamefont
  {Hirschfeld}}, \bibinfo {author} {\bibfnamefont {M.}~\bibnamefont
  {Korshunov}},\ and\ \bibinfo {author} {\bibfnamefont {I.}~\bibnamefont
  {Mazin}},\ }\href {https://doi.org/10.1088/0034-4885/74/12/124508} {\bibfield
   {journal} {\bibinfo  {journal} {Reports on Progress in Physics}\ }\textbf
  {\bibinfo {volume} {74}},\ \bibinfo {pages} {124508} (\bibinfo {year}
  {2011}{\natexlab{a}})}\BibitemShut {NoStop}%
\bibitem [{\citenamefont {Stewart}(2011)}]{stewart_rmp_2011}%
  \BibitemOpen
  \bibfield  {author} {\bibinfo {author} {\bibfnamefont {G.~R.}\ \bibnamefont
  {Stewart}},\ }\href {https://doi.org/10.1103/RevModPhys.83.1589} {\bibfield
  {journal} {\bibinfo  {journal} {Rev. Mod. Phys.}\ }\textbf {\bibinfo {volume}
  {83}},\ \bibinfo {pages} {1589} (\bibinfo {year} {2011})}\BibitemShut
  {NoStop}%
\bibitem [{\citenamefont {Wang}\ and\ \citenamefont
  {Lee}(2011)}]{wang_science_2011}%
  \BibitemOpen
  \bibfield  {author} {\bibinfo {author} {\bibfnamefont {F.}~\bibnamefont
  {Wang}}\ and\ \bibinfo {author} {\bibfnamefont {D.-H.}\ \bibnamefont {Lee}},\
  }\href {https://doi.org/10.1126/science.1200182} {\bibfield  {journal}
  {\bibinfo  {journal} {Science}\ }\textbf {\bibinfo {volume} {332}},\ \bibinfo
  {pages} {200} (\bibinfo {year} {2011})}\BibitemShut {NoStop}%
\bibitem [{\citenamefont {Chubukov}(2012)}]{chubukov_arcmp_2012}%
  \BibitemOpen
  \bibfield  {author} {\bibinfo {author} {\bibfnamefont {A.}~\bibnamefont
  {Chubukov}},\ }\href
  {https://doi.org/10.1146/annurev-conmatphys-020911-125055} {\bibfield
  {journal} {\bibinfo  {journal} {Annu. Rev. Condens. Matter Phys.}\ }\textbf
  {\bibinfo {volume} {3}},\ \bibinfo {pages} {57} (\bibinfo {year}
  {2012})}\BibitemShut {NoStop}%
\bibitem [{\citenamefont {Chubukov}\ and\ \citenamefont
  {Hirschfeld}(2015)}]{chubukov_pt_2015}%
  \BibitemOpen
  \bibfield  {author} {\bibinfo {author} {\bibfnamefont {A.}~\bibnamefont
  {Chubukov}}\ and\ \bibinfo {author} {\bibfnamefont {P.~J.}\ \bibnamefont
  {Hirschfeld}},\ }\href {https://doi.org/10.1063/PT.3.2818} {\bibfield
  {journal} {\bibinfo  {journal} {Physics today}\ }\textbf {\bibinfo {volume}
  {68}},\ \bibinfo {pages} {46} (\bibinfo {year} {2015})}\BibitemShut {NoStop}%
\bibitem [{\citenamefont {Fu}\ and\ \citenamefont {Kane}(2008)}]{fu_prl_2008}%
  \BibitemOpen
  \bibfield  {author} {\bibinfo {author} {\bibfnamefont {L.}~\bibnamefont
  {Fu}}\ and\ \bibinfo {author} {\bibfnamefont {C.~L.}\ \bibnamefont {Kane}},\
  }\href {https://doi.org/10.1103/PhysRevLett.100.096407} {\bibfield  {journal}
  {\bibinfo  {journal} {Phys. Rev. Lett.}\ }\textbf {\bibinfo {volume} {100}},\
  \bibinfo {pages} {096407} (\bibinfo {year} {2008})}\BibitemShut {NoStop}%
\bibitem [{\citenamefont {Day}\ \emph {et~al.}(2021)\citenamefont {Day},
  \citenamefont {Na}, \citenamefont {Zingl}, \citenamefont {Zwartsenberg},
  \citenamefont {Michiardi}, \citenamefont {Levy}, \citenamefont {Schneider},
  \citenamefont {Wong}, \citenamefont {Dosanjh}, \citenamefont {Pedersen},
  \citenamefont {Gorovikov}, \citenamefont {Chi}, \citenamefont {Liang},
  \citenamefont {Hardy}, \citenamefont {Bonn}, \citenamefont {Zhdanovich},
  \citenamefont {Elfimov},\ and\ \citenamefont {Damascelli}}]{day2021three}%
  \BibitemOpen
  \bibfield  {author} {\bibinfo {author} {\bibfnamefont {R.~P.}\ \bibnamefont
  {Day}}, \bibinfo {author} {\bibfnamefont {M.}~\bibnamefont {Na}}, \bibinfo
  {author} {\bibfnamefont {M.}~\bibnamefont {Zingl}}, \bibinfo {author}
  {\bibfnamefont {B.}~\bibnamefont {Zwartsenberg}}, \bibinfo {author}
  {\bibfnamefont {M.}~\bibnamefont {Michiardi}}, \bibinfo {author}
  {\bibfnamefont {G.}~\bibnamefont {Levy}}, \bibinfo {author} {\bibfnamefont
  {M.}~\bibnamefont {Schneider}}, \bibinfo {author} {\bibfnamefont
  {D.}~\bibnamefont {Wong}}, \bibinfo {author} {\bibfnamefont {P.}~\bibnamefont
  {Dosanjh}}, \bibinfo {author} {\bibfnamefont {T.~M.}\ \bibnamefont
  {Pedersen}}, \bibinfo {author} {\bibfnamefont {S.}~\bibnamefont {Gorovikov}},
  \bibinfo {author} {\bibfnamefont {S.}~\bibnamefont {Chi}}, \bibinfo {author}
  {\bibfnamefont {R.}~\bibnamefont {Liang}}, \bibinfo {author} {\bibfnamefont
  {W.~N.}\ \bibnamefont {Hardy}}, \bibinfo {author} {\bibfnamefont {D.~A.}\
  \bibnamefont {Bonn}}, \bibinfo {author} {\bibfnamefont {S.}~\bibnamefont
  {Zhdanovich}}, \bibinfo {author} {\bibfnamefont {I.~S.}\ \bibnamefont
  {Elfimov}},\ and\ \bibinfo {author} {\bibfnamefont {A.}~\bibnamefont
  {Damascelli}},\ }\href {https://arxiv.org/abs/2109.13276} {\bibfield
  {journal} {\bibinfo  {journal} {arXiv preprint arXiv:2109.13276}\ } (\bibinfo
  {year} {2021})}\BibitemShut {NoStop}%
\bibitem [{\citenamefont {K\"onig}\ and\ \citenamefont
  {Coleman}(2019)}]{konig_prl_2019}%
  \BibitemOpen
  \bibfield  {author} {\bibinfo {author} {\bibfnamefont {E.~J.}\ \bibnamefont
  {K\"onig}}\ and\ \bibinfo {author} {\bibfnamefont {P.}~\bibnamefont
  {Coleman}},\ }\href {https://doi.org/10.1103/PhysRevLett.122.207001}
  {\bibfield  {journal} {\bibinfo  {journal} {Phys. Rev. Lett.}\ }\textbf
  {\bibinfo {volume} {122}},\ \bibinfo {pages} {207001} (\bibinfo {year}
  {2019})}\BibitemShut {NoStop}%
\bibitem [{\citenamefont {Qin}\ \emph {et~al.}(2019{\natexlab{a}})\citenamefont
  {Qin}, \citenamefont {Hu}, \citenamefont {Le}, \citenamefont {Zeng},
  \citenamefont {Zhang}, \citenamefont {Fang},\ and\ \citenamefont
  {Hu}}]{qin_prl_2019}%
  \BibitemOpen
  \bibfield  {author} {\bibinfo {author} {\bibfnamefont {S.}~\bibnamefont
  {Qin}}, \bibinfo {author} {\bibfnamefont {L.-H.}\ \bibnamefont {Hu}},
  \bibinfo {author} {\bibfnamefont {C.}~\bibnamefont {Le}}, \bibinfo {author}
  {\bibfnamefont {J.}~\bibnamefont {Zeng}}, \bibinfo {author} {\bibfnamefont
  {F.-C.}\ \bibnamefont {Zhang}}, \bibinfo {author} {\bibfnamefont
  {C.}~\bibnamefont {Fang}},\ and\ \bibinfo {author} {\bibfnamefont
  {J.}~\bibnamefont {Hu}},\ }\href
  {https://doi.org/10.1103/PhysRevLett.123.027003} {\bibfield  {journal}
  {\bibinfo  {journal} {Phys. Rev. Lett.}\ }\textbf {\bibinfo {volume} {123}},\
  \bibinfo {pages} {027003} (\bibinfo {year} {2019}{\natexlab{a}})}\BibitemShut
  {NoStop}%
\bibitem [{\citenamefont {Hosur}\ \emph {et~al.}(2011)\citenamefont {Hosur},
  \citenamefont {Ghaemi}, \citenamefont {Mong},\ and\ \citenamefont
  {Vishwanath}}]{hosur_prl_2011}%
  \BibitemOpen
  \bibfield  {author} {\bibinfo {author} {\bibfnamefont {P.}~\bibnamefont
  {Hosur}}, \bibinfo {author} {\bibfnamefont {P.}~\bibnamefont {Ghaemi}},
  \bibinfo {author} {\bibfnamefont {R.~S.~K.}\ \bibnamefont {Mong}},\ and\
  \bibinfo {author} {\bibfnamefont {A.}~\bibnamefont {Vishwanath}},\ }\href
  {https://doi.org/10.1103/PhysRevLett.107.097001} {\bibfield  {journal}
  {\bibinfo  {journal} {Phys. Rev. Lett.}\ }\textbf {\bibinfo {volume} {107}},\
  \bibinfo {pages} {097001} (\bibinfo {year} {2011})}\BibitemShut {NoStop}%
\bibitem [{\citenamefont {Chiu}\ \emph {et~al.}(2011)\citenamefont {Chiu},
  \citenamefont {Gilbert},\ and\ \citenamefont {Hughes}}]{chiu_prb_2011}%
  \BibitemOpen
  \bibfield  {author} {\bibinfo {author} {\bibfnamefont {C.-K.}\ \bibnamefont
  {Chiu}}, \bibinfo {author} {\bibfnamefont {M.~J.}\ \bibnamefont {Gilbert}},\
  and\ \bibinfo {author} {\bibfnamefont {T.~L.}\ \bibnamefont {Hughes}},\
  }\href {https://doi.org/10.1103/PhysRevB.84.144507} {\bibfield  {journal}
  {\bibinfo  {journal} {Phys. Rev. B}\ }\textbf {\bibinfo {volume} {84}},\
  \bibinfo {pages} {144507} (\bibinfo {year} {2011})}\BibitemShut {NoStop}%
\bibitem [{\citenamefont {Chiu}\ \emph {et~al.}(2012)\citenamefont {Chiu},
  \citenamefont {Ghaemi},\ and\ \citenamefont {Hughes}}]{chiu_prl_2012}%
  \BibitemOpen
  \bibfield  {author} {\bibinfo {author} {\bibfnamefont {C.-K.}\ \bibnamefont
  {Chiu}}, \bibinfo {author} {\bibfnamefont {P.}~\bibnamefont {Ghaemi}},\ and\
  \bibinfo {author} {\bibfnamefont {T.~L.}\ \bibnamefont {Hughes}},\ }\href
  {https://doi.org/10.1103/PhysRevLett.109.237009} {\bibfield  {journal}
  {\bibinfo  {journal} {Phys. Rev. Lett.}\ }\textbf {\bibinfo {volume} {109}},\
  \bibinfo {pages} {237009} (\bibinfo {year} {2012})}\BibitemShut {NoStop}%
\bibitem [{\citenamefont {Li}\ \emph {et~al.}(2014)\citenamefont {Li},
  \citenamefont {Zhang},\ and\ \citenamefont {Wang}}]{li_sr_2014}%
  \BibitemOpen
  \bibfield  {author} {\bibinfo {author} {\bibfnamefont {Z.-Z.}\ \bibnamefont
  {Li}}, \bibinfo {author} {\bibfnamefont {F.-C.}\ \bibnamefont {Zhang}},\ and\
  \bibinfo {author} {\bibfnamefont {Q.-H.}\ \bibnamefont {Wang}},\ }\href
  {https://doi.org/10.1038/srep06363} {\bibfield  {journal} {\bibinfo
  {journal} {Scientific reports}\ }\textbf {\bibinfo {volume} {4}},\ \bibinfo
  {pages} {1} (\bibinfo {year} {2014})}\BibitemShut {NoStop}%
\bibitem [{\citenamefont {Hu}\ \emph {et~al.}(2016)\citenamefont {Hu},
  \citenamefont {Li}, \citenamefont {Xu}, \citenamefont {Zhou},\ and\
  \citenamefont {Zhang}}]{hu_prb_2016}%
  \BibitemOpen
  \bibfield  {author} {\bibinfo {author} {\bibfnamefont {L.-H.}\ \bibnamefont
  {Hu}}, \bibinfo {author} {\bibfnamefont {C.}~\bibnamefont {Li}}, \bibinfo
  {author} {\bibfnamefont {D.-H.}\ \bibnamefont {Xu}}, \bibinfo {author}
  {\bibfnamefont {Y.}~\bibnamefont {Zhou}},\ and\ \bibinfo {author}
  {\bibfnamefont {F.-C.}\ \bibnamefont {Zhang}},\ }\href
  {https://doi.org/10.1103/PhysRevB.94.224501} {\bibfield  {journal} {\bibinfo
  {journal} {Phys. Rev. B}\ }\textbf {\bibinfo {volume} {94}},\ \bibinfo
  {pages} {224501} (\bibinfo {year} {2016})}\BibitemShut {NoStop}%
\bibitem [{\citenamefont {Yan}\ \emph {et~al.}(2020)\citenamefont {Yan},
  \citenamefont {Wu},\ and\ \citenamefont {Huang}}]{yan_prl_2020}%
  \BibitemOpen
  \bibfield  {author} {\bibinfo {author} {\bibfnamefont {Z.}~\bibnamefont
  {Yan}}, \bibinfo {author} {\bibfnamefont {Z.}~\bibnamefont {Wu}},\ and\
  \bibinfo {author} {\bibfnamefont {W.}~\bibnamefont {Huang}},\ }\href
  {https://doi.org/10.1103/PhysRevLett.124.257001} {\bibfield  {journal}
  {\bibinfo  {journal} {Phys. Rev. Lett.}\ }\textbf {\bibinfo {volume} {124}},\
  \bibinfo {pages} {257001} (\bibinfo {year} {2020})}\BibitemShut {NoStop}%
\bibitem [{\citenamefont {Chiu}\ \emph {et~al.}(2020)\citenamefont {Chiu},
  \citenamefont {Machida}, \citenamefont {Huang}, \citenamefont {Hanaguri},\
  and\ \citenamefont {Zhang}}]{chiu_sa_2020}%
  \BibitemOpen
  \bibfield  {author} {\bibinfo {author} {\bibfnamefont {C.-K.}\ \bibnamefont
  {Chiu}}, \bibinfo {author} {\bibfnamefont {T.}~\bibnamefont {Machida}},
  \bibinfo {author} {\bibfnamefont {Y.}~\bibnamefont {Huang}}, \bibinfo
  {author} {\bibfnamefont {T.}~\bibnamefont {Hanaguri}},\ and\ \bibinfo
  {author} {\bibfnamefont {F.-C.}\ \bibnamefont {Zhang}},\ }\href
  {https://doi.org/10.1126/sciadv.aay0443} {\bibfield  {journal} {\bibinfo
  {journal} {Science advances}\ }\textbf {\bibinfo {volume} {6}},\ \bibinfo
  {pages} {eaay0443} (\bibinfo {year} {2020})}\BibitemShut {NoStop}%
\bibitem [{\citenamefont {Ghazaryan}\ \emph {et~al.}(2020)\citenamefont
  {Ghazaryan}, \citenamefont {Lopes}, \citenamefont {Hosur}, \citenamefont
  {Gilbert},\ and\ \citenamefont {Ghaemi}}]{ghazaryan_prb_2020}%
  \BibitemOpen
  \bibfield  {author} {\bibinfo {author} {\bibfnamefont {A.}~\bibnamefont
  {Ghazaryan}}, \bibinfo {author} {\bibfnamefont {P.~L.~S.}\ \bibnamefont
  {Lopes}}, \bibinfo {author} {\bibfnamefont {P.}~\bibnamefont {Hosur}},
  \bibinfo {author} {\bibfnamefont {M.~J.}\ \bibnamefont {Gilbert}},\ and\
  \bibinfo {author} {\bibfnamefont {P.}~\bibnamefont {Ghaemi}},\ }\href
  {https://doi.org/10.1103/PhysRevB.101.020504} {\bibfield  {journal} {\bibinfo
   {journal} {Phys. Rev. B}\ }\textbf {\bibinfo {volume} {101}},\ \bibinfo
  {pages} {020504} (\bibinfo {year} {2020})}\BibitemShut {NoStop}%
\bibitem [{\citenamefont {Li}\ \emph {et~al.}(2021)\citenamefont {Li},
  \citenamefont {Luo}, \citenamefont {Chen}, \citenamefont {Liu}, \citenamefont
  {Zhang},\ and\ \citenamefont {Liu}}]{li_arxiv_2021}%
  \BibitemOpen
  \bibfield  {author} {\bibinfo {author} {\bibfnamefont {C.}~\bibnamefont
  {Li}}, \bibinfo {author} {\bibfnamefont {X.-J.}\ \bibnamefont {Luo}},
  \bibinfo {author} {\bibfnamefont {L.}~\bibnamefont {Chen}}, \bibinfo {author}
  {\bibfnamefont {D.~E.}\ \bibnamefont {Liu}}, \bibinfo {author} {\bibfnamefont
  {F.-C.}\ \bibnamefont {Zhang}},\ and\ \bibinfo {author} {\bibfnamefont
  {X.}~\bibnamefont {Liu}},\ }\href
  {https://ui.adsabs.harvard.edu/abs/2021arXiv210711562L} {\bibfield  {journal}
  {\bibinfo  {journal} {arXiv preprint arXiv:2107.11562}\ } (\bibinfo {year}
  {2021})}\BibitemShut {NoStop}%
\bibitem [{\citenamefont {Qin}\ \emph {et~al.}(2019{\natexlab{b}})\citenamefont
  {Qin}, \citenamefont {Hu}, \citenamefont {Wu}, \citenamefont {Dai},
  \citenamefont {Fang}, \citenamefont {Zhang},\ and\ \citenamefont
  {Hu}}]{qin_sb_2019}%
  \BibitemOpen
  \bibfield  {author} {\bibinfo {author} {\bibfnamefont {S.}~\bibnamefont
  {Qin}}, \bibinfo {author} {\bibfnamefont {L.-H.}\ \bibnamefont {Hu}},
  \bibinfo {author} {\bibfnamefont {X.}~\bibnamefont {Wu}}, \bibinfo {author}
  {\bibfnamefont {X.}~\bibnamefont {Dai}}, \bibinfo {author} {\bibfnamefont
  {C.}~\bibnamefont {Fang}}, \bibinfo {author} {\bibfnamefont {F.-C.}\
  \bibnamefont {Zhang}},\ and\ \bibinfo {author} {\bibfnamefont
  {J.}~\bibnamefont {Hu}},\ }\href {https://doi.org/10.1016/j.scib.2019.07.011}
  {\bibfield  {journal} {\bibinfo  {journal} {Science Bulletin}\ }\textbf
  {\bibinfo {volume} {64}},\ \bibinfo {pages} {1207} (\bibinfo {year}
  {2019}{\natexlab{b}})}\BibitemShut {NoStop}%
\bibitem [{\citenamefont {Borisenko}\ \emph {et~al.}(2016)\citenamefont
  {Borisenko}, \citenamefont {Evtushinsky}, \citenamefont {Liu}, \citenamefont
  {Morozov}, \citenamefont {Kappenberger}, \citenamefont {Wurmehl},
  \citenamefont {Buchner}, \citenamefont {Yaresko}, \citenamefont {Kim},
  \citenamefont {Hoesch} \emph {et~al.}}]{Borisenko2016}%
  \BibitemOpen
  \bibfield  {author} {\bibinfo {author} {\bibfnamefont {S.}~\bibnamefont
  {Borisenko}}, \bibinfo {author} {\bibfnamefont {D.}~\bibnamefont
  {Evtushinsky}}, \bibinfo {author} {\bibfnamefont {Z.-H.}\ \bibnamefont
  {Liu}}, \bibinfo {author} {\bibfnamefont {I.}~\bibnamefont {Morozov}},
  \bibinfo {author} {\bibfnamefont {R.}~\bibnamefont {Kappenberger}}, \bibinfo
  {author} {\bibfnamefont {S.}~\bibnamefont {Wurmehl}}, \bibinfo {author}
  {\bibfnamefont {B.}~\bibnamefont {Buchner}}, \bibinfo {author} {\bibfnamefont
  {A.}~\bibnamefont {Yaresko}}, \bibinfo {author} {\bibfnamefont
  {T.}~\bibnamefont {Kim}}, \bibinfo {author} {\bibfnamefont {M.}~\bibnamefont
  {Hoesch}}, \emph {et~al.},\ }\href {https://doi.org/10.1038/nphys3594}
  {\bibfield  {journal} {\bibinfo  {journal} {Nature Physics}\ }\textbf
  {\bibinfo {volume} {12}},\ \bibinfo {pages} {311} (\bibinfo {year}
  {2016})}\BibitemShut {NoStop}%
\bibitem [{\citenamefont {Zhang}\ \emph {et~al.}(2009)\citenamefont {Zhang},
  \citenamefont {Liu}, \citenamefont {Qi}, \citenamefont {Dai}, \citenamefont
  {Fang},\ and\ \citenamefont {Zhang}}]{zhang_np_2009}%
  \BibitemOpen
  \bibfield  {author} {\bibinfo {author} {\bibfnamefont {H.}~\bibnamefont
  {Zhang}}, \bibinfo {author} {\bibfnamefont {C.-X.}\ \bibnamefont {Liu}},
  \bibinfo {author} {\bibfnamefont {X.-L.}\ \bibnamefont {Qi}}, \bibinfo
  {author} {\bibfnamefont {X.}~\bibnamefont {Dai}}, \bibinfo {author}
  {\bibfnamefont {Z.}~\bibnamefont {Fang}},\ and\ \bibinfo {author}
  {\bibfnamefont {S.-C.}\ \bibnamefont {Zhang}},\ }\href
  {https://doi.org/10.1038/nphys1270} {\bibfield  {journal} {\bibinfo
  {journal} {Nature physics}\ }\textbf {\bibinfo {volume} {5}},\ \bibinfo
  {pages} {438} (\bibinfo {year} {2009})}\BibitemShut {NoStop}%
\bibitem [{\citenamefont {Liu}\ \emph {et~al.}(2010)\citenamefont {Liu},
  \citenamefont {Qi}, \citenamefont {Zhang}, \citenamefont {Dai}, \citenamefont
  {Fang},\ and\ \citenamefont {Zhang}}]{liu_prb_2010}%
  \BibitemOpen
  \bibfield  {author} {\bibinfo {author} {\bibfnamefont {C.-X.}\ \bibnamefont
  {Liu}}, \bibinfo {author} {\bibfnamefont {X.-L.}\ \bibnamefont {Qi}},
  \bibinfo {author} {\bibfnamefont {H.}~\bibnamefont {Zhang}}, \bibinfo
  {author} {\bibfnamefont {X.}~\bibnamefont {Dai}}, \bibinfo {author}
  {\bibfnamefont {Z.}~\bibnamefont {Fang}},\ and\ \bibinfo {author}
  {\bibfnamefont {S.-C.}\ \bibnamefont {Zhang}},\ }\href
  {https://doi.org/10.1103/PhysRevB.82.045122} {\bibfield  {journal} {\bibinfo
  {journal} {Phys. Rev. B}\ }\textbf {\bibinfo {volume} {82}},\ \bibinfo
  {pages} {045122} (\bibinfo {year} {2010})}\BibitemShut {NoStop}%
\bibitem [{\citenamefont {Armitage}\ \emph {et~al.}(2018)\citenamefont
  {Armitage}, \citenamefont {Mele},\ and\ \citenamefont
  {Vishwanath}}]{armitage_rmp_2018}%
  \BibitemOpen
  \bibfield  {author} {\bibinfo {author} {\bibfnamefont {N.~P.}\ \bibnamefont
  {Armitage}}, \bibinfo {author} {\bibfnamefont {E.~J.}\ \bibnamefont {Mele}},\
  and\ \bibinfo {author} {\bibfnamefont {A.}~\bibnamefont {Vishwanath}},\
  }\href {https://doi.org/10.1103/RevModPhys.90.015001} {\bibfield  {journal}
  {\bibinfo  {journal} {Rev. Mod. Phys.}\ }\textbf {\bibinfo {volume} {90}},\
  \bibinfo {pages} {015001} (\bibinfo {year} {2018})}\BibitemShut {NoStop}%
\bibitem [{\citenamefont {Lv}\ \emph {et~al.}(2021)\citenamefont {Lv},
  \citenamefont {Qian},\ and\ \citenamefont {Ding}}]{lv_rmp_2021}%
  \BibitemOpen
  \bibfield  {author} {\bibinfo {author} {\bibfnamefont {B.~Q.}\ \bibnamefont
  {Lv}}, \bibinfo {author} {\bibfnamefont {T.}~\bibnamefont {Qian}},\ and\
  \bibinfo {author} {\bibfnamefont {H.}~\bibnamefont {Ding}},\ }\href
  {https://doi.org/10.1103/RevModPhys.93.025002} {\bibfield  {journal}
  {\bibinfo  {journal} {Rev. Mod. Phys.}\ }\textbf {\bibinfo {volume} {93}},\
  \bibinfo {pages} {025002} (\bibinfo {year} {2021})}\BibitemShut {NoStop}%
\bibitem [{foo()}]{footnote}%
  \BibitemOpen
  \href@noop {} {}\bibinfo {note} {Most bulk iron-based superconductors,
  including these tFeSc candiddates focused in this work, are commonly believed
  to carry an $s_{\pm}$-wave pairing symmetry~\cite{Hirschfeld_2011}. Since the
  $s_{\pm}$-wave pairing will contribute to the vortex topology in the same way
  as an isotropic $s$-wave pairing, we choose to use the $s$-wave pairing in
  our model for simpliciy.}\BibitemShut {Stop}%
\bibitem [{\citenamefont {Chiu}\ \emph {et~al.}(2016)\citenamefont {Chiu},
  \citenamefont {Teo}, \citenamefont {Schnyder},\ and\ \citenamefont
  {Ryu}}]{chiu_rmp_2016}%
  \BibitemOpen
  \bibfield  {author} {\bibinfo {author} {\bibfnamefont {C.-K.}\ \bibnamefont
  {Chiu}}, \bibinfo {author} {\bibfnamefont {J.~C.~Y.}\ \bibnamefont {Teo}},
  \bibinfo {author} {\bibfnamefont {A.~P.}\ \bibnamefont {Schnyder}},\ and\
  \bibinfo {author} {\bibfnamefont {S.}~\bibnamefont {Ryu}},\ }\href
  {https://doi.org/10.1103/RevModPhys.88.035005} {\bibfield  {journal}
  {\bibinfo  {journal} {Rev. Mod. Phys.}\ }\textbf {\bibinfo {volume} {88}},\
  \bibinfo {pages} {035005} (\bibinfo {year} {2016})}\BibitemShut {NoStop}%
\bibitem [{sup()}]{supplementary}%
  \BibitemOpen
  \href@noop {} {}\bibinfo {note} {See Supplemental Material at [URL] for
  numerical details on the vortex spectrum calculations, analytical analysis on
  how the DSM bands will renormalize the TI bands in the infinite $\delta_{so}$
  limit, the low-energy projection for the effective vortex Hamiltonians, the
  $C_4$-breaking vortex Hamiltonian by tilting magnetic field, the effect of
  curved vortex line induced by bulk impurities, and a discussion of vortex
  topological phase diagram for CaKFe$_4$As$_4$, which includes
  Ref.~[\onlinecite{hosur_prl_2011,li_scpma_2019,qin_prl_2019,zhangST_prb_2019,brandt_rpp_1995,kitaev2009periodic,ryu2010topological,fang_prl_2014,liu_prb_2014,Kobayashi_prb_2020,liu_nc_2020}].}\BibitemShut
  {Stop}%
\bibitem [{\citenamefont {Cheng}\ and\ \citenamefont
  {Tu}(2011)}]{cheng2011Majorana}%
  \BibitemOpen
  \bibfield  {author} {\bibinfo {author} {\bibfnamefont {M.}~\bibnamefont
  {Cheng}}\ and\ \bibinfo {author} {\bibfnamefont {H.-H.}\ \bibnamefont {Tu}},\
  }\href {https://doi.org/10.1103/PhysRevB.84.094503} {\bibfield  {journal}
  {\bibinfo  {journal} {Phys. Rev. B}\ }\textbf {\bibinfo {volume} {84}},\
  \bibinfo {pages} {094503} (\bibinfo {year} {2011})}\BibitemShut {NoStop}%
\bibitem [{\citenamefont {Zhang}\ and\ \citenamefont
  {Liu}(2018)}]{zhang2018crystalline}%
  \BibitemOpen
  \bibfield  {author} {\bibinfo {author} {\bibfnamefont {R.-X.}\ \bibnamefont
  {Zhang}}\ and\ \bibinfo {author} {\bibfnamefont {C.-X.}\ \bibnamefont
  {Liu}},\ }\href {https://doi.org/10.1103/PhysRevLett.120.156802} {\bibfield
  {journal} {\bibinfo  {journal} {Phys. Rev. Lett.}\ }\textbf {\bibinfo
  {volume} {120}},\ \bibinfo {pages} {156802} (\bibinfo {year}
  {2018})}\BibitemShut {NoStop}%
\bibitem [{\citenamefont {Gygi}\ and\ \citenamefont
  {Schl\"uter}(1991)}]{gygi_prb_1991}%
  \BibitemOpen
  \bibfield  {author} {\bibinfo {author} {\bibfnamefont {F.~m.~c.}\
  \bibnamefont {Gygi}}\ and\ \bibinfo {author} {\bibfnamefont {M.}~\bibnamefont
  {Schl\"uter}},\ }\href {https://doi.org/10.1103/PhysRevB.43.7609} {\bibfield
  {journal} {\bibinfo  {journal} {Phys. Rev. B}\ }\textbf {\bibinfo {volume}
  {43}},\ \bibinfo {pages} {7609} (\bibinfo {year} {1991})}\BibitemShut
  {NoStop}%
\bibitem [{\citenamefont {Zhang}\ \emph
  {et~al.}(2019{\natexlab{d}})\citenamefont {Zhang}, \citenamefont {Yin},
  \citenamefont {Dai}, \citenamefont {Zheng}, \citenamefont {Chang},
  \citenamefont {Belopolski}, \citenamefont {Wang}, \citenamefont {Lin},
  \citenamefont {Wang}, \citenamefont {Jin},\ and\ \citenamefont
  {Hasan}}]{zhangST_prb_2019}%
  \BibitemOpen
  \bibfield  {author} {\bibinfo {author} {\bibfnamefont {S.~S.}\ \bibnamefont
  {Zhang}}, \bibinfo {author} {\bibfnamefont {J.-X.}\ \bibnamefont {Yin}},
  \bibinfo {author} {\bibfnamefont {G.}~\bibnamefont {Dai}}, \bibinfo {author}
  {\bibfnamefont {H.}~\bibnamefont {Zheng}}, \bibinfo {author} {\bibfnamefont
  {G.}~\bibnamefont {Chang}}, \bibinfo {author} {\bibfnamefont
  {I.}~\bibnamefont {Belopolski}}, \bibinfo {author} {\bibfnamefont
  {X.}~\bibnamefont {Wang}}, \bibinfo {author} {\bibfnamefont {H.}~\bibnamefont
  {Lin}}, \bibinfo {author} {\bibfnamefont {Z.}~\bibnamefont {Wang}}, \bibinfo
  {author} {\bibfnamefont {C.}~\bibnamefont {Jin}},\ and\ \bibinfo {author}
  {\bibfnamefont {M.~Z.}\ \bibnamefont {Hasan}},\ }\href
  {https://doi.org/10.1103/PhysRevB.99.161103} {\bibfield  {journal} {\bibinfo
  {journal} {Phys. Rev. B}\ }\textbf {\bibinfo {volume} {99}},\ \bibinfo
  {pages} {161103} (\bibinfo {year} {2019}{\natexlab{d}})}\BibitemShut
  {NoStop}%
\bibitem [{\citenamefont {Giwa}\ and\ \citenamefont
  {Hosur}(2021)}]{giwa_prl_2021}%
  \BibitemOpen
  \bibfield  {author} {\bibinfo {author} {\bibfnamefont {R.}~\bibnamefont
  {Giwa}}\ and\ \bibinfo {author} {\bibfnamefont {P.}~\bibnamefont {Hosur}},\
  }\href {https://doi.org/10.1103/PhysRevLett.127.187002} {\bibfield  {journal}
  {\bibinfo  {journal} {Phys. Rev. Lett.}\ }\textbf {\bibinfo {volume} {127}},\
  \bibinfo {pages} {187002} (\bibinfo {year} {2021})}\BibitemShut {NoStop}%
\bibitem [{\citenamefont {Giwa}\ and\ \citenamefont
  {Hosur}(2022)}]{giwa_arxiv_2022}%
  \BibitemOpen
  \bibfield  {author} {\bibinfo {author} {\bibfnamefont {R.}~\bibnamefont
  {Giwa}}\ and\ \bibinfo {author} {\bibfnamefont {P.}~\bibnamefont {Hosur}},\
  }\href {https://arxiv.org/abs/2203.06893} {\bibfield  {journal} {\bibinfo
  {journal} {arXiv preprint arXiv:2203.06893}\ } (\bibinfo {year}
  {2022})}\BibitemShut {NoStop}%
\bibitem [{\citenamefont {Li}\ \emph {et~al.}(2022)\citenamefont {Li},
  \citenamefont {Li}, \citenamefont {Cao}, \citenamefont {Zhou}, \citenamefont
  {Wang}, \citenamefont {Jin}, \citenamefont {Chiu}, \citenamefont {Pennycook},
  \citenamefont {Wang},\ and\ \citenamefont {Gao}}]{li_nature_2022}%
  \BibitemOpen
  \bibfield  {author} {\bibinfo {author} {\bibfnamefont {M.}~\bibnamefont
  {Li}}, \bibinfo {author} {\bibfnamefont {G.}~\bibnamefont {Li}}, \bibinfo
  {author} {\bibfnamefont {L.}~\bibnamefont {Cao}}, \bibinfo {author}
  {\bibfnamefont {X.}~\bibnamefont {Zhou}}, \bibinfo {author} {\bibfnamefont
  {X.}~\bibnamefont {Wang}}, \bibinfo {author} {\bibfnamefont {C.}~\bibnamefont
  {Jin}}, \bibinfo {author} {\bibfnamefont {C.-K.}\ \bibnamefont {Chiu}},
  \bibinfo {author} {\bibfnamefont {S.~J.}\ \bibnamefont {Pennycook}}, \bibinfo
  {author} {\bibfnamefont {Z.}~\bibnamefont {Wang}},\ and\ \bibinfo {author}
  {\bibfnamefont {H.-J.}\ \bibnamefont {Gao}},\ }\href
  {https://doi.org/10.1038/s41586-022-04744-8} {\bibfield  {journal} {\bibinfo
  {journal} {Nature}\ ,\ \bibinfo {pages} {1}} (\bibinfo {year}
  {2022})}\BibitemShut {NoStop}%
\bibitem [{\citenamefont {Liu}\ \emph {et~al.}(2021)\citenamefont {Liu},
  \citenamefont {Hu}, \citenamefont {Wang}, \citenamefont {Zhong},
  \citenamefont {Yang}, \citenamefont {Kong}, \citenamefont {Cao},
  \citenamefont {Li}, \citenamefont {Okazaki}, \citenamefont {Kondo} \emph
  {et~al.}}]{liu_arxiv_2021}%
  \BibitemOpen
  \bibfield  {author} {\bibinfo {author} {\bibfnamefont {W.}~\bibnamefont
  {Liu}}, \bibinfo {author} {\bibfnamefont {Q.}~\bibnamefont {Hu}}, \bibinfo
  {author} {\bibfnamefont {X.}~\bibnamefont {Wang}}, \bibinfo {author}
  {\bibfnamefont {Y.}~\bibnamefont {Zhong}}, \bibinfo {author} {\bibfnamefont
  {F.}~\bibnamefont {Yang}}, \bibinfo {author} {\bibfnamefont {L.}~\bibnamefont
  {Kong}}, \bibinfo {author} {\bibfnamefont {L.}~\bibnamefont {Cao}}, \bibinfo
  {author} {\bibfnamefont {G.}~\bibnamefont {Li}}, \bibinfo {author}
  {\bibfnamefont {K.}~\bibnamefont {Okazaki}}, \bibinfo {author} {\bibfnamefont
  {T.}~\bibnamefont {Kondo}}, \emph {et~al.},\ }\href
  {https://arxiv.org/abs/2111.03786} {\bibfield  {journal} {\bibinfo  {journal}
  {arXiv preprint arXiv:2111.03786}\ } (\bibinfo {year} {2021})}\BibitemShut
  {NoStop}%
\bibitem [{\citenamefont {Hirschfeld}\ \emph
  {et~al.}(2011{\natexlab{b}})\citenamefont {Hirschfeld}, \citenamefont
  {Korshunov},\ and\ \citenamefont {Mazin}}]{Hirschfeld_2011}%
  \BibitemOpen
  \bibfield  {author} {\bibinfo {author} {\bibfnamefont {P.~J.}\ \bibnamefont
  {Hirschfeld}}, \bibinfo {author} {\bibfnamefont {M.~M.}\ \bibnamefont
  {Korshunov}},\ and\ \bibinfo {author} {\bibfnamefont {I.~I.}\ \bibnamefont
  {Mazin}},\ }\href {https://doi.org/10.1088/0034-4885/74/12/124508} {\bibfield
   {journal} {\bibinfo  {journal} {Reports on Progress in Physics}\ }\textbf
  {\bibinfo {volume} {74}},\ \bibinfo {pages} {124508} (\bibinfo {year}
  {2011}{\natexlab{b}})}\BibitemShut {NoStop}%
\bibitem [{\citenamefont {Li}\ \emph {et~al.}(2019)\citenamefont {Li},
  \citenamefont {Hu},\ and\ \citenamefont {Zhang}}]{li_scpma_2019}%
  \BibitemOpen
  \bibfield  {author} {\bibinfo {author} {\bibfnamefont {C.}~\bibnamefont
  {Li}}, \bibinfo {author} {\bibfnamefont {L.-H.}\ \bibnamefont {Hu}},\ and\
  \bibinfo {author} {\bibfnamefont {F.-C.}\ \bibnamefont {Zhang}},\ }\href
  {https://doi.org/10.1007/s11433-019-9391-7} {\bibfield  {journal} {\bibinfo
  {journal} {SCIENCE CHINA Physics, Mechanics \& Astronomy}\ }\textbf {\bibinfo
  {volume} {62}},\ \bibinfo {pages} {1} (\bibinfo {year} {2019})}\BibitemShut
  {NoStop}%
\bibitem [{\citenamefont {Brandt}(1995)}]{brandt_rpp_1995}%
  \BibitemOpen
  \bibfield  {author} {\bibinfo {author} {\bibfnamefont {E.~H.}\ \bibnamefont
  {Brandt}},\ }\href {https://doi.org/10.1088/0034-4885/58/11/003} {\bibfield
  {journal} {\bibinfo  {journal} {Reports on Progress in Physics}\ }\textbf
  {\bibinfo {volume} {58}},\ \bibinfo {pages} {1465} (\bibinfo {year}
  {1995})}\BibitemShut {NoStop}%
\bibitem [{\citenamefont {Kitaev}(2009)}]{kitaev2009periodic}%
  \BibitemOpen
  \bibfield  {author} {\bibinfo {author} {\bibfnamefont {A.}~\bibnamefont
  {Kitaev}},\ }in\ \href {https://doi.org/10.1063/1.3149495} {\emph {\bibinfo
  {booktitle} {AIP conference proceedings}}},\ Vol.\ \bibinfo {volume} {1134}\
  (\bibinfo {organization} {American Institute of Physics},\ \bibinfo {year}
  {2009})\ pp.\ \bibinfo {pages} {22--30}\BibitemShut {NoStop}%
\bibitem [{\citenamefont {Ryu}\ \emph {et~al.}(2010)\citenamefont {Ryu},
  \citenamefont {Schnyder}, \citenamefont {Furusaki},\ and\ \citenamefont
  {Ludwig}}]{ryu2010topological}%
  \BibitemOpen
  \bibfield  {author} {\bibinfo {author} {\bibfnamefont {S.}~\bibnamefont
  {Ryu}}, \bibinfo {author} {\bibfnamefont {A.~P.}\ \bibnamefont {Schnyder}},
  \bibinfo {author} {\bibfnamefont {A.}~\bibnamefont {Furusaki}},\ and\
  \bibinfo {author} {\bibfnamefont {A.~W.}\ \bibnamefont {Ludwig}},\ }\href
  {https://doi.org/10.1088/1367-2630/12/6/065010} {\bibfield  {journal}
  {\bibinfo  {journal} {New Journal of Physics}\ }\textbf {\bibinfo {volume}
  {12}},\ \bibinfo {pages} {065010} (\bibinfo {year} {2010})}\BibitemShut
  {NoStop}%
\bibitem [{\citenamefont {Fang}\ \emph {et~al.}(2014)\citenamefont {Fang},
  \citenamefont {Gilbert},\ and\ \citenamefont {Bernevig}}]{fang_prl_2014}%
  \BibitemOpen
  \bibfield  {author} {\bibinfo {author} {\bibfnamefont {C.}~\bibnamefont
  {Fang}}, \bibinfo {author} {\bibfnamefont {M.~J.}\ \bibnamefont {Gilbert}},\
  and\ \bibinfo {author} {\bibfnamefont {B.~A.}\ \bibnamefont {Bernevig}},\
  }\href {https://doi.org/10.1103/PhysRevLett.112.106401} {\bibfield  {journal}
  {\bibinfo  {journal} {Phys. Rev. Lett.}\ }\textbf {\bibinfo {volume} {112}},\
  \bibinfo {pages} {106401} (\bibinfo {year} {2014})}\BibitemShut {NoStop}%
\bibitem [{\citenamefont {Liu}\ \emph {et~al.}(2014)\citenamefont {Liu},
  \citenamefont {He},\ and\ \citenamefont {Law}}]{liu_prb_2014}%
  \BibitemOpen
  \bibfield  {author} {\bibinfo {author} {\bibfnamefont {X.-J.}\ \bibnamefont
  {Liu}}, \bibinfo {author} {\bibfnamefont {J.~J.}\ \bibnamefont {He}},\ and\
  \bibinfo {author} {\bibfnamefont {K.~T.}\ \bibnamefont {Law}},\ }\href
  {https://doi.org/10.1103/PhysRevB.90.235141} {\bibfield  {journal} {\bibinfo
  {journal} {Phys. Rev. B}\ }\textbf {\bibinfo {volume} {90}},\ \bibinfo
  {pages} {235141} (\bibinfo {year} {2014})}\BibitemShut {NoStop}%
\bibitem [{\citenamefont {Kobayashi}\ and\ \citenamefont
  {Furusaki}(2020)}]{Kobayashi_prb_2020}%
  \BibitemOpen
  \bibfield  {author} {\bibinfo {author} {\bibfnamefont {S.}~\bibnamefont
  {Kobayashi}}\ and\ \bibinfo {author} {\bibfnamefont {A.}~\bibnamefont
  {Furusaki}},\ }\href {https://doi.org/10.1103/PhysRevB.102.180505} {\bibfield
   {journal} {\bibinfo  {journal} {Phys. Rev. B}\ }\textbf {\bibinfo {volume}
  {102}},\ \bibinfo {pages} {180505} (\bibinfo {year} {2020})}\BibitemShut
  {NoStop}%
\end{thebibliography}%

\clearpage
\appendix
\onecolumngrid

\begin{center}
	\bf	Supplemental Material for ``Competing Vortex Topologies in Iron-based Superconductors"
\end{center}

\tableofcontents

\section{Appendix A: 6-band Model Hamiltonian and Spectrum for Nodal Vortex Phase}

In this section, we study the topological vortex Majorana bound states (vMBSs) in topological iron-based superconductors (tFeSCs), whose normal band structure contains both Dirac semi-metal phase and topological insulator phase.
Therefore, the minimal model to capture the main physics for tFeSCs is a 6-band Hamiltonian, given in Eq.~(1) in the main text.  
The basis function for this 6-band model reads
\begin{align}
	\Psi_{\mathbf{k}} =  \left\{  \vert p_z,\uparrow\rangle, \vert p_z,\downarrow\rangle, \vert d_{xz+iyz},\downarrow\rangle, \vert d_{xz-iyz},\uparrow\rangle, \vert d_{xz+iyz},\uparrow\rangle, \vert d_{xz-iyz},\downarrow\rangle \right\},
\end{align}
which can be rewritten in terms of z-component total angular momentum and the parity of the basis state as,
\begin{align}
	\Psi_{\mathbf{k}} =  \left\{  \vert p_-, +\tfrac{1}{2}\rangle, \vert p_-, -\tfrac{1}{2}\rangle,  \vert d_+, +\tfrac{1}{2}\rangle, \vert d_+, -\tfrac{1}{2}\rangle, \vert d_+, +\tfrac{3}{2}\rangle, \vert d_+, -\tfrac{3}{2}\rangle \right\}.
\end{align}
The normal Hamiltonian reads
\begin{align}\label{eq-six-ham0}
	\mathcal{H}_0 = \begin{pmatrix}
		M_1(\mathbf{k}) & 0 & A_2k_z & -A_1k_- & A_1k_+ & 0 \\ 
		0 & M_1(\mathbf{k}) & A_1k_+ & A_2k_z & 0 & -A_1k_- \\ 
		A_2k_z & A_1k_- & M_2(\mathbf{k}) & 0 & 0 & D^\ast(\mathbf{k}) \\ 
		-A_1k_+ & A_2k_z & 0 & M_2(\mathbf{k}) & D(\mathbf{k}) & 0 \\ 
		A_1k_- & 0 & 0 & D^\ast(\mathbf{k}) & M_2(\mathbf{k})+\delta_{so} & 0 \\ 
		0 & -A_1k_+ & D(\mathbf{k}) & 0 & 0 & M_2(\mathbf{k})+\delta_{so}
	\end{pmatrix}, 
\end{align}
where $k_\pm = k_x \pm ik_y$, $M_i(\mathbf{k})=M_0^{(i)} + M_1^{(i)}(k_x^2+k_y^2)+M_2^{(i)}k_z^2$ and $D(\mathbf{k})=B_1(k_x^2-k_y^2)-iB_2k_xk_y$.
Here, the $D(\mathbf{k})$ term determines that $d_{xz}$ and $d_{yz}$ bands have distinct masses,
leading to two hole pockets in iron-based superconductors.
And $\delta_{so}$ is the spin-orbital coupling (SOC), which leads to the shifting of the 3D Dirac point when varying the SOC splitting of $d$-orbital bands.
We will show it is the most important parameter for vortex topology later.
To simplify the calculation, $k_z\to \sin k_z$ and $k_z^2\to 2(1-\cos k_z) $ will be used.

For Hamiltonian~\eqref{eq-six-ham0}, the important symmetries are
\begin{align}
	C_{4z} = e^{i\tfrac{\pi}{2}J_z}, \quad
	\Theta = \begin{pmatrix}
		-is_y & 0 & 0 \\ 
		0 & is_y & 0 \\ 
		0 & 0 & -is_y
	\end{pmatrix} \mathcal{K}, \quad
	\mathcal{P},
\end{align}
where $J_z=\text{diag}[\frac{1}{2},-\frac{1}{2},\frac{1}{2},-\frac{1}{2},\frac{3}{2},-\frac{3}{2}]$ is the z-component of the total angular momentum, $\mathcal{K}$ is the complex conjugation and $\mathcal{P}=\text{diag}[-1,-1,1,1,1,1]$ is the spatial inversion symmetry. In addition, the system has mirror symmetry $M_z$ with respect to z axis, defied as
\begin{align} \label{sm-eq-mz-normal}
	M_z=i\times\text{diag}[-1,1,1,-1,-1,1].
\end{align}

To study the topological vortex states, we introduce the 1D vortex line with $\pi$-flux inserted along z axis.
Next, we solve the Bogoliubov–de Gennes (BdG) Hamiltonian as,
\begin{align}\label{eq-bdg-ham}
	\mathcal{H}_{BdG} = \begin{pmatrix}
		\mathcal{H}_0(\mathbf{k})  & \mathcal{H}_\Delta \\ 
		\mathcal{H}_\Delta^\dagger & -\mathcal{H}_0^\ast(-\mathbf{k})
	\end{pmatrix} ,
\end{align}
of which we take the Nambu basis $\{ \Psi_{\mathbf{k}}^\dagger, \Psi_{-\mathbf{k}}^T  \}$.
As a result, the particle-hole symmetry operator $\Xi$ is defined as,
\begin{align}
	\Xi = \gamma_x\mathcal{K},
\end{align}
where $\gamma_x$ is Pauli matrix acting on particle-hole subspace and $\mathcal{K}$ is the complex conjugate.
Here the normal Hamiltonian $\mathcal{H}_0(\mathbf{k})$ is given by Eq.~\eqref{eq-six-ham0} with out symmetry breaking perturbations.
The s-wave pairing is considered in this work,
\begin{align}
	\mathcal{H}_\Delta = \Delta_0\begin{pmatrix}
		is_y & 0 & 0 \\
		0 & -is_y & 0 \\
		0 & 0 & is_y
	\end{pmatrix}
	=\begin{pmatrix}
		0 & \Delta_0 & 0 & 0 & 0 & 0 \\ 
		-\Delta_0 & 0 & 0 & 0 & 0 & 0 \\ 
		0 & 0 & 0 & -\Delta_0 & 0 & 0 \\ 
		0 & 0 & \Delta_0 & 0 & 0 & 0 \\ 
		0 & 0 & 0 & 0 & 0 & \Delta_0 \\ 
		0 & 0 & 0 & 0 & -\Delta_0 & 0
	\end{pmatrix} ,
\end{align}
where the pairing profile in the real space is give by $\Delta_0\to\Delta_0\tanh(r/\xi_0) e^{i\theta}$.
Since the vortex line is orienting along z direction, the 3D Hamiltonian calculation is reduced to a 2D problem by treating $k_z$ as a parameter.
To solve the 2D BdG Hamiltonian with fixed $k_z$ for vortex bound states (VBSs), we take the disc geometry with natural boundary condition. 
In the polar coordinate system $(r,\theta)$, the momentum operators $k_\pm = k_x\pm i k_y$ can be expressed as,
\begin{align}
	k_{+}=e^{i\theta}\left\lbrack -i\frac{\partial}{\partial r}+\frac{1}{r}\frac{\partial}{\partial \theta} \right\rbrack,
	\text{ and }  k_{-}=e^{-i\theta}\left\lbrack -i\frac{\partial}{\partial r}-\frac{1}{r}\frac{\partial}{\partial\theta} \right\rbrack, 
\end{align}
which satisfies that
\begin{align}
	k_{+}\left(e^{in\theta}J_n(\alpha r)\right)  = i\alpha e^{i(n+1)\theta}J_{n+1}(\alpha r),
	\text{ and }   k_{-}\left(e^{in\theta}J_n(\alpha r)\right)  = -i\alpha e^{i(n-1)\theta}J_{n-1}(\alpha r),
\end{align}
where $n$ is an integer and $J_n$ is the Bessel function of the first kind. 
Given that the vortex line has winding number $+1$, the eigenfunctions of the reduced BdG equations for Eq.\eqref{eq-bdg-ham} take the following general forms,
\begin{align}
	\mathcal{H}_{BdG}(k_z) &= \oplus_{n\in\mathcal{Z}} \mathcal{H}_{n}(k_z),\\
	\mathcal{H}_{n}(k_z) \vert E_j(n,k_z) \rangle &= E_j(n,k_z) \vert E_j(n,k_z) \rangle, \\
	\vert E_j(n,k_z) \rangle &= (u_{j,k_z}(n,r,\theta),v_{j,k_z}(n,r,\theta))^T, 
\end{align}
which labels the j$^{th}$ solution in the $n$-subspace with fixed $k_z$, and $u$ (electron wave functions) and $v$ (hole wave functions) are expressed as,
\begin{align}
	u_{j,k_z}(n,r,\theta) &= e^{in\theta} \left(u_1(n,r), u_2(n+1,r)e^{i\theta}, u_3(n,r), u_4(n+1,r)e^{i\theta}, u_5(n-1,r)e^{-i\theta}, u_6(n+2,r)e^{2i\theta} \right), \\
	v_{j,k_z}(n,r,\theta) &= e^{in\theta} \left(v_1(n,r), v_2(n-1,r)e^{-i\theta}, v_3(n,r), v_4(n-1,r)e^{-i\theta}, v_5(n+1,r)e^{i\theta}, v_6(n-2,r)e^{-2i\theta} \right), 
\end{align}
where the components $u_{i}(n,r)$ and $v_{i}(n,r)$ with $i=1,2,3,4,5,6$ can be both expanded by the normalized Bessel function as,
\begin{align}
	\begin{split}
		u(n,r) = \sum_{k=1}^{N} c_{k,n} \Phi(n,r,\alpha_k), \\
		v(n,r) = \sum_{k=1}^{N} c_{k,n}^\prime \Phi(n,r,\alpha_k), 
	\end{split}
\end{align}
where $\Phi(n,r,\alpha_k)=\frac{\sqrt{2}}{R J_{n+1}(\alpha_k)}J_n(\alpha_k r/R)$.
Please note that $n$ used here is $l_z$ used in the main text.
Here, $c$ and $c^\prime$ are the expansion coefficients, \(\alpha_k\) is the \(k^{\text{th}}\) zero of \(J_n(r)\),  and $R$ is the radius of the disc. 
In our calculation, $\xi_0=1$ and $R=120$ are used. And the truncation number for Bessel zeros are $N=140$. In this setting, finite size effect is weak enough for the low-energy VBSs.

\begin{figure}[t]
	\includegraphics[width=0.85\textwidth]{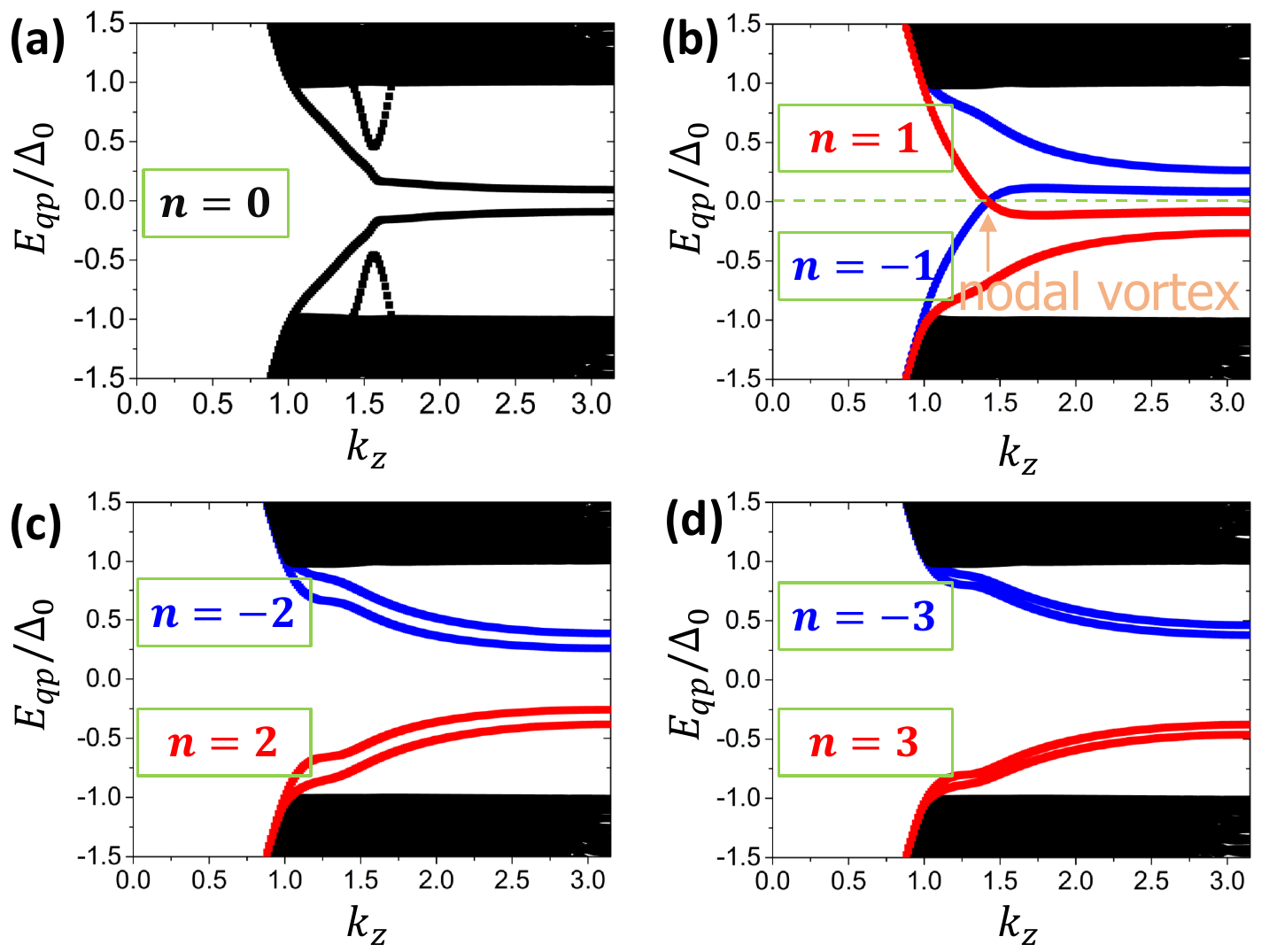}
	\caption{The BdG quasi-particle spectrum of the superconducting vortex line in $n=0,\pm 1, \pm 2,\pm 3$ subspaces. The blue line shows the 1D gapless vMBSs in (b).}
	\label{sm-fig1}
\end{figure}

\begin{figure}[t]
	\includegraphics[width=0.7\textwidth]{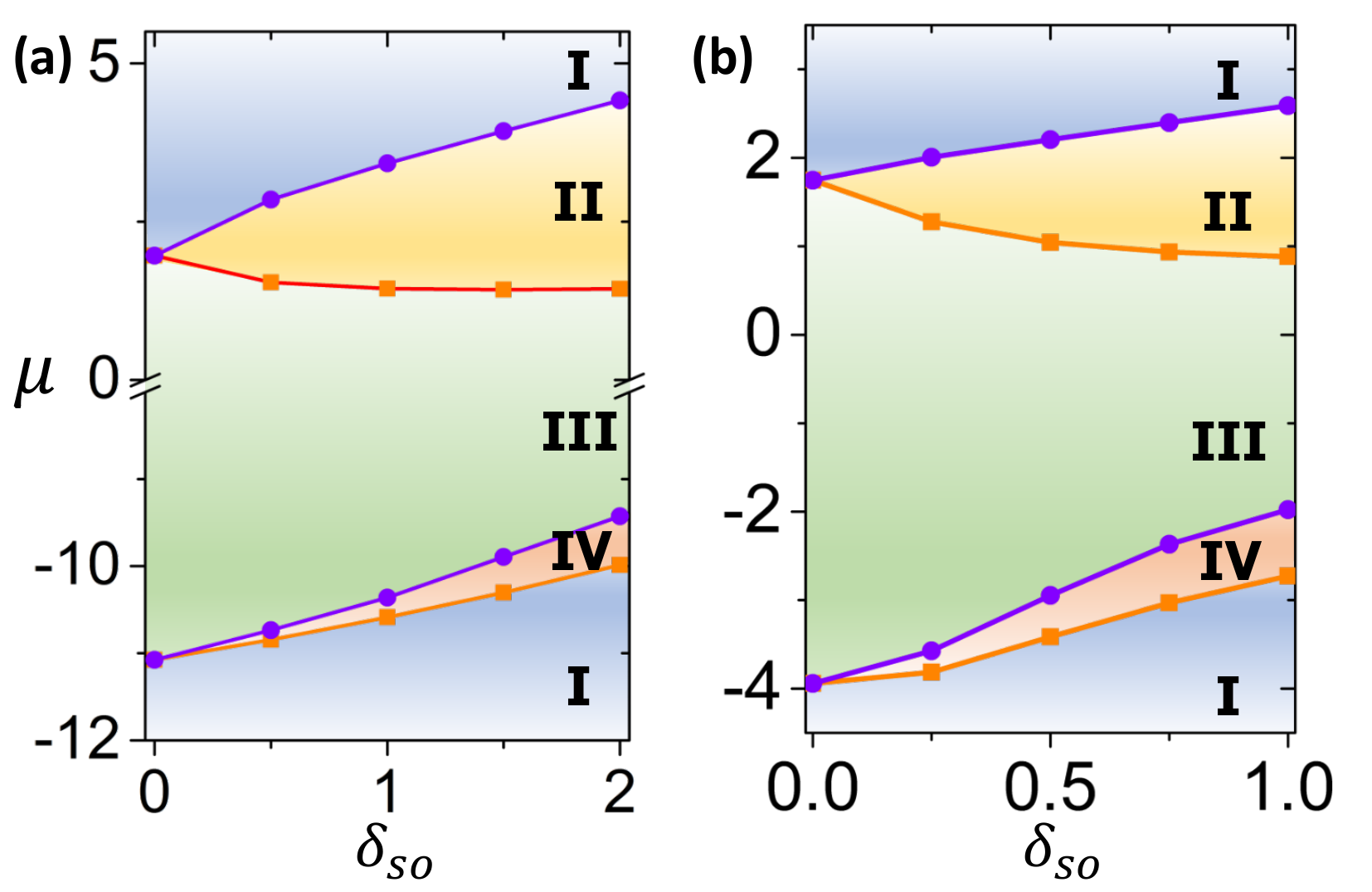}
	\caption{
		The effect of velocity of topological states on the vortex topological phase diagram for a minimal tFeSC model, in (a) $A_1=2$ and in (b) $A_1=1$.
		The phase diagram is mapped out as a function of chemical potential $\mu$ and $\delta_{so}$, the separation energy between TI gap and bulk Dirac nodes. It contains four phases: (I) trivial vortex; (II) nodal vortex; (III) hybrid vortex with both 0D end-localized vMBSs (red cone) and 1D gapless channels. (IV)  Kitaev vortex with 0D end-localized vMBSs. Similar to Fig.~(1) in the main text ($A_1=0.5$ is used there).	
	}
	\label{sm-fig2}
\end{figure}

Therefore, there are two types of numerical results,
\begin{itemize}
	\item[1.)] {\it Searching for the nodal vortex phase.} \\
	By fixing the chemical potential $\mu$ and $\delta_{so}$, we calculate the vortex line spectrum as a function of $k_z$ for the $n$-subspace. Normally, the 1D nodal vortex lives in $\vert n\vert\ge1$ subspaces. The results are shown in Fig.~\ref{sm-fig1}.
	\item[2.)] {\it Mapping out topological phase diagram.} \\
	Fix $k_z$ at TR-invariant planes with band inversion (in our six-band model, band inversion happens at $k_z=\pi$), then calculate the vortex spectrum as a function of $\mu$. The gap closing at $\mu_c$ indicates topological phase transition.  
	The results are shown in Fig.~(1) in the main text. 
	By increasing in-plane velocity of the surface state (i.e. $A_1$), the hybrid vortex phase is significantly enlarged, shown in Fig.~\ref{sm-fig2}.
\end{itemize}

For the phase diagrams in Fig.~\ref{sm-fig2}, we have adopted the following parameters that are slightly different the ones in the main text, with 
\begin{align}\label{eq-parameter-toy-model}
	\begin{split}
		&M_0^{(2)} = -2, M_1^{(2)} = -1, M_2^{(2)}=1, \\
		&M_0^{(1)} = 2, M_1^{(1)} = 1, M_2^{(1)} = -1, \\
		&A_2=0.1, B_1=0.5, B_2=-2B_1.
	\end{split}
\end{align}

\section{Appendix B: Analytical Results on the Projected TI model with Infinite $\delta_{so}$}

In this section, let us check the TI limit from the six-band model in Eq.~\eqref{eq-six-ham0} by taking $\delta_{sc}\to\infty$.
It implies that the bulk Dirac cone is far away from the TI surface Dirac cone, so that we can eliminate the high-energy bands from $\vert d_+, +\tfrac{3}{2}\rangle, \vert d_+, -\tfrac{3}{2}\rangle$.
Therefore, the six-band model is reduced to a four-band model which describes the topological insulator.
The vortex phase transition is studied by P.~Hosur in Ref.~[\onlinecite{hosur_prl_2011}], and they find the critical chemical potential for topological vMBSs is,
\begin{align}
	M_1(k_\parallel,k_z=\pi) = 0 \; \Rightarrow k_\parallel = \sqrt{-(M_0^1+4M_2^1)/M_1^1},
\end{align}
which indicates that $\mu_c^+ = -\mu_c^- = A_1k_\parallel = A_1 \sqrt{-(M_0^1+4M_2^1)/M_1^1} $. 
Here $k_\parallel = \sqrt{k_x^2+k_y^2}$ is in-plane momentum.
The analytical result is consistent with numerical calculation in Ref.~[\onlinecite{hosur_prl_2011,li_scpma_2019}].
Please note that the results are only for the $\delta_{so}\to\infty$ limit in the 6-band model.

Numerically, we find that the critical chemical potential $\mu_c^-$ varies very rapidly and obtains a large negative value for a small $\delta_{so}$. We use perturbative method to provide a semiquantitative understanding about this phenomenon.
Then, let us analyze the case with finite but large enough $\delta_{so}$ by perturbation theory. 
We also take the high-energy bands from $\vert d_+, +\tfrac{3}{2}\rangle, \vert d_+, -\tfrac{3}{2}\rangle$ as perturbation terms,
\begin{align}
	\mathcal{H}_0 &= \begin{pmatrix}
		H_{TI}     & V \\
		V^\dagger  & H_p
	\end{pmatrix}, \\
	H_{TI} = \begin{pmatrix}
		M_1(\mathbf{k}) & 0 & A_2k_z & -A_1k_- \\
		0 & M_1(\mathbf{k}) & A_1k_+ & A_2k_z \\
		A_2k_z & A_1k_- & M_2(\mathbf{k}) & 0  \\
		-A_1k_+ & A_2k_z & 0 & M_2(\mathbf{k})
	\end{pmatrix} , 
	V &= \begin{pmatrix}
		A_1k_+ & 0 \\ 
		0 & -A_1k_- \\ 
		0 & D^\ast(\mathbf{k}) \\ 
		D(\mathbf{k}) & 0 \\ 
	\end{pmatrix},
	H_p = \begin{pmatrix}
		M_2(\mathbf{k})+\delta_{so} & 0 \\
		0 & M_2(\mathbf{k})+\delta_{so}
	\end{pmatrix}.
\end{align}
The projected four-by-four effective TI Hamiltonian is $H_{eff} = H_{TI} - V H_p^{-1} V^\dagger$ as
\begin{align}
	H_{\text{eff}} =
	\begin{pmatrix}
		M_1(\mathbf{k}) & 0 & A_2k_z & -A_1k_- \\
		0 & M_1(\mathbf{k}) & A_1k_+ & A_2k_z \\
		A_2k_z & A_1k_- & M_2(\mathbf{k}) & 0  \\
		-A_1k_+ & A_2k_z & 0 & M_2(\mathbf{k})
	\end{pmatrix} - \frac{1}{M_2(\mathbf{k})+\delta_{so}}
	\begin{pmatrix}
		A_1^2k_\parallel^2 & 0 & 0 & A_1k_+D_k^\ast \\
		0 & A_1^2k_\parallel^2 & -A_1k_-D_k & 0 \\
		0 & -A_1k_+D_k^\ast & D_k^\ast D_k & 0 \\
		A_1k_-D_k & 0 & 0 & D_k^\ast D_k
	\end{pmatrix},
\end{align}
where $k_\pm = k_x \pm i k_y$. 
Then we take the approximation around Fermi energy 
\begin{align}
	\frac{1}{M_2(\mathbf{k})+\delta_{so}} \sim \frac{1}{\delta_{so}}.
\end{align}
Hereafter, we only keep the $k_\parallel^2$ order terms so that we ignore the $D_k$-terms.
Then, we can use the analytical criteria derived by P.~Hosur for the topological vortex phase transition for $H_{eff}$, 
\begin{align}
	H_{eff} = - \frac{A_1^2}{2\delta_{so}}k_\parallel^2 + 
	\begin{pmatrix}
		M_1'(\mathbf{k}) & 0 & A_2k_z & -A_1k_- \\
		0 & M_1'(\mathbf{k}) & A_1k_+ & A_2k_z \\
		A_2k_z & A_1k_- & -M_1'(\mathbf{k}) & 0  \\
		-A_1k_+ & A_2k_z & 0 & -M_1'(\mathbf{k})
	\end{pmatrix},
\end{align}
where $M_1'(\mathbf{k}) = M_1(\mathbf{k}) - \frac{A_1^2}{2\delta_{so}}k_\parallel^2 = (M_0^1 +4M_2^1) + (M_1^1- \frac{A_1^2}{2\delta_{so}})k_\parallel^2 $.
Making $M_1'(\mathbf{k})=0$ so that $k_\parallel = \sqrt{-\frac{M_0^1 +4M_2^1}{M_1^1- \frac{A_1^2}{2\delta_{so}}} }$.
So the critical chemical potential $\mu_{\text{eff},c}^\pm$ is given by,
\begin{align}
	\mu_{\text{eff},c}^- =   - \frac{A_1^2}{2\delta_{so}}\left( -\frac{M_0^1 +4M_2^1}{M_1^1- \frac{A_1^2}{2\delta_{so}}} \right)  - A_1\sqrt{-\frac{M_0^1 +4M_2^1}{M_1^1- \frac{A_1^2}{2\delta_{so}}} }.
\end{align}
Now let us decrease $\delta_{so}$ from infinity to a finite value (assume $\delta_{so}>0$ for simplicity), but we still assume $ M_1^1- \frac{A_1^2}{2\delta_{so}}>0$ to keep the validness of the perturbation theory. Therefore, we have
\begin{align}
	\begin{split}
		\delta_{so} \text{ decreases}  \quad &\Rightarrow \quad M_1^1- \frac{A_1^2}{2\delta_{so}} \text{ decreases}   \\
		& \Rightarrow \quad \sqrt{-\frac{M_0^1 +4M_2^1}{M_1^1- \frac{A_1^2}{2\delta_{so}}} } \text{ decreases} \\
		& \Rightarrow \quad \mu_{\text{eff},c}^- \text{ decreases} .
	\end{split}
\end{align}
The above analysis explains why $\mu_{eff,c}^- (<0)$ is negative large for a small $\delta_{so}$ case.
For a simple comparison,
\begin{align}
	\begin{cases}
		\delta_{so}\to \infty \text{(TI limit): } \mu_{\text{eff},c}^- \to -2.82843, \\
		\delta_{so}=2.5 \text{(TI+DSM): }  \mu_{\text{eff},c}^- = -10.32456,
	\end{cases}
\end{align}
which shows the significant affect on $\mu_c^-$ by changing $\delta_{so}$.
This clearly explains the domination of hybrid vortex in the superconducting vortex line phase diagram of tFeSCs, shown in Fig.~(1) in the main text.

\section{Appendix C: Low-energy Projection for Effective Vortex Hamiltonians}

In this section, we derive the low-energy effective Hamiltonian for the superconducting vortex line states, including the Kitaev vortex in the $n=0$ subspace and the nodal vortex in the $n=\pm 1$ subspace. This severs as the starting point to address the symmetry breaking effects induced by tilting external magnetic fields (see Sec.~\ref{sm-section-iv}) or bulk impurity induced curved vortex line (see Sec.~\ref{sm-section-iv}).

\subsection{Low-energy effective Hamiltonian for Kitaev vortex}
Before that, let us briefly discuss the low-energy states.
As for the $n=0$ subspace, we firstly set two lowest eigen-states of the 2D BdG Hamiltonian at $k_z=\pi$,
\begin{align}
	\mathcal{H}_{n=0}(k_z=\pi) \vert f_\pm \rangle = \pm m_0 \vert f_\pm \rangle, 
\end{align}	
with
\begin{align}
	\Xi \vert f_\pm \rangle = \left(\vert f_\mp\rangle \right)^\ast.
\end{align}
Next, we take $k_z\neq0$ terms expanded around $k_z=\pi$ for Eq.~\eqref{eq-bdg-ham} as perturbations, including the $\mathcal{H}_{n=0}(k_z^2)\propto k_z^2$-term and the $\mathcal{H}_{n=0}(k_z)\propto k_z$-term, which obey
\begin{align}
	\{ \Xi , \mathcal{H}_{n=0}(k_z^2)\} =0, \text{ and } [\Xi , \mathcal{H}_{n=0}(k_z)]=0.
\end{align}
Without symmetry breaking terms, the BdG Hamiltonian also preserve the Mirror symmetry
\begin{align}
	M_z^{BdG}=\begin{pmatrix}
		M_z & 0 \\
		0 & M_z^\ast \\
	\end{pmatrix},
\end{align}
where $M_z$ is defined in Eq.~\eqref{sm-eq-mz-normal}. And we can set $M_z^{BdG} \vert f_\pm\rangle = \pm i\vert f_\pm\rangle$. For the perturbation terms,
\begin{align}
	[ M_z^{BdG}, \mathcal{H}_{n=0}(k_z^2)] =0, \text{ and } \{M_z^{BdG}, \mathcal{H}_{n=0}(k_z)\}=0.
\end{align}
Thus, the eigen-states are simplified as,
\begin{subequations} \label{sm-eq-kitaev-basis-fpm}
	\begin{align}
		\vert f_+\rangle &=\left(0,u_2(1,r)e^{i\theta}, u_3(0,r),0,0, u_6(2,r)e^{2i\theta}, 
		v_1(0,r),0,0,v_4(-1,r)e^{-i\theta},v_5(1,r)e^{i\theta},0 \right),\\
		\vert f_-\rangle &=\left(u_1(0,r),0,0,u_4(1,r)e^{i\theta},u_5(-1,r)e^{-i\theta},0 , 0,v_2(-1,r)e^{-i\theta}, v_3(0,r),0,0, v_6(-2,r)e^{-2i\theta} \right). 
	\end{align}
\end{subequations}
Here $u_i=v_i^\ast$ is required by the particle-hole symmetry.
Please notice that all the symmetry constraints are numerically checked.
According to symmetry constraints of $\Xi $ and $M_z^{BdG}$, we have
\begin{align}
	\langle f_+ \vert \mathcal{H}_{n=0}(k_z^2) \vert f_+\rangle &= -\langle f_- \vert \mathcal{H}_{n=0}(k_z^2) \vert f_-\rangle = m_1 k_z^2 , \text{ and } \langle f_\pm \vert \mathcal{H}_{n=0}(k_z^2) \vert f_\mp \rangle=0, \\
	\langle f_+ \vert \mathcal{H}_{n=0}(k_z) \vert f_-\rangle  &= \langle f_- \vert \mathcal{H}_{n=0}(k_z) \vert f_+\rangle = m_2 k_z, \text{ and } \langle f_\pm \vert \mathcal{H}_{n=0}(k_z) \vert f_\pm \rangle=0,
\end{align}
which leads to,
\begin{align} \label{sm-eq-kitaev-ham}
	\mathcal{H}_{\text{eff},n=0} = (m_0+m_1k_z^2)\sigma_z + m_2 k_z \sigma_x,
\end{align}
which describes the lowest vortex line states in the $n=0$ subspace, resembling the 1D topological Kitaev chain.
As a result, we call it Kitaev vortex in the main text.
Once $m_0m_1<0$ and $m_2\neq0$, the vortex line is topological and there exists a single 0D Majorana zero mode (MZM) at each end of the vortex line. Here we take an example with $\mu=0,\delta_{so}=0.5$, and the numerically calculated $m_i$ are
\begin{align}
	m_0=0.0178, m_1 = -0.0041, m_2 = 0.0154.
\end{align}
It confirms the topological Kitaev vortex phase discussed in the main text.

\subsection{Low-energy effective Hamiltonian for Nodal vortex}
Next, let us derive the effective Hamiltonian for $n=\pm1$ subspaces for the nodal vortex phase, which satisfy
\begin{align}
	\mathcal{H}_{n=\pm 1}(k_z=\pi) \vert f'_\pm \rangle = \pm m_0 \vert f'_\pm \rangle, 
\end{align}	
with
\begin{align}	
	\Xi \vert f'_\pm \rangle &= \left(\vert f'_\mp\rangle \right)^\ast, \\
	M_z^{BdG} \vert f'_\pm\rangle &= \pm i\vert f'_\pm\rangle.
\end{align}
The symmetry constraints for $\vert f'_\pm \rangle$ are express as,
\begin{subequations} \label{sm-eq-nodal-fpm}
	\begin{align}
		\vert f'_+ \rangle &= e^{i\theta}\left(0,u_2(2,r)e^{i\theta}, u_3(1,r),0,0, u_6(3,r)e^{2i\theta}, 
		v_1(1,r),0,0,v_4(0,r)e^{-i\theta},v_5(2,r)e^{i\theta},0 \right),\\
		\vert f'_- \rangle &= e^{-i\theta}\left(u_1(-1,r),0,0,u_4(0,r)e^{i\theta},u_5(-2,r)e^{-i\theta},0 , 0,v_2(-2,r)e^{-i\theta}, v_3(-1,r),0,0, v_6(-3,r)e^{-2i\theta} \right). 
	\end{align}
\end{subequations}
In the $ \{ \vert f'_+ \rangle, \vert f'_- \rangle \} $ subspace, we project $\vert \mathcal{H}_{n=0}(k_z^2)$ term and $\vert \mathcal{H}_{n=0}(k_z)$ term into this lowest energy basis. As a result,
\begin{align}
	\langle f'_+ \vert \mathcal{H}_{n=0}(k_z^2) \vert f'_+\rangle &= -\langle f'_- \vert \mathcal{H}_{n=0}(k_z^2) \vert f'_-\rangle = m_1 k_z^2 , \text{ and } \langle f'_\pm \vert \mathcal{H}_{n=0}(k_z^2) \vert f'_\mp \rangle=0, \\
	\langle f'_+ \vert \mathcal{H}_{n=0}(k_z) \vert f'_-\rangle  &= \langle f'_- \vert \mathcal{H}_{n=0}(k_z) \vert f'_+\rangle  = \langle f'_\pm \vert \mathcal{H}_{n=0}(k_z) \vert f'_\pm \rangle=0,
\end{align}
Since there is a $+2$ difference in the total angular momentum, therefore, the off-diagonal $\langle f'_i \vert \mathcal{H}_{n=0}(k_z) \vert f'_j \rangle =0$.
Please notice that $\mathcal{H}_{n=0}(k_z)$ couples electrons with the same angular momentum.
Mathematically, the integral over $\theta$ vanishes. 
It leads to,
\begin{align} \label{sm-eq-nodal-ham}
	\mathcal{H}_{\text{eff},n=\pm 1} = (m_0'+m_1'k_z^2)\sigma_z,
\end{align}
which indicates topological nodal vortex phase under criteria $m_0'm_1'<0$. 
As a result, we call it nodal vortex in the main text.

\section{Appendix D: Tilting External Magnetic Field}
\label{sm-section-iv}

In this section, we use both an analytical second-order perturbation theory and numerical calculations to explore the topological physics of a tilted vortex line in tFeSCs. We expect the vortex line tilting to break $C_4$ rotation symmetry explicitly and further qualitatively update our vortex topological phase diagram with the tilting angle being a new Majorana control knob.  We also note that field-line tilting and its impact on the vortex configurations have been achieved experimentally in LiFeAs~\cite{zhangST_prb_2019}. Therefore, it is quite conceivable that the results discussed in this session will soon be experimentally supported in the near future.

\begin{figure*}[t]
	\includegraphics[width=0.3\textwidth]{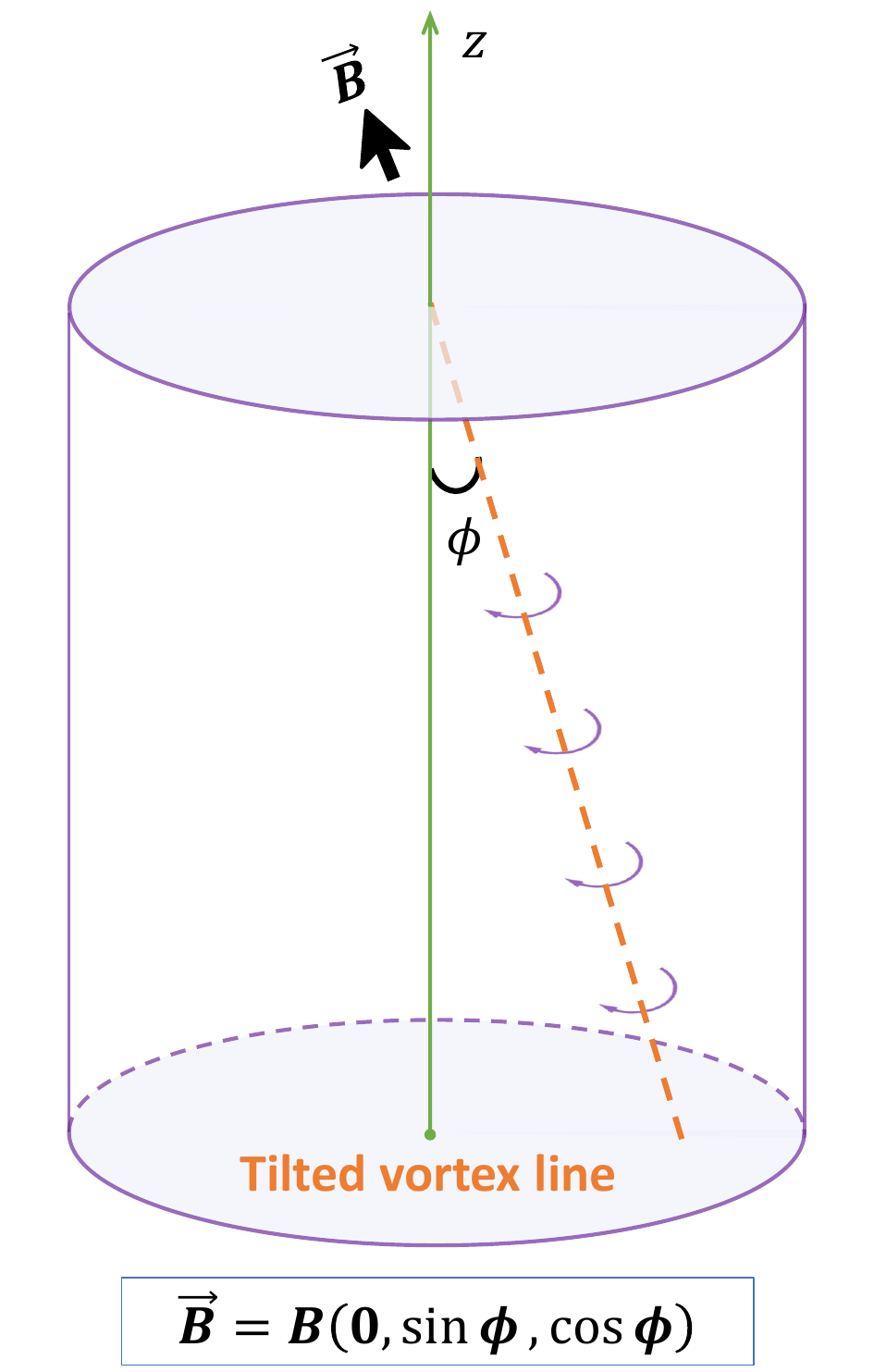}
	\caption{ It shows the tilted vortex line (green dashed line) is induced by tilting external magnetic field. 
	}
	\label{sm-fig3}
\end{figure*}

\subsection{BdG Hamiltonian with Tilted Vortex Lines}

We consider the external magnetic field being misaligned from z-axis with a small tilting angle $\phi$. Here we assume that the superconducting vortex line is a 1D rigid body [see Fig.~\ref{sm-fig3}], and this assumption remains valid only in a disorder-free system. We will relax this assumption later when we consider the curvature of vortex line induced by bulk impurities in Sec.~\ref{sm-section-v}.

In the continuum limit, the vortex line tilting induced by $\vec{B}=B(0,0,1)\to B(0,\sin\phi,\cos\phi)$ is equivalent to a coordinate transformation for the BdG Hamiltonian with 
\begin{align}
	\mathcal{H}_{BdG}(k_x,k_y,k_z,x,y,z) \to \mathcal{H}_{BdG} (k_x,k_y',k_z',x,y',z',\phi),
\end{align}
where $k_z'$ is along the tilted vortex line direction and
\begin{subequations}
	\begin{align}
		(x,y,z) &\to (x, y'\cos\phi - z'\sin\phi, y'\sin\phi + z'\cos\phi) , \\
		(k_x,k_y,k_z) &\to (k_x, k_y'\cos\phi - k_z'\sin\phi, k_y'\sin\phi + k_z'\cos\phi).
	\end{align}
\end{subequations}
In the small $\phi\ll\frac{\pi}{2}$ limit, 
\begin{subequations} \label{sm-eq-ky-kz}
	\begin{align}
		k_y = k_y' \cos\phi - k_z'\sin\phi = k_y' + \left\lbrack k_y'(\cos\phi-1) - k_z'\sin\phi \right\rbrack, \\
		k_z = k_y' \sin\phi + k_z'\cos\phi = k_z' + \left\lbrack k_y'\sin\phi + k_z'(\cos\phi -1) \right\rbrack.
	\end{align}
\end{subequations}
We then find that the tilted normal-state Hamiltonian with $\phi$ consists of two parts 
\begin{align}
	\mathcal{H}_0(k_x,k_y,k_z) = \mathcal{H}_0(k_x,k_y',k_z') + \mathcal{H}_0'(k_x,k_y',k_z',\phi),
\end{align}
where the first part is formally identical to the original one [i.e., $\phi=0$ case], but up to some renormalization of model parameters. The second part $H_0' (k_x,k_y',k_z',\phi)$ is solely caused by the non-zero tilting angle $\phi$ and can be treated as a perturbative Hamiltonian. Crucially, $H_0' (k_x,k_y',k_z',\phi)$ accounts for breaking the $C_4$ rotational symmetry about the $z'$-axis. 
\begin{align} \label{eq-six-ham0-phi}
	\mathcal{H}_0'(k_x,k_y',k_z',\phi) = \begin{pmatrix}
		M_1' & 0 & A_2' & -i A_1' & -i A_1' & 0 \\ 
		0 & M_1' & -i A_1' & A_2' & 0 & -iA_1' \\ 
		A_2' & i A_1' & M_2' & 0 & 0 & 0 \\ 
		i A_1' & A_2' & 0 & M_2' & 0 & 0 \\ 
		i A_1' & 0 & 0 & 0 & M_2' & 0 \\ 
		0 & i A_1' & 0 & 0 & 0 & M_2'
	\end{pmatrix}, 
\end{align}
where the off-diagonal $k^2$-order is ignored and the other terms are given by, 
\begin{subequations}
	\begin{align}
		M_1' &= \left( -m_{1}^{(1)} + m_{2}^{(1)} \right) \sin\phi \left( 2k_y' k_z' \cos \phi + 
		(k_y'+k_z')(k_y'-k_z')\sin\phi \right) ,\\
		M_2' &= \left( -m_{1}^{(2)} + m_{2}^{(2)} \right) \sin\phi \left( 2k_y' k_z' \cos \phi + 
		(k_y'+k_z')(k_y'-k_z')\sin\phi \right) , \\
		A_1' &= A_1 \left\lbrack k_y'(\cos\phi-1) - k_z'\sin\phi  \right\rbrack, \\
		A_2' &= A_2 \left\lbrack  k_z'(\cos\phi-1) + k_y' \sin \phi  \right\rbrack.
	\end{align}
\end{subequations}

\subsection{Symmetry Construction of Effective $C_4$-breaking Vortex Hamiltonian}

While the physical out-of-plane mirror symmetry $M_{z'}$ is explicitly broken by the vortex line tilting, it can be restored if we require $\phi$ to flip its sign under $M_{z'}$ as well,
\begin{align}
	M_{z'} \mathcal{H}_0' (k_x, k_y',k_z',\phi) M_{z'}^{-1} = \mathcal{H}_0'(k_x, k_y', -k_z',-\phi),
\end{align}
In fact, the above ``modified" mirror symmetry is the key for us to construct the low-energy effective tilted vortex Hamiltonian, just based on symmetry principles. Hereafter, we will also use $k_y$ and $k_z$ instead of $k_y'$ and $k_z'$ for simplification. 

Next, we discuss the effective low-energy vortex Hamiltonian $h_{\text{vortex}}(k_z,\phi)$ up to the $\phi^2$-order to capture the main physics by considering the mirror symmetry and particle-hole symmetry. The basis is taken from the solution of the unperturbed Hamiltonian at $k_z'=\pi$ for the four low-energy vortex states, including two $n=0$ vortex states [i.e.,~$\vert f_\pm\rangle$ given by Eq.~\eqref{sm-eq-kitaev-basis-fpm}] and two $n=\pm1$ vortex states [i.e.,~$\vert f_\pm'\rangle$ given by Eq.~\eqref{sm-eq-nodal-fpm}]. These four unperturbed vortex states are listed as follows,

\begin{table}[!htbp]
		\begin{tabular}{c|cc|cc} \hline\colrule
			& $\vert f_+\rangle$ & $\vert f_-\rangle$ & $\vert f_+'\rangle$  & $\vert f_-'\rangle$   \\ \hline\colrule
			
			$\mathcal{H}_{BdG}(k_x,k_y,\pi)$ &  $m_0$  &  $-m_0$  & $m_0'$  & $-m_0'$     \\
			$M_z$                  &  $i$    &  $-i$ &  $i$ & $-i$ \\ \hline\colrule 
		\end{tabular}
	\end{table} 
	
	Based on the above vortex-state basis $\{ \vert f_+\rangle, \vert f_-\rangle,  \vert f_+'\rangle, \vert f_-'\rangle \}$, the one-dimensional $\vec{k}\cdot\vec{p}$-type low-energy effective vortex Hamiltonian $h_{\text{vortex}}(k_z,\phi)$ up to the $\phi^2$-order is generally given by
	\begin{align} \label{sm-eq-vortex-ham-total}
		h_{\text{vortex}}(k_z,\phi) = h_{\text{vortex}}^{(0)}(k_z) + h_{\text{vortex}}^{(1)}(k_z,\phi),
	\end{align}
	where the first term is the unperturbed vortex Hamiltonian that is just a direct sum of Kitaev vortex Hamiltonian in Eq.~\eqref{sm-eq-kitaev-ham} and nodal vortex Hamiltonian in Eq.~\eqref{sm-eq-nodal-ham}, namely,
	\begin{align} \label{sm-eq-ham-vortex-zero}
		h_{\text{vortex}}^{(0)}(k_z) = \begin{pmatrix}
			m_{k_z} & m_2 k_z & 0 & 0 \\
			m_2 k_z & -m_{k_z} & 0 & 0 \\
			0 & 0 & m_{k_z}' & 0 \\
			0 & 0 & 0 & -m_{k_z}' 
		\end{pmatrix},
	\end{align}
	where  $m_{k_z}=m_0+m_1 k_z^2$ and $m_{k_z}'=m_0'+m_1' k_z^2$.  And the second term in Eq.~\eqref{sm-eq-vortex-ham-total} is given by
	\begin{align} \label{sm-eq-votex-ham-h1}
		h_{\text{vortex}}^{(1)}(k_z,\phi) = \begin{pmatrix}
			m_5\phi^2 + m_6k_z^2\phi^2  & 0 & m_3k_z\phi & m_4\phi \\
			0 & -m_5\phi^2 - m_6k_z^2\phi^2 & -m_4^\ast \phi & m_3^\ast k_z\phi \\
			m_3^\ast k_z\phi & -m_4\phi & m_5'\phi^2 + m_6'k_z^2\phi^2 & m_2'^\ast k_z\phi^2 \\
			m_4^\ast \phi & m_3 k_z\phi & m_2'^\ast k_z\phi^2 & -m_5' \phi^2 - m_6'k_z^2\phi^2 
		\end{pmatrix}.
	\end{align}
	All these terms are in principle allowed by the modified mirror symmetry ($M_z$) and particle-hole symmetry ($\Xi$). Their representations are
	\begin{align}
		\mathcal{D}[M_z] = \begin{pmatrix}
			i & 0 & 0 & 0 \\
			0 & -i & 0 & 0 \\
			0 & 0 & i & 0 \\
			0 & 0 & 0 & -i
		\end{pmatrix}, \; 
		\mathcal{D}[\Xi] = \begin{pmatrix}
			0 & 1 & 0 & 0 \\
			1 & 0 & 0 & 0 \\
			0 & 0 & 0 & 1\\
			0 & 0 & 1 & 0
		\end{pmatrix} \mathcal{K}.
	\end{align}
	And one can easily check that 
	\begin{subequations}
		\begin{align}
			\mathcal{D}[M_z] h_{\text{vortex}}(k_z,\phi)  \left( \mathcal{D}[M_z] \right)^{-1} &= h_{\text{vortex}}(-k_z,-\phi), \\
			\mathcal{D}[\Xi] h_{\text{vortex}}(k_z,\phi)  \left( \mathcal{D}[\Xi] \right)^{-1} &= -h_{\text{vortex}}(-k_z,\phi).
		\end{align}
	\end{subequations}
	In the $\phi=0$ decoupling limit, the $m_0 m_1<0$ case describes the topological Kitaev vortex phase ($n=0$ states) and $m_0' m_1'<0$ is for the nodal vortex phase ($n=\pm1$ states). And $h_{\text{vortex}}^{(0)}(k_z)$ describes the unperturbed hybrid vortex phase, while the perturbation Hamiltonian $h_{\text{vortex}}^{(1)}(k_z,\phi)$ represents three effects:
	\begin{itemize}
		\item  Parameter renormalizations: 
		\begin{eqnarray}
			&& m_0 \to m_0+m_5 \phi^2,\ \ m_0' \to m_0'+m_5'\phi^2 \nonumber \\
			&& m_1\to m_1 + m_6\phi^2,\ \ m_1'\to m_1'+m_6'\phi^2.
		\end{eqnarray}
		The above updates modify the topological conditions for vortex phases
		\begin{subequations}
			\begin{align}
				\text{Kitaev vortex: } m_0m_1 <0 \to (m_0+m_5 \phi^2)(m_1 + m_6\phi^2)<0 , \\
				\text{Nodal vortex: } m_0'm_1' <0 \to (m_0'+m_5' \phi^2)(m_1' + m_6'\phi^2)<0.
			\end{align}
		\end{subequations}
		As a result, the topological phase boundaries among all distinct vortex phases will depend on the value of $\phi$. 
		\item  The topological gap opening for the nodal vortex due to the spinless $p$-wave-like pairing term $m_2' k_z \phi^2$. This could make the nodal vortex as another topological Kitaev vortex.
		\item  The hybridization between the Kitaev vortex and the nodal vortex is caused by those off-diagonal terms: $m_3 k_z\phi$ and $m_4\phi$.
	\end{itemize}

	\subsection{Second-order Perturbation Theory for $C_4$-breaking Vortex Hamiltonian}
	
	We now justify  this vortex Hamiltonian $h_{\text{vortex}}^{(1)}(k_z,\phi) $ via the second-order perturbation theory [i.e., $\phi$ is the perturbation parameter]. The unperturbed vortex basis include those four vortex states with lowest energy $\{ \vert f_+\rangle, \vert f_-\rangle,  \vert f_+'\rangle, \vert f_-'\rangle \}$ and others denoted as $\vert g\rangle$ with energy $E_g$. And the perturbation Hamiltonian is 
	\begin{align}
		\mathcal{H}_{BdG}' = \begin{pmatrix}
			\mathcal{H}_0'(k_x,k_y,k_z,\phi) & 0 \\
			0 & -(\mathcal{H}_0'(-k_x,-k_y,-k_z,\phi) )^\ast
		\end{pmatrix}.
	\end{align}
	where the perturbation Hamiltonian $\mathcal{H}_0'(k_x,k_y,k_z,\phi)$ is defined in Eq.~\eqref{eq-six-ham0-phi}. The zero-order vortex Hamiltonian is just $h_{\text{vortex}}^{(0)}(k_z)$ in Eq.~\eqref{sm-eq-ham-vortex-zero}. And the first-order perturbation gives rise to
	\begin{align}
		\left\lbrack h_{\text{vortex}}^{(1)}(k_z,\phi) \right\rbrack_{i,j} =  \langle f_i \vert  \mathcal{H}_{BdG}' \vert f_j \rangle,
	\end{align}
	where $\vert f_i\rangle \in \{ \vert f_+\rangle, \vert f_-\rangle,  \vert f_+'\rangle, \vert f_-'\rangle \}$ and $i,j=1,2,3,4$. We will show later the first-order results can reproduce all the terms in Eq.~\eqref{sm-eq-votex-ham-h1} expect the $m_2'k_z\phi^2$ term due to the vanishing of $\langle f_+'\vert \mathcal{H}_{BdG}' \vert f_-'\rangle$. Thus, we have to consider the second-order corrections,
	\begin{align}
		\left\lbrack h_{\text{vortex}}^{(2)}(k_z,\phi) \right\rbrack_{3,4} = \sum_{i=1}^{2}
		\frac{\langle f_+'  \vert  \mathcal{H}_{BdG}'  \vert f_i \rangle \langle f_i \vert  \mathcal{H}_{BdG}'  \vert f_-'\rangle}{(-m_0'-m_0')(-m_0'-E_i)} + \sum_{\vert g\rangle}
		\frac{\langle f_+'  \vert  \mathcal{H}_{BdG}'  \vert g \rangle \langle g \vert  \mathcal{H}_{BdG}'  \vert f_-'\rangle}{(-m_0'-m_0')(-m_0'-E_g)}.
	\end{align}
	
	To do the above perturbation integrals, we first need to expand $\mathcal{H}_0'(k_x,k_y,k_z,\phi)$ up to the $\phi^2$-order,
	\begin{align} 
		\mathcal{H}_0'(k_x,k_y,k_z,\phi) = \begin{pmatrix}
			M_1' & 0 & A_2' & -i A_1' & -i A_1' & 0 \\ 
			0 & M_1' & -i A_1' & A_2' & 0 & -iA_1' \\ 
			A_2' & i A_1' & M_2' & 0 & 0 & 0 \\ 
			i A_1' & A_2' & 0 & M_2' & 0 & 0 \\ 
			i A_1' & 0 & 0 & 0 & M_2' & 0 \\ 
			0 & i A_1' & 0 & 0 & 0 & M_2'
		\end{pmatrix} + \mathcal{O}(\phi^3) , 
	\end{align}
	where 
	\begin{subequations}
		\begin{align}
			M_1' &= \left( -m_{1}^{(1)} + m_{2}^{(1)} \right)  \left( 2k_y k_z \phi + 
			(k_y^2-k_z^2) \phi^2 \right) , \nonumber \\ 
			&\equiv \left( -m_{1}^{(1)} + m_{2}^{(1)} \right)  \left( -i(k_+-k_-) k_z \phi 
			-\frac{1}{4}(k_+^2 + k_-^2 - 2k_+k_-)\phi^2 - k_z^2 \phi^2 \right)  ,\\
			M_2' &= \left( -m_{1}^{(2)} + m_{2}^{(2)} \right)  \left( 2k_y k_z \phi + 
			(k_y^2-k_z^2) \phi^2 \right)  ,\nonumber \\
			&\equiv \left( -m_{1}^{(2)} + m_{2}^{(2)} \right)  \left( -i(k_+-k_-) k_z \phi 
			-\frac{1}{4}(k_+^2 + k_-^2 - 2k_+k_-)\phi^2 - k_z^2 \phi^2 \right)  ,\\
			A_1' &= A_1\left\lbrack k_y(-\frac{1}{2}\phi^2) - k_z \phi  \right\rbrack
			\equiv  A_1 \left\lbrack \frac{i}{4}(k_+ - k_-)\phi^2 - k_z\phi  \right\rbrack, \\
			A_2' &= A_2\left\lbrack k_z(-\frac{1}{2}\phi^2) + k_y \phi  \right\rbrack
			\equiv A_2 \left\lbrack -\frac{1}{2} k_z \phi^2 - \frac{i}{2} (k_+ - k_-)\phi \right\rbrack.
		\end{align}
	\end{subequations}
	where $k_y=\frac{1}{2i}(k_+-k_-)$ have been used with $k_\pm = k_x \pm i k_y$. Then, we will be able to derive the expressions for the coefficients in the vortex Hamiltonian,
	\begin{itemize}
		\item The {\bf first-order} perturbation for the diagonal terms for Kitaev vortex: 
		\begin{align}
			\left\lbrack h_{\text{vortex}}^{(1)}(k_z,\phi) \right\rbrack_{1,1} = -\left\lbrack h_{\text{vortex}}^{(1)}(k_z,\phi) \right\rbrack_{2,2} = m_5\phi^2 + m_6 k_z^2\phi^2,
		\end{align}
		where the coefficients $m_5$ and $m_6$ are given by
		\begin{subequations}
			\begin{align} 
				m_5 &=  \frac{m_{2}^{(2)}-m_{1}^{(2)}}{2} \lbrack \langle u_3(0,r) \vert k_+k_- \vert u_3(0,r) \rangle  
				+  \langle u_6(2,r) \vert k_+k_- \vert u_6(2,r) \rangle  -  \langle v_4(-1,r) \vert k_+k_- \vert v_4(-1,r) \rangle    \nonumber \\
				& -  \langle v_5(1,r) \vert k_+k_- \vert v_5(1,r) \rangle  \rbrack  + 
				\frac{m_{2}^{(1)}-m_{1}^{(1)}}{2} \lbrack \langle u_2(1,r) \vert k_+k_- \vert u_2(1,r) \rangle  - \langle v_1(0,r) \vert k_+k_- \vert v_1(0,r) \rangle \rbrack \nonumber  \\
				& + \frac{A_1}{4} \text{Re} \left\lbrack  \langle u_2(1,r)e^{i\theta} \vert  k_+ \vert u_3(0,r)\rangle 
				-  \langle u_6(2,r)e^{i2\theta} \vert k_+\vert u_2(1,r) e^{i\theta} \rangle  \right\rbrack  , \\
				m_6 &= -(m_{2}^{(2)}-m_{1}^{(2)} ) \lbrack \langle u_3(0,r)  \vert u_3(0,r) \rangle  
				+  \langle u_6(2,r) \vert  u_6(2,r) \rangle  -  \langle v_4(-1,r) \vert  v_4(-1,r) \rangle  -  \langle v_5(1,r) \vert v_5(1,r) \rangle  \rbrack  \nonumber \\
				&  -(m_{2}^{(1)}-m_{1}^{(1)} ) \lbrack \langle u_2(1,r) \vert  u_2(1,r) \rangle  - \langle v_1(0,r) \vert  v_1(0,r) \rangle \rbrack ,
			\end{align}
		\end{subequations}
		Hereafter, we will use this notation for integral over $r$, $\langle u(n,r) \vert \mathcal{P} \vert u(n,r)\rangle = \int_{0}^{R}r dr\, \left\lbrack u^\ast(n,r) \mathcal{P} u(n,r) \right\rbrack$.
		\item The {\bf first-order} perturbation for the diagonal terms for nodal vortex: 
		\begin{align}
			\left\lbrack h_{\text{vortex}}^{(1)}(k_z,\phi) \right\rbrack_{3,3} = -\left\lbrack h_{\text{vortex}}^{(1)}(k_z,\phi) \right\rbrack_{4,4} = m_5'\phi^2 + m_6' k_z^2\phi^2 ,
		\end{align}
		where the coefficients $m_5'$ and $m_6'$ are given by 
		\begin{subequations}
			\begin{align} 
				m_5' &=  \frac{m_{2}^{(2)}-m_{1}^{(2)}}{2} \lbrack \langle u_3(1,r) \vert k_+k_- \vert u_3(1,r) \rangle  
				+  \langle u_6(3,r) \vert k_+k_- \vert u_6(3,r) \rangle  -  \langle v_4(0,r) \vert k_+k_- \vert v_4(0,r) \rangle    \nonumber \\
				& -  \langle v_5(2,r) \vert k_+k_- \vert v_5(2,r) \rangle  \rbrack  + 
				\frac{m_{2}^{(1)}-m_{1}^{(1)}}{2} \lbrack \langle u_2(2,r) \vert k_+k_- \vert u_2(2,r) \rangle  - \langle v_1(1,r) \vert k_+k_- \vert v_1(1,r) \rangle \rbrack \nonumber  \\
				& + \frac{A_1}{4} \text{Re} \left\lbrack  \langle u_2(2,r)e^{i2\theta} \vert  k_+ \vert u_3(1,r) e^{i\theta} \rangle 
				-  \langle u_6(3,r)e^{i3\theta} \vert k_+\vert u_2(2,r) e^{i2\theta} \rangle  \right\rbrack  , \\
				m_6' &= -(m_{2}^{(2)}-m_{1}^{(2)} ) \lbrack \langle u_3(0,r)  \vert u_3(0,r) \rangle  
				+  \langle u_6(2,r) \vert  u_6(2,r) \rangle  -  \langle v_4(-1,r) \vert  v_4(-1,r) \rangle  -  \langle v_5(1,r) \vert v_5(1,r) \rangle  \rbrack  \nonumber \\
				&  -(m_{2}^{(1)}-m_{1}^{(1)} ) \lbrack \langle u_2(1,r) \vert  u_2(1,r) \rangle  - \langle v_1(0,r) \vert  v_1(0,r) \rangle \rbrack  .
			\end{align}
		\end{subequations}
		\item The {\bf first-order} perturbation for the off-diagonal term: 
		\begin{align}
			\left\lbrack h_{\text{vortex}}^{(1)}(k_z,\phi) \right\rbrack_{1,3} = \left(\left\lbrack h_{\text{vortex}}^{(1)}(k_z,\phi) \right\rbrack_{2,4} \right) ^\ast = m_3 k_z\phi ,
		\end{align}
		where the coefficient $m_3$ is given by
		\begin{align}
			\begin{split}
				m_3 &= (m_{2}^{(1)}-m_{1}^{(1)} )  \left\lbrack \langle u_2(1,r) e^{i\theta} \vert i k_- \vert u_2(2,r) e^{i2\theta} \rangle - \langle v_1(0,r) \vert ik_- \vert v_1(1,r) e^{i\theta} \rangle  \right\rbrack  \\
				&+ (m_{2}^{(2)}-m_{1}^{(2)} ) \lbrack  \langle u_3(0,r)  \vert ik_- \vert  u_3(1,r)e^{i\theta} \rangle +  \langle u_6(2,r)e^{i2\theta} \vert ik_- \vert u_6(3,r) e^{i3\theta} \rangle   \\
				&- \langle v_4(-1,r)e^{-i\theta} \vert i k_- \vert v_4(0,r) \rangle - \langle v_5(1,r) e^{i\theta} \vert ik_- \vert v_5(2,r) e^{i2\theta} \rangle    \rbrack \\
				&+ i A_1 \left\lbrack \langle u_2(1,r) \vert u_3(1,r) \rangle  - \langle u_6(2,r) \vert u_2(2,r)\rangle - \langle v_1(0,r)\vert v_4(0,r)\rangle + \langle v_5(1,r) \vert v_1(1,r)\rangle  \right\rbrack.
			\end{split}
		\end{align}
		Here we take the first $\theta$ integral ($\int_0^{2\pi} d\theta $) as an example to show the integration does not vanish because of 
		\begin{align}
			\begin{split}
				k_- \vert u_2(2,r) e^{i2\theta} \rangle  &= k_- \left(  \sum_{k=1}^{N} c_{k,2} \frac{\sqrt{2}}{R J_{3}(\alpha_k)}J_2(\alpha_k r/R) e^{i2\theta} \right) \\
				&= e^{i\theta} \sum_{k=1}^{N} (-i\alpha_k/R) c_{k,2} \frac{\sqrt{2}}{R J_{3}(\alpha_k)}J_1(\alpha_k r/R) \\
				&\triangleq -i e^{i\theta} \tilde{u}_2(1,r),
			\end{split} 
		\end{align}
		which gives rise to
		\begin{align}
			\langle u_2(1,r) e^{i\theta} \vert i k_- \vert u_2(2,r) e^{i2\theta} \rangle  
			=  \langle u_2(1,r) \vert  \tilde{u}_2(1,r) \rangle \neq 0 .
		\end{align}
		Similar analysis can be also done for the other integral. 
		\item The {\bf first-order} perturbation for the off-diagonal term: 
		\begin{align}
			\left\lbrack h_{\text{vortex}}^{(1)}(k_z,\phi) \right\rbrack_{1,4} = -\left(\left\lbrack h_{\text{vortex}}^{(1)}(k_z,\phi) \right\rbrack_{2,3} \right)^\ast = m_4 \phi ,
		\end{align}
		where the coefficient $m_4$ is given by   
		\begin{align}
			\begin{split}
				m_4 = -\frac{A_2}{2} \lbrack &\langle u_2(1,r)e^{i\theta} \vert ik_+ \vert u_4(0,r)\rangle + \langle u_3(0,r) \vert ik_+ \vert u_1(-1,r)e^{-i\theta} \rangle \\
				+& \langle v_1(0,r) \vert ik_+ \vert v_3(-1,r) e^{-i\theta} \rangle + \langle v_4(-1,r)e^{-i\theta} \vert ik_+ \vert v_2(-2,r) e^{-i2\theta}\rangle \rbrack.
			\end{split}
		\end{align}
		\item The {\bf second-order} perturbation for the off-diagonal term:
		\begin{align}
			\left\lbrack h_{\text{vortex}}^{(1)}(k_z,\phi) \right\rbrack_{3,4}  = m_2'k_z \phi^2 ,
		\end{align}
		where the first-part contribution to the coefficient $m_2'$ is given by $\sum_{i=1}^{2}
		\frac{\langle f_+'  \vert  \mathcal{H}_{BdG}'  \vert f_i \rangle \langle f_i \vert  \mathcal{H}_{BdG}'  \vert f_-'\rangle}{(-m_0'-m_0')(-m_0'-E_i)} $,
		\begin{align}
			\begin{split}
				m_2' = m_3^\ast m_4 \left\lbrack  \frac{1}{2m_0'(m_0 + m_0')} + \frac{1}{2m_0'(m_0 - m_0')}  \right\rbrack + \cdots,
			\end{split}
		\end{align}
		where the ``$\cdots$'' part means that there are other contributions from $\sum_{\vert g\rangle}
		\frac{\langle f_+'  \vert  \mathcal{H}_{BdG}'  \vert g \rangle \langle g \vert  \mathcal{H}_{BdG}'  \vert f_-'\rangle}{(-m_0'-m_0')(-m_0'-E_g)}$.
	\end{itemize}
	
	In a brief conclusion, the second-order perturbation theory is necessary to derive the vortex Hamiltonian in Eq.~\eqref{sm-eq-votex-ham-h1}.

	\subsection{Numerical Simulations and Application to LiFeAs}
	
	Next, we perform numerical simulations for the BdG Hamiltonian with tilted vortex lines. Since the $C_4$ symmetry is broken, the simulation based on the Bessel function expansion is no longer valid. Instead, we can still use tight-binding model for the simulation by taking
	\begin{align}
		k_x \to \sin k_x, k_y \to \sin k_y \text{ and } k_x^2 \to 2(1-\cos k_x), k_y^2 \to 2(1-\cos k_y).
	\end{align}
	Note that $k_z$ is still preserved. For the low-energy vortex states, we expect our approximation is valid at relatively small $\phi$. This enables to calculate:
	\begin{itemize}
		\item The vortex spectrum $E_{qp}(k_z)$ at fixed $\mu$, $\delta_{so}$ and $\phi$. 
		\item The evolution of the minimal gap of $E_{qp}(k_z)$ by varying $\phi$.
		\item Topological vortex phase diagram as a function of $\mu$ and $\phi$. 
	\end{itemize}

	\begin{figure*}[t]
		\includegraphics[width=\textwidth]{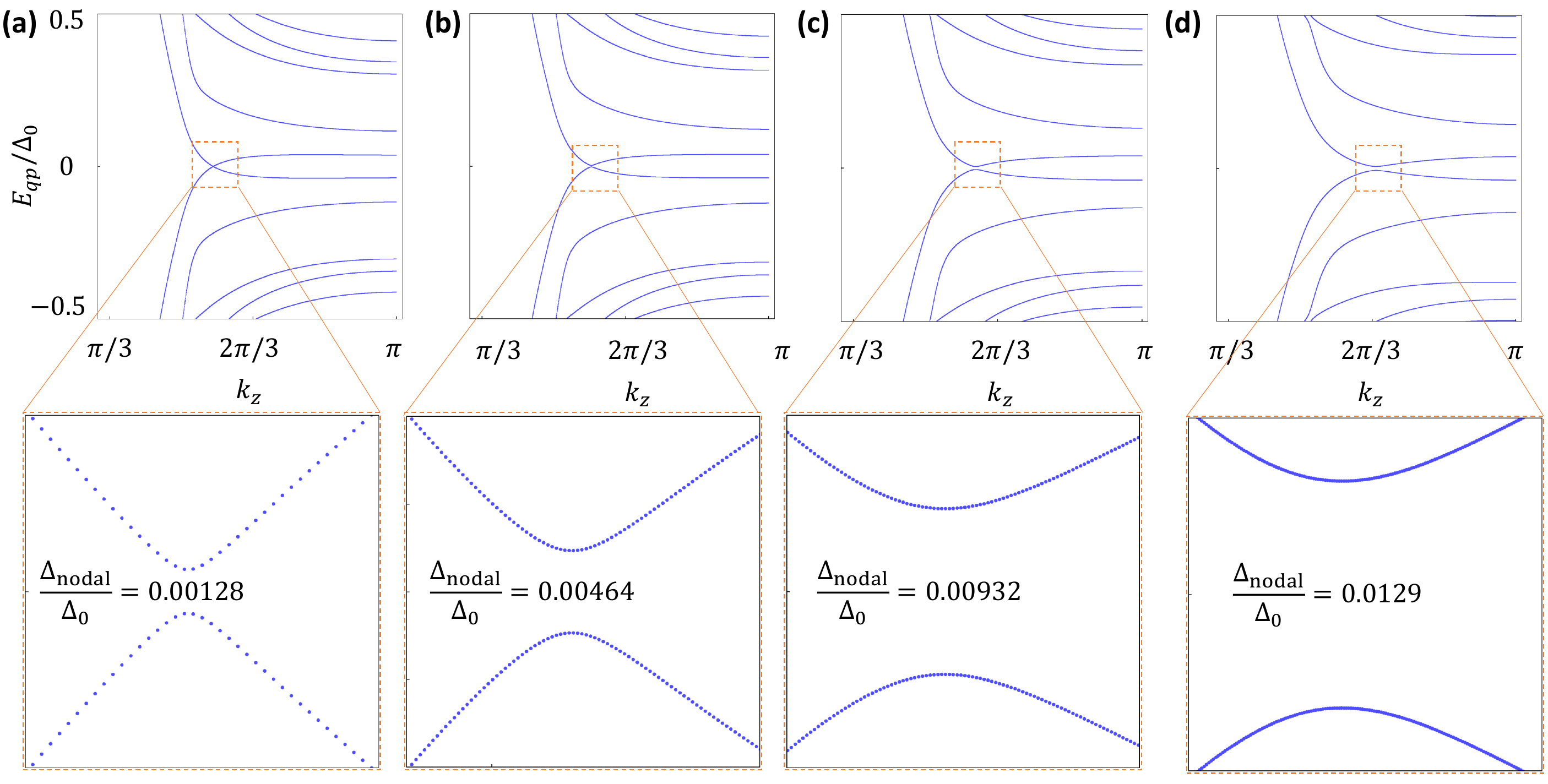}
		\caption{ The numerically calculated vortex spectrum $E_{qp}(k_z)/\Delta_0$. From these spectra, we can obtain the topological gap $\Delta_{\text{nodal}}/\Delta_0$ as a function tilting angle $\phi$ of external magnetic field. Here, $\phi$ is chosen as $5^\circ$ in (a), $10^\circ$ in (b), $15^\circ$ in (c) and $20^\circ$ in (d). The zoom-in figures show the topological gap opening for the nodal vortex due to the vortex-line tilting.
		}
		\label{sm-fig4}
	\end{figure*}

	First, we study the vortex spectrum. The vortex-line band structure evolution of $E_{qp}(k_z )$ by varying $\phi$ from $0^\circ$ to $20^\circ$ is shown in Fig.~\ref{sm-fig4}. Clearly, the zoom in figures in Fig.~\ref{sm-fig4} shows that $\Delta_{\text{nodal}}/\Delta_0$ increases by increasing $\phi$.
	
\begin{figure*}[!htbp]
		\includegraphics[width=0.5\textwidth]{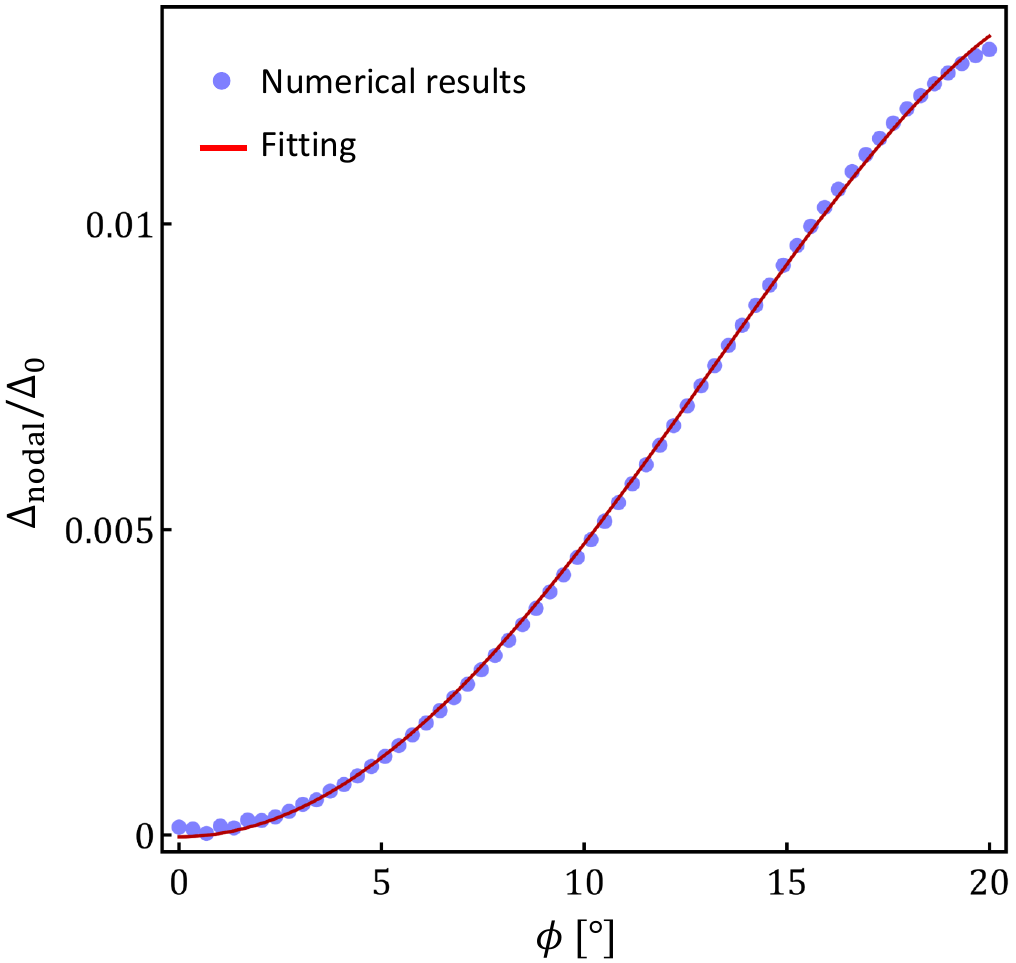}
		\caption{ The numerical calculated topological gap $\Delta_{\text{nodal}}/\Delta_0$ as a function tilting angle $\phi$. The calculated numerical results are labeled by the blue solid circles, while the red line is the fitting curve by $-3\times10^{-5}+5.3\times10^{-5}\phi^2-5\times10^{-8}\phi^4$. Note that the gap should be exactly zero at $\phi$=0, but a tiny numerical value is shown in this figure. 
		}
		\label{sm-fig5}
	\end{figure*}

	From these results, we find that $\Delta_{\text{nodal}}/\Delta_0 \approx  \phi^2$ as what we expect.	 Motivated by this observation, we are also able to figure out a more exacter relationship between $\Delta_{\text{nodal}}/\Delta_0$ and $\phi$. The numerical results is shown in Fig.~\ref{sm-fig5} for $\phi\in [0^\circ,20^\circ]$. The fitting (red line) shows good agreement with numerical results (blue solid circles),
	\begin{align}
		\frac{\Delta_{\text{nodal}}(\phi)}{\Delta_0} =	-3\times10^{-5}+5.3\times10^{-5}\phi^2-5\times10^{-8}\phi^4.
	\end{align}
	The tiny constant $-3\times10^{-5}$ is likely due to the numerical error and the $\phi^4$ dependence can be ignored at small $\phi$. This directly confirms the validity of our perturbation theory that $\Delta_\text{nodal}\sim \phi^2$. 
	
	As we mentioned earlier, $\Delta_\text{hybrid}\sim \phi$ implies that its energy scale should be one order of magnitude larger than $\Delta_{\text{nodal}}$. This leads to an estimate that $\Delta_{\text{hybrid}}$ should reach $0.1\Delta_0$ for a relatively small $\phi\sim 15\deg$. We note that $\Delta_0$ for LiFeAs is around $5$ meV, which directly leads to an estimate of surface Majorana splitting of $2\Delta_{\text{hybrid}}~1$ meV. This clearly is an observable thanks to the state-of-the-art STM resolution.
	
	Moreover, we can also semi-quantitatively discuss the hybridization strength between the Kitaev vortex states and nodal vortex states caused by the off-diagonal terms $m_3k_z\phi$ and $m_4\phi$ in Eq.~\eqref{sm-eq-votex-ham-h1} obtained from the first-order perturbation. We denote the hybridization energy scale as $\Delta_{\text{hybrid}}$, which is difficult to be directly figured out. However, one can use $\Delta_{\text{nodal}}$ as an energy-scale reference for $\Delta_{\text{hybrid}}$ at small $\phi$, since both of them are obtained form perturbation theory or symmetry construction. Furthermore, we have shown that the energy-scale for $\Delta_{\text{nodal}}$ is about $0.01\Delta_0$, which could imply that $\Delta_{\text{hybrid}}$ is order of $0.1\Delta_0$ that is comparable to the energy resolution in experiments ($\sim0.05\Delta_0$), such as the scanning tunneling microscope (STM). This $C_4$-breaking mechanism by tilting external magnetic field is the driving force for the trivialization of a hybrid vortex, thus providing a more concrete explanation to the experimental observation of LiFeAs (non-MZM is detected in a free vortex).
	
	With the above understanding of the role of tilting magnetic field on the vortex spectrum, we now study a new topological vortex phase diagram as a function of $\mu$ and $\phi$. And these two parameters can be feasibly controlled in experiments. Please notice that it has been experimentally demonstrated for the superconducting vortex matter in LiFeAs by utilizing a combination of vector magnetic field and scanning tunneling microscopy [e.g., see Fig. 3 and Fig. 4 in Ref.~\cite{zhangST_prb_2019}]. The results are shown in the main text, indicating that
	\begin{itemize}
		\item At non-zero $\phi$, the nodal vortex becomes topological Kitaev vortex. Moreover, the hybrid vortex becomes trivial vortex, because of the cancellation of two Kitaev vortices.  
		\item At small $\phi$, the critical chemical potentials for the topological vortex phase transitions have small modifications, while large $\phi$ has strong effects and even eliminates all the Kitaev vortex phases.
	\end{itemize}

	\section{Appendix E: Impurity-Induced Curved Vortex Line}
	\label{sm-section-v}
	
	In this section, we will discuss how a near-surface impurity will cause a curvature of vortex line and its impact on the vortex topology. This effect breaks $C_4$ and the translational symmetries simultaneously. Note that such impurities, if hidden underneath the surface, can be both invisible and inevitable. As a result, $\phi$ could locally reach a relatively large value, even if the magnetic field is carefully aligned with $\hat{z}$ direction.

	\begin{figure*}[t]
		\includegraphics[width=0.3\textwidth]{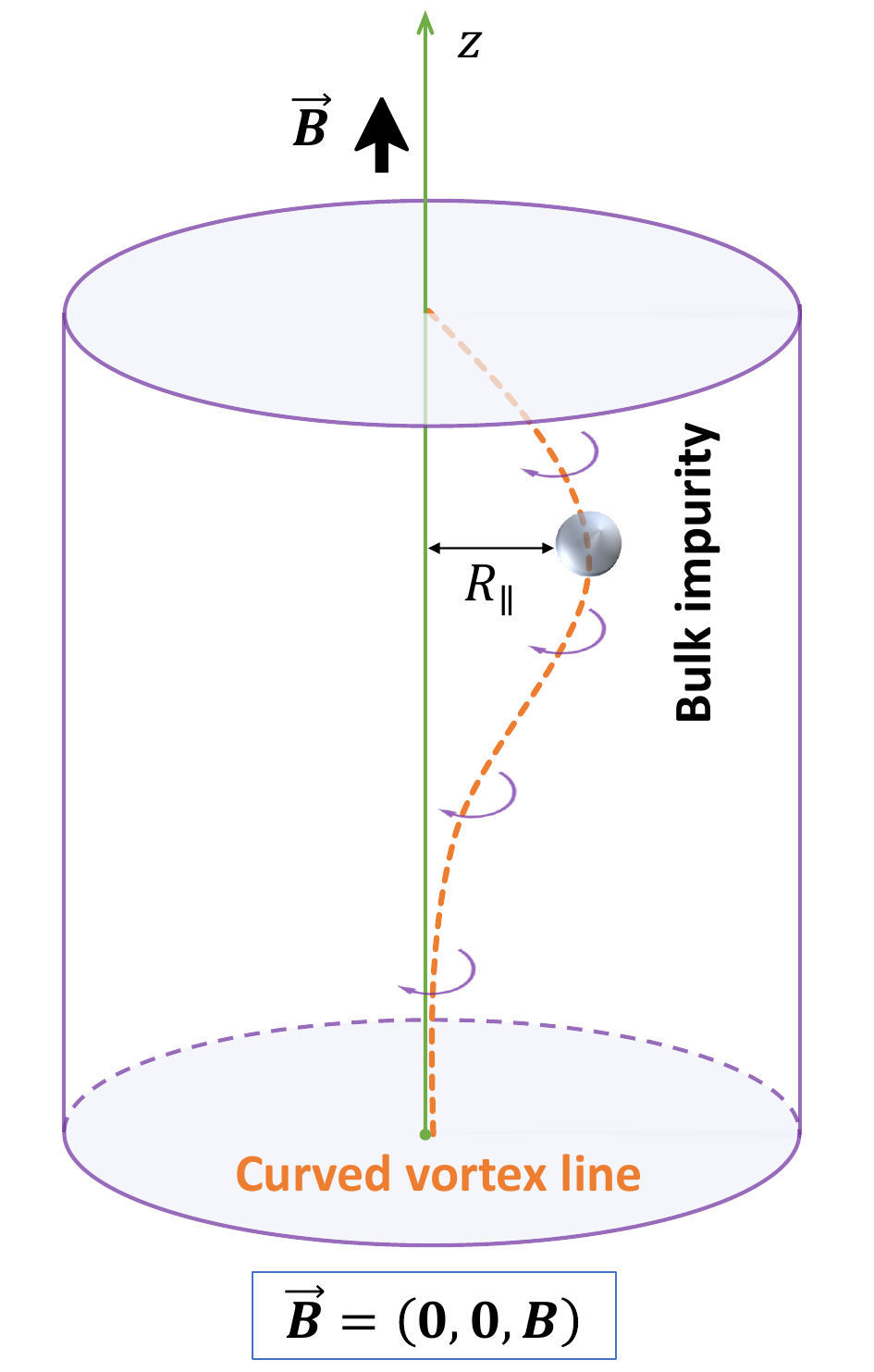}
		\caption{It shows the curved vortex line induced by bulk impurity due to its pinning effect. The vortex line (orange dashed line) is curved inside the bulk system, the deviation from the z-axis (green line) is represented by $R_\parallel$ and it reaches maximum around the bulk impurity (blue sphere). 
		}
		\label{sm-fig8}
	\end{figure*}

	\subsection{Effective Model of a Curved Vortex Line}
	
\begin{figure*}[t]
		\includegraphics[width=\textwidth]{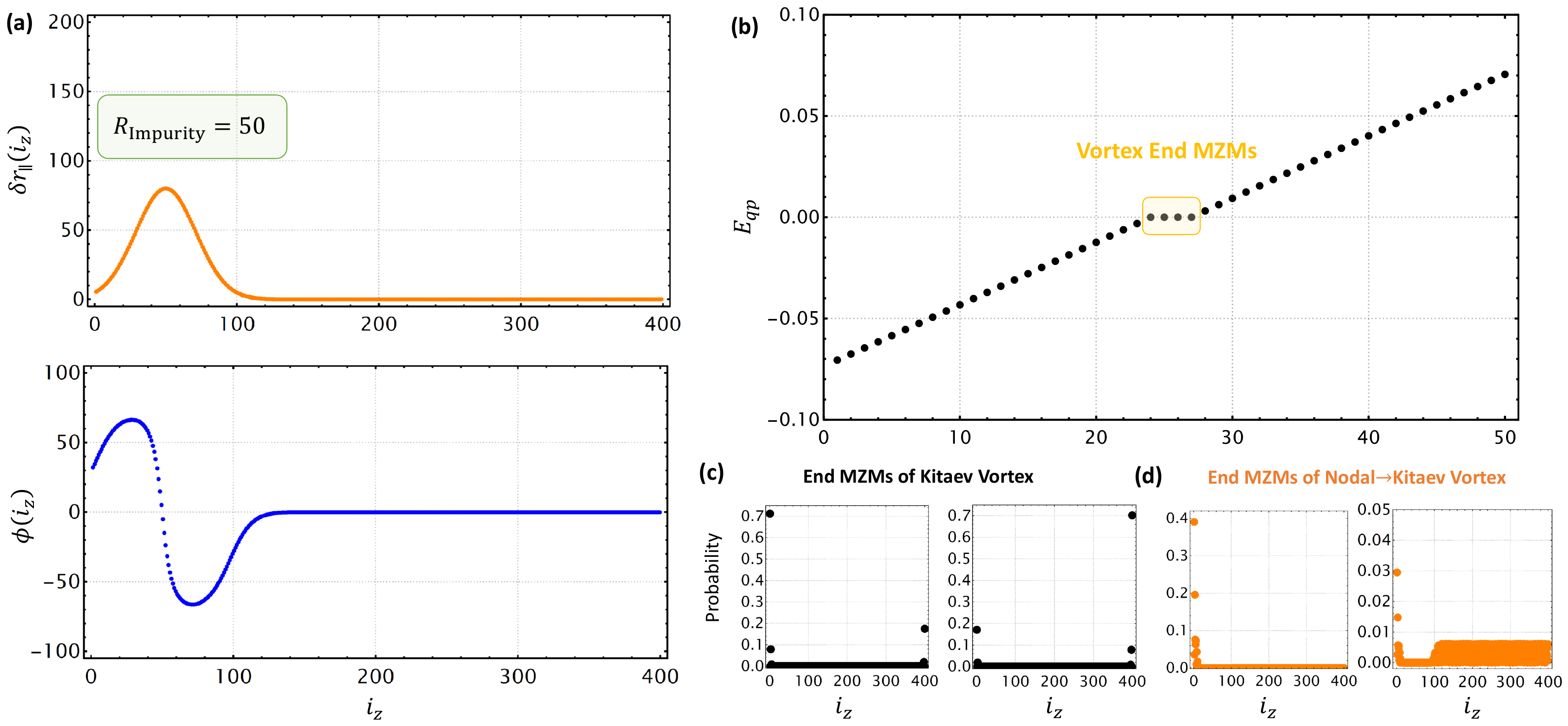}
		\caption{ The simulation of curved vortex line on vortex Majorana zero modes in the decoupling limit.  (a) shows the vortex-line distribution as a function of site $i_z$ for a bulk impurity illustrated in Fig.XXX. Solving the 1D lattice model, the energy spectrum with four Majorana zero modes is shown in (b). And the corresponding wave function distributions are plotted in (c) for $h_{\text{Kitaev}}^{n=0}$ and in (d) for $h_{\text{nodal}\to\text{Kitaev}}^{n=\pm 1}$.
		}
		\label{sm-fig6}
	\end{figure*}

	As illustrated in Fig.~\ref{sm-fig8}, the vortex line can be elastically distorted by a point pining force due to the bulk impurity~\cite{brandt_rpp_1995}. In this case, the vortex line at fixed $z$ deviates from the $z$-axis with an in-plane distance $\delta r_\parallel$, which reaches the maximal when it reaches the bulk impurity. And the curved vortex line may be parameterized by
	\begin{align} \label{sm-eq-vortex-line-z}
		\delta r_\parallel (z) = A e^{-\frac{(z-R_{\text{impurity}})^2}{B^2}}
	\end{align}
	where $R_{\text{impurity}}$ is the position of one bulk impurity, $A$ and $B$ is the position of one bulk impurity. Besides, the curved vortex line is also characterized by the inhomogeneous tilting angle $\phi$ even though the external magnetic field is strictly along the z-axis, which in turn breaks the translation symmetry. Based on Eq.~\eqref{sm-eq-vortex-line-z}, $\phi(z)$ is given by
	\begin{align}
		\phi(i_z) = \text{ArcTan}[ \delta r_\parallel (i_z + 1) - \delta r_\parallel (i_z)]
		\label{sm-eq-iz-phi}
	\end{align}
	where $i_z$ is the site for the 1D vortex line taken from $i_z\in [1,N_z]$ with $N_z$ is the number of layers. The lattice constant along z-axis is set to be $1$. For example, this leads to the spatial distribution of curved vortex line profile in Fig.~\ref{sm-fig6} (a) by using parameters 
	\begin{align}
		R_{\text{impurity}} = 50, N_z =400, A=80, B=30.
	\end{align}
	Since the impurity is near the top surface, the vortex line is curved with a non-zero $\phi$ only for $i_z\in[1,120]$ and remains vertical (i.e., $\phi=0$) for $i_z\in[120,400]$. To model the effects of curved vortex line on the vortex Majorana physics, we can map the 3D lattice simulation into a 1D lattice problem. We note that while this simulation approach contains many approximations, it is of great computational efficiency, and we do expect it to qualitatively capture the essential topological physics of the curved vortex line. In particular, we expect it to answer the following two questions:
	\begin{itemize}
		\item Starting from a nodal vortex, do translational-symmetry-breaking perturbations favor Kitaev or trivial vortex states?
		\item Can impurity-induced vortex-line distortion lead to a surface Majorana splitting for a hybrid vortex state?    
	\end{itemize}
	
	The 1D effective lattice model for a curved vortex line is given by, 
	\begin{align}
		\mathcal{H}_{1D} = \sum_{i_z}  \Psi^\dagger(i_z) [T_0(\phi)]
		\Psi(i_z)  + \Psi^\dagger(i_z) [T_z(\phi)] \Psi(i_z+1) + \text{h.c.},
	\end{align}
	where $\Psi(i_z) = \left\{ C_{f_+}(i_z) , C_{f_-}^\dagger(i_z), C_{f_+'}(i_z) , C_{f_-'}^\dagger(i_z)  \right\}^T$, and the on-site and hopping matrices are given by
	\begin{subequations}
		\begin{align}
			T_0(\phi) &=  \begin{pmatrix} 
				\chi_0 & 0 & 0 & m_4\phi \\
				0 & -\chi_0 &  -m_4^\ast\phi & 0 \\
				0 & -m_4\phi & \chi_1  & 0 \\
				m_4^\ast \phi & 0 & 0 & -\chi_1 \\
			\end{pmatrix}, \\ 
			T_z(\phi) &= \begin{pmatrix}
				-(m_1+m_6\phi^2) & \frac{m_2}{2} & \frac{m_3}{2}\phi & 0 \\
				\frac{m_2}{2}  & m_1+m_6\phi^2 & 0 & \frac{m_3^\ast}{2}\phi  \\
				\frac{m_3^\ast}{2}\phi  & 0 & -(m_1'+m_6'\phi^2) &  \frac{m_2'}{2}\phi^2 \\
				0 & \frac{m_3}{2}\phi & \frac{m_2'^\ast}{2} \phi^2 & m_1'+m_6'\phi^2  \\
			\end{pmatrix},
		\end{align}
	\end{subequations}
	where $\chi_0 = m_0+2m_1 + (m_5+ 2m_6)\phi^2 $ and $\chi_1=m_0'+2m_1' + (m_5'+ 2m_6')\phi^2$. Notice that $\phi$ now depends on $i_z$ for curved vortex line following Eq.~\ref{sm-eq-vortex-line-z} and~\ref{sm-eq-iz-phi}.

	\subsection{Gapping and Majorana Hybridization from Vortex Line Curvature}
	
	To answer the first question, we consider turning off the hybridization terms between $l_z=0$ and $L_z=\pm 1$ sectors, just to focus on whether a translational-symmetry-breaking vortex-line geometry can transform a nodal vortex into a Kitaev one. This is achieved by choosing model parameters as $m_0=1,m_1=-0.5,m_2=0.5,m_0'=0.5,m_1'=-0.3$, and $m_3=0,m_4=0,m_2'=0.2$ to semi-quantitatively study the vortex topology analysis. We also take $m_5=m_6=m_5'=m_6'=0$ for simplicity. In this case, the off-diagonal terms vanish so that $h_{\text{vortex}}(k_z,\phi)$ is simplified as 
	\begin{align}
		h_{\text{vortex}}(k_z,\phi) = \begin{pmatrix}
			h_{\text{Kitaev}}^{n=0} & 0 \\
			0 & h_{\text{nodal}\to\text{Kitaev}}^{n=\pm 1} \\
		\end{pmatrix},
	\end{align}
	Fig.~\ref{sm-fig6} (b) shows our simulated energy spectra, where we indeed find four Majorana zero modes, two for each end of the vortex line. Since the impurity we considered is near the left end, it only induces an energy gap locally, and that’s why we still have a continuous gapless spectrum that arises from the right-half of the nodal vortex. We summarize the features below:
	\begin{itemize}
		\item 
		The continues energy spectrum is due to the nodal vortex in the $i_z\in[120,400]$ range where $\phi(i_z)=0$.
		\item 
		Four Majorana zero modes: one pair is due to the Kitaev vortex for the $n=0$ subspace, and the other pair is attributed to the $n=±1$ subspace for the $i_z\in[1,120]$ range. 
	\end{itemize}
	
	We also plot the wave function distributions for those Majorana zero modes in Fig.~\ref{sm-fig6} (c) and (d). Due to the decoupling between $n=0$ vortex states and $n=±1$ vortex states, the $n=0$ subspace $h_{\text{Kitaev}}^{n=0}$ leads to two Majorana zero modes that are localized at $i_z=1$ and $i_z=N_z$, respectively, in Fig.~\ref{sm-fig6} (c). On the other hand, the $n=1$ subspace $h_{\text{nodal}\to\text{Kitaev}}^{n=\pm 1}$ becomes Kitaev vortex near the top surface ($i_z\in[1,120]$) so that a zero-dimensional Majorana zero mode is localized at $i_z=1$ in Fig.~\ref{sm-fig6} (d). The other zero mode is an extended mode in the $i_z\in[120,N_z]$ region, where the nodal vortex persists due to the vanishing $\phi$.

	\begin{figure*}[t]
		\includegraphics[width=0.85\textwidth]{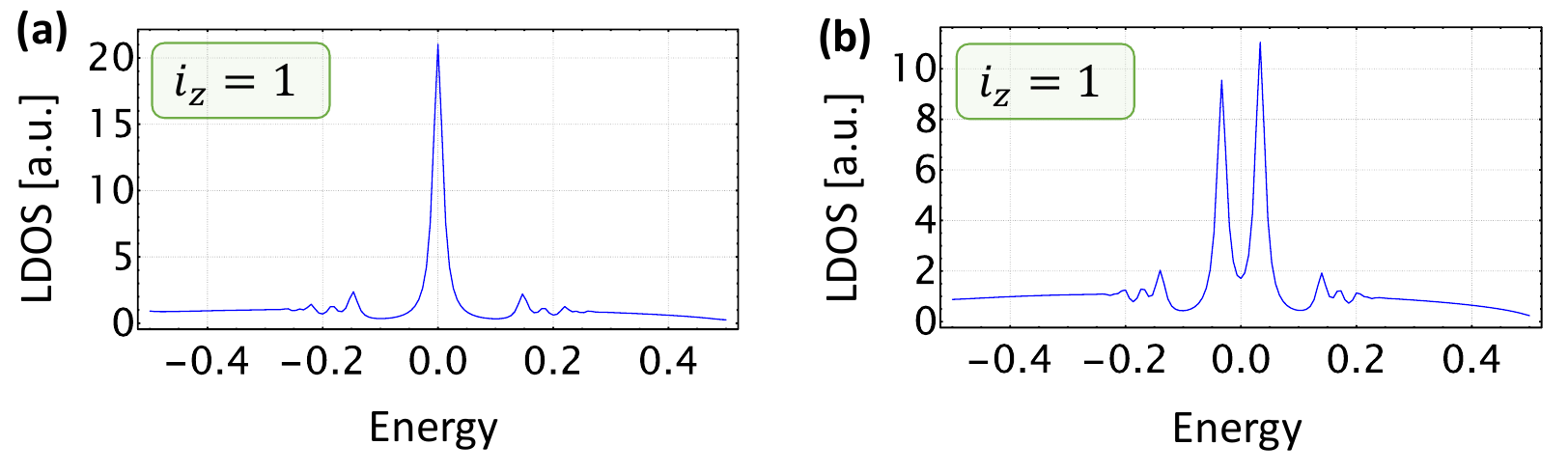}
		\caption{ The top-surface local density of states (LDOS) at $i_z=1$. The decoupling limit is shown in (a), while the coupling case is shown in (b). Comparing (a) and (b), the trivialization of a topological hybrid vortex is induced by the curved vortex line. The two sub-gap peaks in (b) indicates the spoiling of Majorana signals.
		}
		\label{sm-fig7}
	\end{figure*}

	Furthermore, we also calculate the local density of states (LDOS) via
	\begin{align}
		\text{LDOS}(i_z,\omega) = -\frac{1}{\pi} \text{Tr}[G(i_z,\omega)] ,
	\end{align}
	where the green function is given by $G(i_z,\omega)=\frac{1}{i\eta - \mathcal{H}_{1D}}$ with $\eta=0.01$ at zero temperature. The numerical result for the end-localized (i.e., surface for a 3D sample) LDOS at $i_z=1$ is shown in Fig.~\ref{sm-fig7} (a). The zero-bias peak in Fig.~\ref{sm-fig7} (a) shows the two decoupled Majorana zero modes, confirming that 
	\begin{itemize}
		\item Impurity-induced vortex line bending is capable of locally gapping the nodal vortex state and further leads to a Kitaev vortex state.
	\end{itemize}
	
	Now we are ready to turn on the coupling term between $l_z=0$ and $l_z=\pm1$ sectors, and test if the two end Majorana modes yield a hybridization. The calculated local density of states are plotted in Fig. Fig.~\ref{sm-fig7} (b) for $i_z=1$. Clearly, trivialization of a topological hybrid vortex is demonstrated by the split finite-bias peaks of the LDOS near the vortex-line end. This split peak pattern exactly reproduces the observed ``trivial" surface vortex signal in undoped LiFeAs~\cite{kong_nc_2021}. We thus conclude that
	\begin{itemize}
		\item When an impurity bends a hybrid topological vortex line near the surface, it can lead to an energy splitting of surface vortex Majorana modes.
	\end{itemize}

	\section{Appendix F: Vortex Topology in CaKFe$_4$As$_4$}
	
\begin{figure*}[t]
		\includegraphics[width=0.98\textwidth]{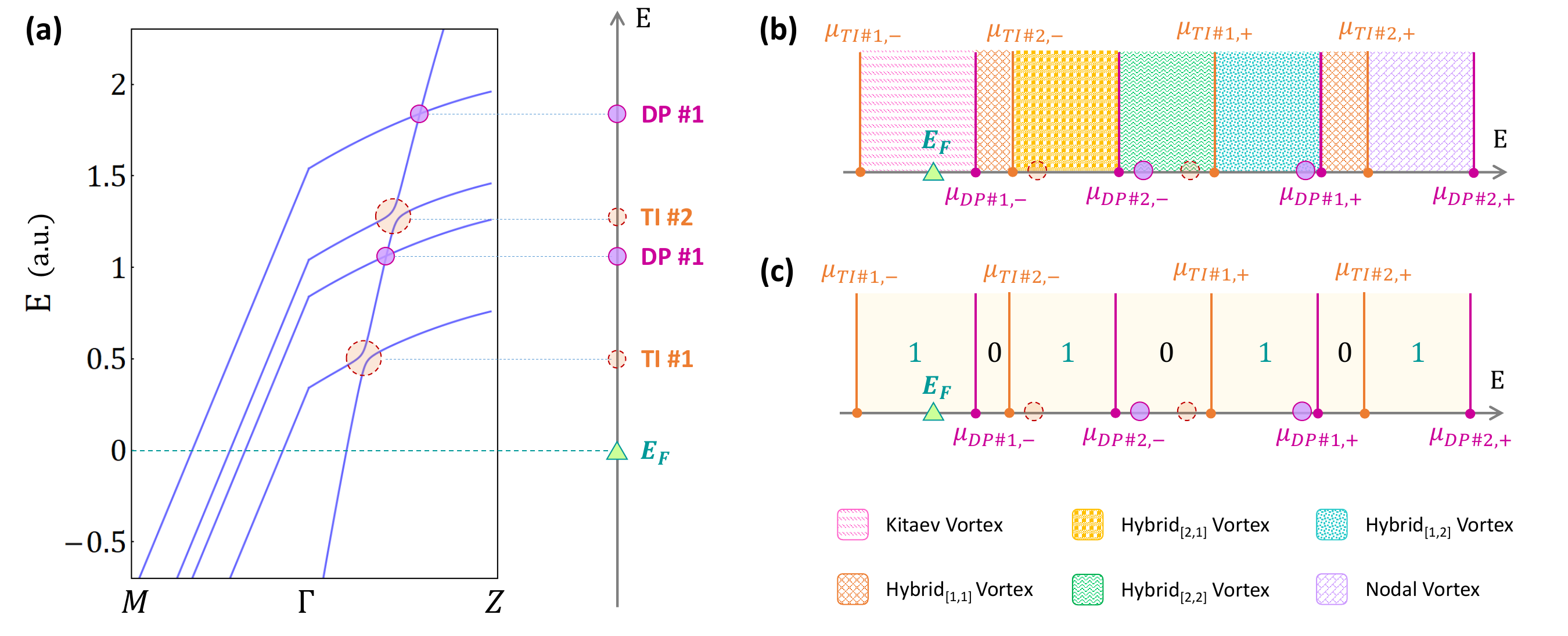}
		\caption{A schematic plot for CaKFe$_4$As$_4$ band structure is shown in (a), where both bulk Dirac nodes and TI gaps are highlighted. A possible vortex topological phase diagram for CaKFe$_4$As$_4$ is shown in (b). The corresponding vMBS number for each vortex phase is shown in (c), where $C_4$ breaking effect has been taken into account.}
		\label{sm-fig9}
	\end{figure*}

	In this section, we discuss the vortex topological phase diagram for CaKFe$_4$As$_4$. As pointed out by Ref.~\cite{liu_nc_2020}, CaKFe$_4$As$_4$ features an interesting bilayer Fe-As structure and manifests as an alternate stacking of CaFe$_2$As$_2$ and KFe$_2$As$_2$. Such a bilayer structure of CaKFe$_4$As$_4$ leads to a Brillouin zone folding along $k_z$ when comparing to its parent compounds Ca(K)Fe$_2$As$_2$, and further enables two copies of TI and DSM bands. A schematic sketch for normal-state band structure for CaKFe$_4$As$_4$ is shown in Fig.~\ref{sm-fig9} (a), following the DFT+DMFT calculation in Ref.~\cite{liu_nc_2020}. Note that TI \#2 and DP \#2 are essentially a duplicate of TI \#1 and DP \#2, thanks to the Brillouin zone folding, where ``DP" is short for bulk Dirac points. 
	
	The key to map out the vortex phase diagram lies in the identification of phase boundaries. As we have discussed in Fig.~3 in the main text, each set of TI or DSM bands will contribute to a pair of phase boundaries. As a result, the vortex phase diagram for CaKFe$_4$As$_4$ necessarily consists of 8 critical chemical potentials as the phase boundaries, which we denote as $\mu_{\xi,\pm}$ and $\xi\in\{\text{TI\#1,\ DP\#1,\ TI\#2,\ DP\#2}\}$. Therefore, the phase diagram is completely determined by the energy sequence of all eight $\mu_{\xi,\pm}$s.

	\begin{table}[!htbp]
		\begin{ruledtabular}
			\begin{tabular}{cccc }
				Material & $\delta^T_{so}$ (meV) & $\delta^E_{so}$ (meV) & Vortex Phase    \\
				\colrule
				FeTe$_{0.5}$Se$_{0.5}$ &  70  &   35  & Kitaev      \\
				LiFeAs                  & 45    &  10.7 & Hybrid \\
				CaKFe$_4$As$_4$         & 50      & N/A & Kitaev \\
			\end{tabular}
		\end{ruledtabular}
		\caption{\label{soc} Effective model related parameters from theoretical calculations and experimental measurements for typical tFeSCs. $\delta^{T/E}_{so}$ denotes the theoretical/experimental splitting of $d_{xz/yz}$ band owing to spin-orbit coupling. }
	\end{table}

	Notably, such a sequence sensitively depends on the competition among the following energy scales:
	\begin{itemize}
		\item $\delta_{so}$: the energy splitting between TI \#1 and DSM \#1 (or equivalently TI \#2 and DSM \#2);
		\item $\delta_{t}$: the energy splitting between TI \#1 and TI \#2;
		\item $\delta_{\mu}$: the energy difference between $\mu_{\xi,+}$ and $\mu_{\xi,-}$.
	\end{itemize} 
	Practically, it is of technical difficulty for us to obtain accurate values of the above quantities, especially due to the strong electron correlations in CaKFe$_4$As$_4$ and the lack of experimental data. Nonetheless, we can make some rough estimate based on the existing DFT+DMFT calculation and ARPES data. We find that 
	\begin{equation}
		\delta_{so}\sim \delta_{t} \sim 50\text{ meV},\ \delta_{\mu}\geq 20\text{ meV}. 
	\end{equation}
	Values for $\delta_{so}$ of other tFeSC candidates can be found in Table.~\ref{soc}. Here the lower bound for $\delta_{\mu}$ is estimated based on the observation that vMBS signal exists for CaKFe$_4$As$_4$, even though the Fermi level is found to be $20$ meV below the surface Dirac point in experiment. We emphasize that a concrete prediction of $\delta_{\mu}$ will require a first-principles-based vortex spectrum calculation with correlation effects being carefully included, which is beyond the scope of our work. Nonetheless, based on the large $\delta_{\mu}$ found in our six-band minimal model (see Fig.~1 in the main text) and the bandwidth of TI bands in CaKFe$_4$As$_4$, we expect that $\delta_{\mu}$ should be much greater than 20 meV in the actual material.

	Assuming $\delta_{\mu}>\delta_{so}\sim \delta_{t}$, we schematically show in Fig.~\ref{sm-fig9} (b) a possible vortex topological phase diagram for CaKFe$_4$As$_4$, which contains 7 topologically distinct vortex phases. Compared with the phase diagram in the main text, we now have several new types of hybrid topological vortex states termed ``hybrid$_{[m,n]}$ vortex", which is essentially a superposition of $m$ Kitaev vortices and $n$ nodal vortices. According to this new notation, the hybrid vortex in the main text is hybrid$_{[1,1]}$ vortex by definition. In Fig.~\ref{sm-fig9} (c), we list the number of vMBS for each vortex state when $C_4$ breaking effect is considered. Since the Fermi level is below the surface Dirac point of TI \#1, the most probable Majorana-carrying vortex state for CaKFe$_4$As$_4$ is the Kitaev vortex phase, as indicated in Fig.~\ref{sm-fig9} (b) and (c). Remarkably, if we can continuously lift the Fermi level by electron doping, we are expected to see an interesting oscillation of vMBS number as a function of $E_F$.
	
\end{document}